\documentclass[12pt]{article}                              

\usepackage{amsmath,amsthm,amsfonts,amscd,amssymb,eucal,latexsym}
\newlength{\dinwidth}
\newlength{\dinmargin}
\def\rest{\upharpoonright}

\numberwithin{equation}{section}

\def\cA{{\cal A}}
\def\cB{{\cal B}}
\def\cC{{\cal C}}

\def\cF{{\cal F}}
\def\cG{{\cal G}}
\def\cH{{\cal H}}

\def\cK{{\cal K}}
\def\cL{{\cal L}}
\def\cM{{\cal M}}
\def\cN{{\cal N}}
\def\cO{{\cal O}}
\def\Oo{{\cal O}}
\def\cP{{\cal P}}

\def\cS{{\cal S}}
\def\cT{{\cal T}}

\def\cW{{\cal W}}
\def\cX{{\cal X}}

\def\bC{{\mathbb C}}

\def\bN{{\mathbb N}}
\def\NN{{\mathbb N}}
\def\bP{{\mathbb P}}
\def\bR{{\mathbb R}}
\def\RR{{\mathbb R}}
\def\bT{{\mathbb T}}
\def\bZ{{\mathbb Z}}

\def\a{\alpha}
\def\b{\beta}
\def\g{\gamma}        
\def\d{\delta}        \def\D{\Delta}
\def\eps{\varepsilon} 
     
\def\z{\zeta}
\def\th{\vartheta}    
\def\k{\kappa}
\def\l{\lambda}       \def\L{\Lambda}

\def\x{\xi}
\def\p{\pi}
\def\r{\rho}
\def\s{\sigma}
    
\def\S{\Sigma}
\def\t{\tau}

\def\f{\varphi}

\def\o{\omega}        \def\O{\Omega}

\def\fA{{\mathfrak A}}

\def\fP{{\mathfrak P}}

\def\imply{\Rightarrow}

\def\ov{\overline}

\def\Gp{G_+}            
\def\Gpo{G^\uparrow_+}  
\def\Gpnot{G_0}         
\def\Gcov{\widetilde{G}}  
\def\Gd{G_1}            
\def\Gdcov{\widetilde{G}_1} 
\def\j{{\mathbf j}}

\def\h{\mathfrak h}
\def\Wtilde{\widetilde W}
\newtheorem{Thm}{Theorem}[section]
\newtheorem{Cor}[Thm]{Corollary}
\newtheorem{Prop}[Thm]{Proposition}
\newtheorem{Lemma}[Thm]{Lemma}
\theoremstyle{definition}

\theoremstyle{remark}
\newtheorem{rem}[Thm]{Remark} 
 
\begin{document}
${}$ \par ${}$ \vspace*{-2.2cm} ${}$ \par \noindent
\begin{center}
{\Large \bf Charged Sectors, Spin and Statistics\\[10pt] in
    Quantum Field Theory on Curved Spacetimes}
\\[30pt]
{\sc D. Guido$^*$\footnote[1]{Supported by GNAFA and MURST.}, 
R. Longo$^{**1}$, J.E. Roberts$^{**1}$, R. Verch$^{***}$}
\\[24pt]
{\normalsize $^*$ Dipartimento di Matematica, }\\
{\normalsize Universit\`a della Basilicata,}\\
{\normalsize I-85100 Potenza, Italy}\\
{\normalsize e-mail: guido$@$unibas.it}
\\[6pt]
{\normalsize $^{**}$ Dipartimento di Matematica, }\\
{\normalsize Universit\`a di Roma ``Tor Vergata'',}\\
{\normalsize I-00133 Roma, Italy}\\
{\normalsize e-mail: longo$@$mat.uniroma2.it,}\\
{\normalsize roberts$@$mat.uniroma2.it}
\\[6pt]
{\normalsize ${}^{***}$ Institut f\"ur Theoretische Physik,}\\
{\normalsize Universit\"at G\"ottingen,}\\
{\normalsize  D-37073 G\"ottingen, Germany}\\
{\normalsize e-mail: verch$@$theorie.physik.uni-goettingen.de}
\end{center}
\date{\today}
${}$\\[22pt]
\newcommand{\Cin}{C^{\infty}}
\newcommand{\Coin}{C_{0}^{\infty}}
\newcommand{\hh}{{\sf h}}
\newcommand{\rr}{{\sf r}}
\newcommand{\lcrc}{ \mbox{\footnotesize $\circ$}}
\newcommand{\sG}{\mbox{\small $\Gamma$}}

\noindent
{\small  {\bf Abstract:}
The first part of this paper extends the
 Doplicher-Haag-Roberts theory of superselection sectors 
 to quantum field theory on arbitrary globally
 hyperbolic spacetimes. The statistics of a
 superselection sector may be defined as in flat spacetime 
 and each charge has a conjugate charge when the
 spacetime possesses non-compact Cauchy surfaces. In this case,
 the field net and the gauge group can be constructed as in
 Minkowski spacetime.

 The second part of this paper
 derives spin-statistics theorems on spacetimes with
 appropriate symmetries. Two situations are
 considered: First,
 if the spacetime has a bifurcate Killing horizon, as is the 
case in the presence of black holes,
then restricting the observables to the Killing horizon together with
``modular covariance'' for the Killing flow yields a 
 conformally covariant quantum field theory on the circle 
 and a conformal spin-statistics theorem for charged sectors
 localizable on the Killing horizon. Secondly, if
 the spacetime has a rotation and PT symmetry like 
 the Schwarzschild-Kruskal black holes, ``geometric modular action'' 
 of the rotational symmetry leads to a
 spin-statistics theorem for charged covariant sectors where the
 spin is defined via the $SU(2)$-covering of the spatial
 rotation group $SO(3)$.}
\newpage
\noindent
\begin{center}
{\Large \bf Table of Contents}
\end{center}
\begin{tabular}{lr}
1. Introduction & 2\\
${}$\quad \footnotesize 1.1 General Setting & \footnotesize 3\\
${}$\quad \footnotesize 1.2 Superselection Sectors & \footnotesize 5\\
${}$\quad \footnotesize 1.3 Covariant Sectors and Univalence (Spin) &
\footnotesize 7\\
${}$\quad \footnotesize 1.4 Tomita-Takesaki Theory and Symmetry &
\footnotesize 7\\
${}$\quad \footnotesize 1.5 Modular Inclusion and  
Conformal Theories on the Circle &\footnotesize 8\\
${}$\quad \footnotesize 1.6 Description of Contents & \footnotesize 9\\[4pt]
2. Some Spacetime Geometry & 10\\
${}$\quad  \footnotesize 2.1 Generalities & \footnotesize 10\\
${}$\quad \footnotesize 2.2 Appendix to Chapter 2 & \footnotesize 14\\[4pt]
3. Superselection Structure in Curved Spacetimes & 15\\
${}$\quad \footnotesize 3.1 Introduction & \footnotesize 15 \\
${}$\quad \footnotesize 3.2 The Selection Criterion & \footnotesize 17 \\
${}$\quad \footnotesize 3.3 Localized Endomorphisms &  \footnotesize 18\\
${}$\quad \footnotesize 3.4 The Left Inverse and Charge Transfer &
\footnotesize 23\\
${}$\quad \footnotesize 3.5 Sectors of a Fixed-Point Net &
\footnotesize 25\\
${}$\quad \footnotesize 3.6 Appendix to Chapter 3 & \footnotesize 27\\[4pt]
4. The Conformal Spin and Statistics Relation   & {}\\
${}$\quad\, for Spacetimes with Bifucate Killing Horizon & 37\\
${}$\quad \footnotesize 4.1 Spacetimes with bKh & \footnotesize 38\\
${}$\quad \footnotesize 4.2 Conformal Spin-Statistics Relation
& \footnotesize 42\\
${}$\quad \footnotesize 4.3 Appendix to Chapter 4 & \footnotesize 49\\[4pt]
5. The Spin and Statistics Relation for Spacetimes with Rotation
Symmetry & 51\\
${}$\quad \footnotesize 5.1 Geometric Assumptions & \footnotesize 51\\
${}$\quad \footnotesize 5.2 Quantum Field Theories on Spacetimes with
Rotation Symmetry & \footnotesize 56\\
${}$\quad \footnotesize 5.3 Appendix. Equivalence between local and
global intertwiners in Minkowski spacetime & \footnotesize 65\\[4pt]
Acknowledgements & 68\\[4pt]
References & 69
\end{tabular}
 \setcounter{section}{0}
\section{Introduction.}\label{sec:zeroth}

General Relativity is a theory of gravitation with a geometric 
interpretation. A solution to the Einstein--Hilbert equations describes 
a {\it curved} spacetime manifold, whose
curvature is related to the distribution of matter.

Quantum Field Theory on the other hand arose as a theory for 
describing finitely many elementary particles and the underlying mathematical
structure is that of a net of {\it noncommutative} von Neumann algebras of 
local observables. 

There have been many attempts to fuse the two theories to obtain a 
theory of Quantum Gravity but, as is well known, the 
basic problems remain unsolved and their solution would seem 
to be still a long way off.

There is however one theory describing the effects of gravitation on 
quantum systems and this is Quantum Field Theory on a Curved Spacetime, 
where the gravitational field is treated as a background field so 
that the backreaction of the quantum system is ignored.  
Of course, this approximation cannot be expected to remain valid down 
to distances comparable to the Planck length.

Progress in the field was initially hampered not only by 
the difficulties of handling interactions, well known from 
Minkowski space, but also through using a mathematical
 formalism which was
not general enough. Nor did it help that no really 
interesting physical effects were found. 
 This last point changed dramatically with
the advent of Black Hole Thermodynamics and more particularly 
with the well known Hawking effect whereby a quantum effect causes 
a black hole to radiate thermally \cite{Haw,Wald2}.

More recently, the field has evolved rapidly on the mathematical
side, too, primarily thanks  to adopting 
methods and concepts from algebraic quantum field theory as e.g.\ in 
the work of \cite{FreHa,KayWald,Haag,Long4,Wald2}.
 But there have been other important developments, 
too. In particular, the discovery by Radzikowski that the Hadamard 
condition is equivalent to a wavefront set condition
\cite{Rad,BFK} is worth mentioning. This has led
to ambitious rigorous work on perturbative quantum field theory in
curved spacetime by Brunetti and Fredenhagen \cite{BrFre}.  
Very recent work in algebraic quantum field theory \cite{BuFS,Reh}
contributes to clarifying the structure of quantum field theories on
anti-de\,Sitter spacetime and its conformal boundary, an issue which
has nowadays attracted great attention.

The DHR analysis of superselection sectors in Minkowski spacetime is a 
good illustration of the effectiveness of algebraic quantum 
field theory in treating structural and conceptual problems. The aim of 
this paper is to lay the foundations of superselection
theory in quantum field theory on curved spacetimes and to derive some
first results.

We find it advantageous to proceed by recalling, for the benefit of
the non-expert reader, the basic ideas and features of algebraic
quantum field theory relevant to the two main themes
of this paper: the general theory of superselection selection sectors
and the connection between Tomita-Takesaki modular theory of von
Neumann algebras and spacetime symmetries, particularly in the context
of covariant superselection sectors. Our presentation will be 
simplified, with full details appearing in the main body of the paper. 
 Readers familiar with superselection theory and the
relations between modular theory and symmetry in algebraic quantum
field theory may wish to turn directly to the
outline of the contents in Sec.\ 1.6 
where relations to other papers are indicated.
\subsection{ Algebraic Quantum Field Theory on
 Curved Spacetimes: General Setting}
In formulating algebraic quantum field theory on a curved
spacetime one assumes the underlying spacetime 
to be described by a smooth
manifold $M$ (of any dimension $\ge 2$) together with a Lorentzian
metric $g$. The quantum system in question is supposed to be
described by an inclusion preserving map $\cK \owns \cO \mapsto
\cA(\cO)$ assigning to each member $\cO$ in a collection $\cK$ of
subregions of $M$ a $C^*$-algebra $\cA(\cO)$. Usually, $\cK$ is chosen
to be a base for the topology of $M$ (we will specify $\cK$ later on).

The motivating idea is that $\cA(\cO)$ contains the
observables which can be measured at times and locations within
the spacetime region $\cO$ and that the way these algebras relate
to each other for different regions $\cO$ essentially fixes the
physical content of the theory \cite{Haag}

The collection $\cK$ of subregions of $M$ need not be directed 
under inclusion, but we shall nevertheless refer to
$\cK \owns \cO \mapsto \cA(\cO)$ as a {\it net of local
  algebras}. If $\cK$ is directed, then one can form the 
``quasilocal algebra'', i.e.\ the smallest $C^*$-algebra
containing all the local algebras $\cA(\cO)$. It is
the norm closure of the union of the local algebras,
$\overline{\bigcup_{\cO}\cA(\cO)}$. In the generic case where
$\cK$ is not directed, this possibility is denied to us. 
But one can still expect Hilbert space representations of the 
inclusion-preserving map $\cK \owns \cO \mapsto \cA(\cO)$. 
More precisely, we say that a {\it representation} of $\cK \owns \cO \mapsto
\cA(\cO)$ is a consistent family $\{\p_{\cO}\}_{\cO\in \cK}$ of 
representations of the local algebra $\cA(\cO)$ by bounded
operators on a common Hilbert space $\cH^{\p}$, 
i.e.\ $\p_{\cO_1}\rest{\cA(\cO)} =
\p_{\cO}$ whenever $\cO_1 \supset \cO$.

For the known examples of quantum field theories on globally
hyperbolic spacetimes and (conformal) quantum field theories on $S^1$,
such representations exist in abundance. (There are indications to
the contrary for non-globally-hyperbolic spacetimes \cite{KRW,KayRP}.
The present paper is restricted to quantum field theory on
globally hyperbolic spacetimes and the above notion of
representation suffices.) Every
representation $\{\p_{\cO}\}_{\cO \in \cK}$ yields states on
the local algebras $\cA(\cO)$ since each normal state $\o$ on
$\cB(\cH^{\p})$ restricts to a state
$$ \o_{\cO}(A) := \o(\p_{\cO}(A))\,, \quad A \in \cA(\cO)\,.$$
of the local algebra.
Not every consistent family of local states corresponds to a physical 
state of
the system; nor can all representations of the observable net
be considered as physical so one needs criteria to select
physical representations. In practice, one begins with some 
collection of physical representations and uses them to construct others. 
In what follows, we compile a brief list of
criteria to be fulfilled by such an initial collection $\fP$ of physical
representations of the net $\cK \owns \cO \mapsto \cA(\cO)$ of
local observables on a curved spacetime $(M,g)$.
\\[6pt]
1) $\p_{\cO}$, $\cO \in \cK$ is faithful for each $\{\p_{\cO}\}_{\cO
  \in \cK} \in \fP$. Otherwise the description of the system by the
net of local algebras $\cK \in \cO \mapsto \cA(\cO)$ would contain
redundancies.
\\[6pt]
2) {\it Locality: } The algebras $\pi_{\cO}(\cA(\cO))$ and
$\p_{\cO'}(\cA(\cO'))$ commute elementwise if the regions $\cO$ and
$\cO'$ cannot be connected by a causal curve.
\\[6pt]
3) {\it Irreducibility and Duality: } $\fP$ consists of irreducible
representations, i.e.\ representations $\{\p_{\cO}\}_{\cO \in \cK}$
fulfilling \footnote{$\cA' = \{B \in \cB(\cH): BA=AB\ \forall\,A \in
  \cA\}$ denotes the commutant of $\cA \subset \cB(\cH)$.}
 $$ \{\,\bigcup_{\cO \in \cK}\p_{\cO}(\cA(\cO))\,\}' = \bC\,1\,.$$
These representations are required to fulfill {\it essential duality},
i.e.\ the net,
$$ \cK \owns \cO \mapsto \cA_{\p}^d(\cO) := \bigcap_{\cO_1}
\p_{\cO_1}(\cA(\cO_1))'\,, $$
is local, where the intersection is taken over all $\cO_1 \in \cK$
causally disjoint from $\cO$.

This property is stronger than locality but not as strong as
Haag duality which demands that $\p_{\cO}(\cA(\cO))'' =
\cA_{\p}^d(\cO)$ for all $\cO \in \cK$. This latter property means
that the von Neumann algebras $\p_{\cO}(\cA(\cO))''$ cannot 
be enlarged by adding elements of $\cB(\cH^{\p})$ without
violating the locality condition.
\\[6pt]
4) {\it Local Equivalence: } Whenever $\{\p_{\cO}\}_{\cO \in \cK}$ and
$\{\p'_{\cO}\}_{\cO \in \cK}$ are two members of $\fP$, there is for
each $\cO \in \cK$ a unitary $U_{\cO} : \cH^{\p} \to \cH^{\p'}$ such
that 
 $$ U_{\cO} \p_{\cO}(A) = \p'_{\cO}(A)U_{\cO}\,, \quad A \in
 \cA(\cO)\,.$$
5) {\it Covariance: } For each $\{\p_{\cO}\}_{\cO \in \cK} \in \fP$
there is an (anti-)unitary \footnote{That is, $U_{\p}(\g)$ is
  anti-unitary if $\g$ reverses the time-orientation, otherwise it is
  unitary}
$G \owns \g \mapsto U_{\p}(\g)$ of a (subgroup of) the spacetime
isometry group $G$ on $\cH^{\p}$ so that
$$ U_{\p}(\g)\p_{\cO}(\cA(\cO))U_{\p}(\g)^* = \p_{\g
  \cO}(\cA(\g\cO))\,, \quad \g \in G,\ \cO \in \cK\,.$$
Obviously, if the underlying spacetime $(M,g)$ has a
trivial isometry group, this condition is void.

If $(M,g)$ is Minkowski spacetime, there is
typically a distinguished vacuum representation $\p_{\rm
  vac}$ in $\fP$ which is irreducible and covariant and possesses a cyclic
vacuum vector $\O_{\rm vac} \in \cH^{\p_{\rm vac}}$ invariant
under the action of $U_{\p_{\rm vac}}$. Moreover, a vacuum
representation fulfills the spectrum condition, i.e.\ the
time-translations in any Lorentz frame have positive generator. In
more general spacetimes, one can usually not select a distinguished
vacuum representation by similar requirements since, in the absence of a 
sufficiently large isometry group, there is no analogue of the vacuum 
vector nor of the spectrum condition. However, one expects that a
collection of physical representations  $\fP$ can still be selected
in quantum field theory on curved spacetimes, even if 
there is no single preferred representation. For a
Klein-Gordon field on any four-dimensional globally hyperbolic
spacetime the representations induced by pure quasifree Hadamard states have 
been shown to form a collection $\fP$ satisfying the
conditions listed above \cite{Ver1}.
\subsection{Superselection Sectors}
We assume now that a curved spacetime $(M,g)$, 
a net $\cK \owns \cO \mapsto \cA(\cO)$ of local algebras on this
spacetime background and a collection $\fP$ of physical
representations fulfilling the conditions stated above has been
given.
 To simplify notation, we denote a
 representation $\{\p_{\cO}\}_{\cO \in \cK}$ of the net of local
 algebras simply by $\p$.

Picking an irreducible physical representation $\p^0\in\fP$ as
reference, another irreducible representation $\p$ (not
necessarily belonging to $\fP$)
is said to satisfy the {\it selection criterion for localizable
  charges} if, given $\cO \in \cK$, there is a unitary $V_{\cO}$
between the representation Hilbert spaces $\cH^{\p}$ and $\cH^{\p^0}$
such that 
$$ V_{\cO} \p_{\cO_1}(A) = \p_{\cO_1}^0(A)V_{\cO}\,, \quad A \in
{\cA}(\cO_1)\,,$$
for all regions $\cO_1 \in \cK$ causally disjoint from
$\cO$.
Irreducible representations which fulfill this selection criterion and are
globally unitarily equivalent are said to carry the same {\it charge},
or to define the same {\it superselection sector}.

The selection criterion thus selects representations $\p$ 
differing from the reference representation by some ``charges'' which can
be localized in any
 spacetime region $\cO$ (and are then not detectable in
spacetime regions situated acausally to $\cO$). This form of
localizability does not apply to all kinds of charges, e.g.\ electric
charge is not localizable in this way (cf.\ \cite{Haag} and
references therein for further discussion). Yet for
certain general types of charges, like flavours in strong
interactions, this description is appropriate and hence a
useful starting point.

The notion of localized charge and superselection
sector now apparently depends on the
chosen reference representation $\p^0$ (typically the vacuum representation 
in the case of flat spacetime), but as physical representations are 
required to be locally equivalent, the charge structure, being
given by the structure of the space of intertwining operators
of representations fulfilling the selection criterion, is expected 
to be independent of that choice. Here, a bounded operator $T:
\cH^{\p} \to \cH^{\p'}$ is called an {\it intertwiner} for the
representations $\p$ and $\p'$ of $\cK \owns \cO \mapsto \cA(\cO)$ if
$$ T\p_{\cO}(A) = \p_{\cO}'(A)T \,, \quad A \in {\cA}(\cO)\,, \ \
\cO \in \cK\,.$$
 
A crucial point is that the space of intertwiners admits a product
having the formal properties of a tensor product. The {\it
  statistics} of the charges in the theory reflects the
behaviour of this product under interchange of factors.
 Under certain general conditions, e.g.\ if the Cauchy surfaces of the
 spacetime are not compact, each
 charge has a conjugate charge and then the statistics of each charge 
can be characterized by a number,
 its {\it statistics parameter}. This number can be split
into its phase and modulus being, respectively, the {\it
  statistics phase} and the inverse of the {\it statistical
  dimension}. (The latter is defined to be $\infty$ if the
statistics parameter equals 0 and one says that the
superselection sector has infinite statistics. We shall only
consider superselection sectors having finite statistics.) If the
statistics phase takes the values $\pm 1$, then the
(para-~)Bose/Fermi alternative holds in that there is a conventional 
description in terms of Bose and Fermi fields 
commuting or anticommuting  when localized in causally disjoint
regions. This is the generic situation in physical spacetime
dimension. In lower spacetime dimension, braid group
statistics may occur and the statistics phase may take
values different from $\pm 1$. 

In previous papers \cite{DR,DR1} it was shown that, in Minkowski
spacetime, one can construct a field net together with a unitary
action of a compact (global) gauge group containing 
the observable net $\cA$ as fixed points so that the
superselection sectors correspond naturally to 
the equivalence classes of irreducible representations of the gauge
group. A similar result will turn out to hold in curved spacetime as
well.
 As so little input is used (essentially
only the physically motivated selection criterion and local commutativity 
of the observables) this result clearly demonstrates how effective the
operator algebraic approach to quantum field theory can be. 
\subsection{Covariant Sectors and Univalence (Spin)}
Our notion of spin on curved spacetime involves a group $G$ of isometries
although there ought to be a more general notion not involving symmetries.
For this reason, we assume covariance of 
our reference representation $\p^0$.
 
A superselection sector described by a representation $\p$ is {\it covariant} 
if there
exists an (anti-)unitary representation $\widetilde{G} \owns \sG \mapsto
\widetilde{U}_{\p}(\sG)$ of the universal covering group of $G$ on
$\cH^{\p}$ with
$$ \p_{\g\cO} \lcrc \a_{\g} = {\rm Ad}\,\widetilde{U}_{\p}(\sG) \lcrc
\p_{\cO}\,, \quad \sG \in \widetilde{G}\,,\ \ \cO \in \cK\,,$$
where $\sG \mapsto \g$ denotes the covering projection.

We may now consider continuous curves $[0,2\p] \owns t \mapsto
\sG(t)$ whose projection $[0,2\p] \owns t
\mapsto \g(t)$ is a cycle, i.e.\ a closed curve possessing no closed
sub-curves. 
$\widetilde{U}_{\p}(\sG(2\p))$ may be different from 1, but as $\p$ 
is irreducible, $\widetilde{U}_{\p}(\sG(2\p)) =s_{\p}\cdot 1$ where 
$s_{\p}$ is a complex number of modulus 1. When
the cycle $\g([0,2\p])$ has the geometric interpretation of a ``spatial
rotation by $2\p$'', then it is appropriate to refer to the phase
factor $s_{\p}$ as the ``spin'', or more precisely, the {\it
  univalence} of the charge represented by $\p$.
\footnote{We do not wish to discuss how
$s_{\p}$ depends on the different possible ``rotations''. It suffice to
say that in the relevant cases the above procedure 
assigns an invariant $s_{\p}$ to any covariant superselection
sector.}
 Then, the {\it spin-statistics connection} is said to hold if, 
for all covariant
super\-selection sectors of the theory, {\it the univalence equals the
  statistics phase}.
\subsection{Tomita-Takesaki Theory and Symmetry}
Let us next summarize some basic points of the modular theory for von
Neumann algebras by Tomita and Takesaki \cite{Tak}.
Given a
von Neumann algebra $\cN$ on a Hilbert space $\cH$ together with a
cyclic and separating unit vector $\O \in \cH$,
the antilinear operator $S : \cN\O \to \cN\O$ defined by $S(A\O)
:= A^*\O$ admits a minimal closed extension with polar decomposition
$\overline{S} = J \D^{1/2}$ where $J$ is anti-unitary. $J$ is referred
to as {\it modular conjugation} and $\{\D^{it}\}_{t \in \bR}$ as {\it
  modular unitary group} associated with the pair $\cN,\O$; one refers
to ${\rm Ad}J$ as the {\it antilinear modular morphism} associated with
$\cN,\O$ and usually denotes it by $j$.These modular
objects satisfy $J\cN J = \cN'$ and $\D^{it}\cN\D^{-it}
= \cN$, $t \in \bR$. Moreover, a state $\o$ on a $C^*$-algebra $\fA$ 
is a KMS-state (thermal equilibrium state) at inverse temperature
$\b$ with respect to a one-parametric group $\{\a_t\}_{t\in \bR}$ of modular
automorphisms of $\cA$ if and only if 
$$ \p_{\o}\lcrc\a_t = {\rm Ad}\,\D^{-i\b t/2\p}\,\lcrc\,\p_{\o} $$
where $\p_{\o}$ is the GNS-representation of $\o$ and $\{\D^{it}\}_{t
  \in \bR}$ is the modular group of $\p_{\o}({\cA})'',\O_{\o}$,
 $\O_{\o}$ being the GNS-vector. Thus the modular
group may, in certain situations, have a physical (dynamical)
significance.

Furthermore, Bisognano and Wichmann showed \cite{BiWi}
 that, in Wightman's setting of quantum fields in Minkowski spacetime, the
modular objects associated with pairs $\cA(W),\O$, where $\cA(W)$ is
the von Neumann algebra of observables in a certain ``wedge-region''
\footnote{A wedge region is any Poincar\'e transform of the set
  $\{(x_0,\ldots,x_n) : 0 <x_1,\,0\le |x_0| < x_1\}$ in Minkowski spacetime.}
$W$ and $\O$ the vacuum vector, induce spacetime transformations.
That is, if $J_W,\{\D_W^{it}\}_{t \in \bR}$ denote the corresponding
modular objects, then there are elements ${\bf j}_W$, $\L_{W,t}$ in
the Poincar\'e group so that 
\begin{eqnarray}
 {\rm Ad}\,J_W\,\cA(\cO) &=& \a_{{\bf j}_W}(\cA(\cO)) \ = \ \cA({\bf
   j}_W(\cO))\,,\\
{\rm Ad}\,\D^{it}_W\,\cA(\cO) & = & \a_{\L_{W,t}}(\cA(\cO))\  =  \  
    \cA(\L_{W,t}(\cO)),
\end{eqnarray}
for all open subregions $\cO$ of Minkowski spacetime, all $t \in
\bR$ and all wedge-regions $W$.

Further investigations (e.g.\ \cite{Bor1,GuLo1,BuSu1,BGLo2,Bor4})
relate spacetime symmetries and modular objects and
indicate that vacuum states in Minkowski spacetime can possibly be 
characterized through the
geometric meaning of the modular objects associated with $\cA(W),\O$
for a certain class of wedge-regions $W$. This idea has been pursued
in non-flat spacetimes with a sufficiently rich group of
isometries and a suitable class of wedge-regions, such as
de\,Sitter spacetime and, to some extent, Schwarzschild-Kruskal
spacetime, too \cite{Sew,BrMo,BoBu}.
There are indications that physical states of quantum field theory on 
arbitrary spacetime manifolds can be distinguished
by the ``geometrical action''  of the
corresponding modular objects for a certain class of regions, understood
in sufficient generality. The reader is
referred to \cite{BDFS1} and references therein for considerable further
discussion. 

In Minkowski spacetime, the geometric action of the modular objects 
associated with wedge-algebras $\cA(W)$ and the vacuum vector $\O$ 
has important consequences for the relation between spin and statistics. 
It can be derived either from {\it geometric modular action}
\cite{Kuc},
i.e.\ the geometric action of the modular conjugations as in (1.1), or
from {\it modular covariance} \cite{GuLo2},
meaning the geometric action of the
modular group as in (1.2). 
Similarly, for conformal quantum field theories on the circle $S^1$ 
where modular objects and conformal symmetry are intimately 
related, there is a spin-statistic relation, 
as will be briefly summarized in the next section.  
\subsection{Modular Inclusion and Conformal Theories on the Circle}
In this section we summarize the connection between conformally
covariant theories on the circle $S^1$ and halfsided modular
inclusions established by Wiesbrock
\cite{Wies1,Wies4,Wies5,Wies6}.
 
We briefly recall what is meant by a conformally covariant
theory on the circle $S^1$ (see e.g.\ \cite{GuLo1,GuLo3} for further 
details).  This is a net
(or precosheaf) $I \mapsto \cM(I)$ taking proper open
subintervals $I$ of $S^1$ to von Neumann algebras $\cM(I)$ on a
Hilbert space $\cH_{\cM}$ so that locality holds, i.e.\ $\cM(S^1
\backslash \overline{I}) \subset \cM(I)'$. Moreover, there exists a
unitary strongly continuous positive energy representation $U$ of
$PSL(2,\bR)$ acting covariantly, $U(g)\cM(I)U(g)^* = \cM(gI)$, and
preserving a unit vector $\O_{\cM}$, cyclic for
the von Neumann algebra generated by the $\cM(I)$'s.
(In other words, the theory is given in a reference ``vacuum representation''.)

 The theory may be
equivalently described as a net of von~Neumann algebras indexed by
intervals on the real line, identified as the circle
with one point removed. Using the Cayley transform,
conformal transformations on the circle correspond to
fractional linear transformations on the line. Modular transformations 
have a geometric meaning and Haag duality holds for 
any conformal theory on the circle, namely
$\cM(S^1\setminus\overline{I})=\cM(I)'$ \cite{BGLo1}. Haag duality 
on the line, $\cM(\bR\setminus\overline{I}) = \cM(I)'$, holds precisely 
when the net $I \mapsto \cM(I)$ is strongly additive\cite{GLWi1}, i.e.\ if 
$\cM(I) = \cM(I_1) \vee \cM(I_2)$ whenever the union of $I_1$ and
$I_2$ yields $I$ up to at most a single point.

We recall that a $\pm${\it hsm inclusion} $(\cN\subset \cM,\O)$ is 
given by a pair $\cN \subset \cM$ of von Neumann
algebras on some Hilbert space together with a unit vector $\O$,
cyclic and separating for both $\cN$ and $\cM$, such that
$\Delta^{it}\cN\Delta^{-it} \subset \cN$ for all $\mp t \ge 0$, where
$\Delta^{it}$, $t \in \bR$, is the modular group of $\cM,\O$. A
$\pm$hsm inclusion $(\cN \subset \cM,\O)$ is called standard if $\O$
is cyclic  for $\cN'\cap \cM$, too (hsm abbreviates
``half sided modular'').

An interesting result of Wiesbrock (\cite{Wies1,Wies4} see also 
\cite{GLWi1}) asserts that there is a one-to-one correspondence 
between strongly additive conformally covariant theories on $S^1$ 
and standard $\pm$hsm inclusions.  

The rotations of $S^1$ form a subgroup of the covering group of
$PSL(2,\bR)$.
Let $\p$ be a Hilbert space representation of a covariant
superselection sector and $\widetilde{U}_{\p}$ the associated unitary
representation of the covering group of $PSL(2,\bR)$. Assuming that
$\widetilde{U}_{\p}$ has positive energy, the generator of 
rotations in the unitary representation $\widetilde{U}_{\p}$ has a
lowest eigenvalue $L_{\p}$. Then the conformal spin of the
superselection sector, or rather, its univalence, is defined by
 $s_{\p} = {\rm e}^{2\pi i L_{\p}}$. For superselection sectors with 
positive energy in a conformally covariant theory on $S^1$,  
the univalence equals the
statistics phase, which may be any complex number of modulus 1
\cite{GuLo3}.  
%
\subsection{Description of Contents}
We now describe the contents of the subsequent chapters.

In Chapter 2 we summarize several notions of spacetime geometry needed 
here. Lemma 2.2, of relevance to superselection theory,
asserts that the set of pairs of causally separated 
points in a globally hyperbolic spacetime is connected. 

Chapter 3 contains the general framework for superselection theory in
curved spacetimes, patterned conceptually on the DHR analysis in Minkowski
spacetime (\cite{DHR1}, cf.\ also \cite{Haag,R} and references given there).
 It will be formulated for nets
$\cK \owns \cO \mapsto \cA(\cO)$ of operator algebras in a reference
representation with general index sets $\cK$ possessing a partial ordering and
a causal disjointness relation. Thus 
quantum fields on arbitrary globally hyperbolic
spacetimes in any dimensions, with compact or non-compact
Cauchy surfaces, as well as quantum field theory on the circle, can be
treated on an equal footing. The existence of statistics is established 
in this generality. If the index set $\cK$ is directed, all the other basic
results known for superselection theory on Minkowski spacetime,
classification of statistics, existence of
charge conjugation and  construction of field algebra and gauge group
(cf.\ \cite{DR}) can again be shown to hold.

Chapter 4 begins with a summary of the geometry of spacetimes with a
bifurcate Killing horizon following  Kay and Wald
 \cite{KayWald}. We
introduce a family of wedge-regions $R_a$, $a >0$ which are copies of
the canonical right wedge shifted by $a$ in the affine geodesic
parameter on the horizon (a similar construction can be carried out
for the left wedge). We suppose that we are given a net of von Neumann
algebras $\cO \mapsto \cA(\cO)$ in the representation of a state which
is, in restriction to the subnet of observables which are localized on
the horizon,
a KMS-state at Hawking temperature for the Killing flow. 
 Thus on the horizon we have modular covariance and
are consequently in Wiesbrock's situation of half-sided modular
inclusion \cite{Wies1}.
Using Haag duality and additivity of the net, it follows that the maximal
subnet of observables localized on the horizon is a conformally
covariant family of von Neumann algebras.
Restricting the original net of von Neumann algebras
to the Killing horizon thus yields a conformal quantum field
theory on $S^1$. A conformal spin is therefore assigned to a 
superselection sector of the original theory, localizable on the horizon, 
and the conformal spin-statistics connection 
\cite{GuLo3} holds. This approach has, however, the drawback of applying 
only to
horizon-localizable charges, and this may be quite restrictive.

In Chapter 5 we introduce a class of spacetimes with a special
rotation symmetry and certain adapted wedge-regions. Essentially we
assume that there is a group of symmetries, to be viewed as rotations,
generated by pairs of time-reversing wedge-reflections 
mapping wedge-regions onto each other. In the 
Schwarzschild-Kruskal spacetime, for example, these wedge-regions can
be envisaged as the causal completions of ``halves'' of the
canonical Cauchy-surfaces chosen so that 
rotating by $\p$ about a suitable axis maps each such half onto its
causal complement. These wedge-regions differ from the usual
canonical ``right'' and ``left'' wedges ($R$ and $L$ in Chapter 4) and
lie in a sense transversal to the latter.
Then we consider a net of von Neumann algebras $\cO \mapsto \cA(\cO)$ over
such a spacetime in a representation where the full isometry group acts
covariantly. Moreover we suppose that there is
an isometry-invariant state and that the modular
conjugations associated with the vacuum vector and the von Neumann
algebras $\cA(W)$ for the said class of wedges $W$ induce the
geometric action of the wedge-reflections. This form of geometric
modular action will allow us to define the rotational spin of a
covariant superselection sector and to derive the spin and statistics
connection using a variant of arguments presented in
 \cite{Long3}.

%


\section{Some Spacetime Geometry}\label{Sec2}
\setcounter{equation}{0}
\subsection{Generalities}
In the present section we summarize some notions about
causal structure of Lo\-rent\-zian manifolds, thereby establishing our
notation. Standard references for this section 
include \cite{BeemEh,HawEll,ONeill,WaldI}.

We begin by recalling that a curved spacetime $(M,g)$ is 
a $1+s$-dimensional $(s \in \NN)$, Lorentzian manifold. In other words, 
it is a $1+s$-dimensional orientable, Hausdorff, second
countable $C^{\infty}$-manifold equipped with a smooth Lorentzian
metric $g$ having signature $(+,-,\ldots,-)$. 

A continuous, (piecewise) smooth curve $\gamma : I \to M$, defined on a
connected subset $I$ of $\RR$ and having tangent $\dot{\gamma}$,
 is called a timelike curve whenever
$g(\dot{\gamma},\dot{\gamma})> 0 $, a causal curve if 
$g(\dot{\gamma},\dot{\gamma}) \ge 0 $, and a lightlike curve if
$g(\dot{\gamma},\dot{\gamma})= 0 $ while $\dot{\gamma} \neq 0$, for
all parameter values $t$.

A spacetime $(M,g)$ is called time-orientable if there exists a global
timelike (non-vanishing) vectorfield $\xi$ on $M$. Such a vector field 
induces a time-orientation: a causal curve $\gamma$ is called
future-directed 
or past-directed according as $g(\xi,\dot{\gamma}) > 0$ or
$g(\xi,\dot{\gamma}) < 0$. We shall henceforth tacitly assume our 
spacetimes to be time-orientable with a given time-orientation.

A future-directed causal curve $\gamma : I \to M$ is said to have a
future (past)-endpoint if $\gamma(t)$ converges to some point in $M$
as the parameter $t$ approaches $\sup I$ ($\inf I$). Correspondingly
one defines the past (future)-endpoints of past-directed causal
curves. A future (past)-directed causal curve is said to start at a
point $p \in M$ provided that $p$ is the past (future)-endpoint of
$\gamma$. Moreover, one calls a future (past)-directed causal curve
future (past)-inextendible if it possesses no future (past)-endpoint.

For any subset $\Oo$ of $M$ one defines the sets $J^{\pm}(\Oo)$ as
consisting of all points in $M$ lying on future(+)/past(--)-directed
causal curves that start at some point in $\Oo$. Then $J^{\pm}(\Oo)$
are called the causal future(+)/causal past(--) of $\Oo$. The set
$J(\Oo) := J^+(\Oo) \cup J^-(\Oo)$ is then referred to as the causal
set of $\Oo$. The subsets $D^{\pm}(\Oo)$ of $M$ are, for given $\Oo
\subset M$, defined as the collection of all those points $p \in M$
such that every past(+)/future(--)-inextendible causal curve starting
at $p$ meets $\Oo$. One calls $D^{\pm}(\Oo)$ the future(+)/past(--)-domain 
of dependence of $\Oo$, and $D(\Oo) := D^+(\Oo) \cup D^-(\Oo)$
the domain of dependence of $\Oo$. 

One says that two points $p$ and $q$ in $M$ are {\it causally disjoint}, in
symbols $p \perp q$, if there are open neighbourhoods $U$ of $p$ and
$V$ of $q$ such that there is no causal curve connecting $U$ and $V$
(i.e.\ $U \cap J(V) = \emptyset = V \cap J(U)$). Correspondingly one
calls two subsets $P$ and $Q$ of $M$ causally disjoint if $p \perp q$
holds for all pairs $p \in P$ and $q \in Q$; this will be abbreviated
as $P \perp Q$.

In the present paper we will primarily
 be interested in globally hyperbolic spacetimes. A spacetime $(M,g)$
 is {\it globally hyperbolic} if it can  be smoothly foliated in acausal
 Cauchy surfaces.
 Here, an acausal Cauchy surface $C$ is a smooth hypersurface in $M$
 such that each causal curve in $(M,g)$ without endpoints meets $C$
 exactly once. This implies that $C$ is indeed acausal, i.e.\ 
 $p \perp q$ holds for all distinct $p,q \in C$.
 By a (smooth) foliation of $(M,g)$ in
acausal Cauchy surfaces we mean a diffeomorphism $F:\RR \times \Xi \to
M$ where $\Xi$ is an $s$-dimensional smooth manifold such that $F(\{t\}
\times \Xi)$ is, for each $t \in \RR$, an acausal Cauchy surface in
$(M,g)$, and the curves $\RR \owns t \mapsto F(t,q)$, $q \in \Xi$,
are timelike and endpointless. Thus, the foliation-parameter $t$ plays
the role of a ``time-parameter''. 
One may give a broader definition of Cauchy surfaces which are not
 necessarily acausal, by defining a Cauchy surface as a $C^0$
 hypersurface $C$ such that $C \cap {\rm int}\,J^{\pm}(C) = \emptyset$
 and $D(C) = M$. With this definition, a Cauchy surface is allowed to
 have lightlike parts. Such a broader definition of Cauchy surfaces is
 often useful.
However, it is a remarkable fact that the existence of a single, not
 necessarily acausal Cauchy surface in $(M,g)$ already implies that
 $(M,g)$ is globally hyperbolic in the above sense
 \cite{Ger,Dieck,WaldI}.  

 Whilst the question 
of whether physical spacetime models are necessarily globally hyperbolic 
has been discussed in the literature (see \cite{Cla,WaldI,Wald3}
and references given there), 
it is certainly the case that 
a great number of the prominent  spacetime models are
globally hyperbolic, like Schwarzschild-Kruskal,
deSitter, the Robertson-Walker models, and many others, including of
course Minkowski spacetime. One may therefore regard the
class of globally hyperbolic spacetimes as being sufficiently general and
comprising many examples of physical interest.
 Note that global hyperbolicity in no way
presupposes the presence of spacetime symmetries.

At this point we recall some properties of causal sets; for their
proof and further discussion, we refer to the indicated
references. Whenever $N \subset M$ and $(M,g)$ is globally hyperbolic,
then: $N$ compact implies $J^{\pm}(N)$  closed, $N$ compact
implies that $J(N) \cap C$ is  compact for each Cauchy surface $C$,
$N$ compact implies $D(N)$ compact. Furthermore, $J^+(N_+) \cap J^-(N_-)$
is empty or compact for all compact $N_+,N_- \subset M$. Moreover,
in (time-orientable) spacetimes $(M,g)$,
a time-orientation preserving isometry $\tau$ of $(M,g)$ satisfies 
\begin{equation}
                 \tau(J^{\pm}(\Oo)) = J^{\pm}(\tau(\Oo))\,,
\end{equation}
for $\Oo\subset M$.
It is moreover worth mentioning that for any two subsets $P$ and $Q$
of a globally hyperbolic spacetime $(M,g)$ we have $P \perp Q$ if and
only if $P \subset Q^{\perp}$, where the causal complement $Q^{\perp}$
of $Q \subset M$ is defined by $Q^{\perp} := M \backslash
\overline{J(Q)}$, see e.g.\ \cite[Prop.\ 8.1]{Keyl}.

We need to consider special regions of a globally
 hyperbolic spacetime $(M,g)$ namely those causally closed 
regions generated by an open subset of a Cauchy surface. 
More particularly we are interested in {\it regular diamonds}
defined as follows. A set of the form $\cO = {\rm int}\,D(G)$ 
is  a regular diamond  provided $\cO^\perp$ is non-void and 
\begin{itemize}
\item[(i)] $G$ is an open subset of an acausal Cauchy-surface $C$, and
  $\overline{G}$ is compact and contractible to a point in $G$,
\item[(ii)] $\partial G$, the boundary of $G$, is a (possibly multiply
  connected) locally flat embedded, two-sided topological submanifold
  of $C$ which is an embedded $C^{\infty}$-submanifold
near to points in each of its connected components.   
\end{itemize}
We refer to \cite{Brown,Ver2} for the precise definition of locally
flat embeddings and two-sidedness. Intuitively, these two conditions
are substitutes for the existence of an oriented normal vector field
over $\partial G$. These regularity properties serve to prove the
following assertion:
\begin{Lemma}\label{RainerLemma}
Let $\cO$ be a regular diamond and $p \in \cO^{\perp}$. Then there
exists another regular diamond $\cO_1$ with 
$$ \cO \cup \{p\}  \subset \cO_1\,.$$
\end{Lemma}
A rough sketch of the proof will be given in Sec.\ 2.2, the Appendix
to this chapter. The reader is referred to
\cite{Ver2} for a detailed proof.

  A double cone in Minkowski space is, of course, a regular diamond. 
Double cones may be generalized easily to curved spacetime. They are 
sets of the form ${\rm int}\,(J^-(\{v^+\})\cap J^+(\{v^-\}))$ with 
$v^+\in\text{int}J^+(\{v^-\})$. However, double cones need not 
have the property analogous to Lemma 2.1, think e.g.\ of a 
spacelike strip in Minkowski spacetime. Nor is it clear that a 
double cone is a regular diamond. For this reason, it is not 
clear, even for simple free fields, whether duality is satisfied 
for such regions. We expect the requirement of essential duality 
(cf.\ Sec.\ 1.1) to be realistic for regular diamonds, in particular, 
as their bases are assumed contractible. Furthermore, duality 
for regular diamonds has already been established for the 
Klein-Gordon field[57] and can presumably be verified for other 
free fields. For these reasons, we have chosen to use the collection 
$\cK$ of regular diamonds rather than the collection of double cones 
whose causal complement has non-empty interior as an index set
in a globally hyperbolic spacetime.\medskip
 
 Given a spacetime $(M,g)$, we introduce the set 
\begin{equation}
  {\cX}_{M,g} :=\{(x,y) \in M \times M : x \perp  y\}
\end{equation}
of pairs of causally disjoint points in $M$. According to the
definition of causal disjointness, this set is an open
subset of $M \times M$. 
The  subsequent assertion about ${\cX}_{M,g}$ will prove to be
important in discussing the statistics of superselection sectors in
the next chapter. It may be known to experts, but as we have not
found it in the literature, we put it
on record here.
\begin{Lemma}
  Let $(M,g)$ be a globally hyperbolic spacetime then 
${\cal X}_{M,g}$ is pathwise connected except when its Cauchy surfaces are 
noncompact and 1--dimensional in which case there are precisely two
path--components corresponding to $x$ being causally to the left or to the 
right of $y$.\end{Lemma}
\begin{proof}
 Let $F:{\bR}\times\Xi\to M$ be a foliation in acausal
Cauchy surfaces and write $C:=F(\{0\}\times\Xi)$. We first  
show that it suffices to restrict one's attention to the Cauchy 
surface $C$. More precisely, we show that 
$${\cal Y}:=\{(x,y)\in C\times C: x\perp y\}$$ 
is a strong deformation retract of ${\cal X}_{M,g}$. In fact, using $F$ to 
parametrize $M$ and defining 
$h:{\cal X}_{M,g}\times I\to{\cal X}_{M,g}$ by 
$$h(t,\xi;t',\xi';s):=((1-s)(t+s(t'-t)),\xi;(1-s)t',\xi')$$
we have a homotopy of the identity on $\cX_{M,g}$ onto the 
projection, $(t,\xi;t',\xi')\mapsto(0,\xi;0,\xi')$, onto $C$ leaving $C$ 
fixed. The only non--trivial point is to show that the image of $h$ lies in 
$\cX_{M,g}$ and this is where the causal structure enters. However, 
two remarks suffice: first, causal disjointness reduces to disjointness on an 
acausal Cauchy surface and hence is preserved if we pass from one acausal 
Cauchy surface to another by changing the value of $t$. Secondly, if 
we take causally disjoint points $x_i=F(t_i,\xi_i)$,
 $i=1,2$ with distinct values 
of $t$ then the curve $\gamma:[\text{inf}\{t_1,t_2\},\text{sup}\{t_1,t_2\}]\ni 
t\mapsto F(t,\xi_1)$ is timelike and 
connects $x_1$ with that Cauchy surface of 
the foliation containing $x_2$. Its range must lie in $\{x_2\}^\perp$ or there 
would be a causal curve coming arbitrarily 
close to connecting $x_1$ and $x_2$, 
contrary to assumption. We now know that the inclusion of $\cal{Y}$ in 
${\cal X}_{M,g}$ induces an isomorphism in homotopy and, in particular, an 
isomorphism of path-components. Now unless $C$ is one dimensional and 
non--compact, the complement of a point of $C$ 
is path--connected and  ${\cal Y}$ is then also path--connected. If $C$ is one 
dimensional and non--compact it is isomorphic to ${\bR}$ so that $\cal{Y}$ 
has two path--components.  
\end{proof}

  When ${\cX}_{M,g}$ has two components, we use the foliation
 $F:\bR\times\bR
\to M$ into acausal Cauchy surfaces to distinguish the ``right'' 
component from the ``left'' component as that containing pairs $(x,y)$, where 
the spatial component of $y$ is greater than that of $x$. In fact, this 
distinction depends only on the nowhere vanishing spacelike vector field 
$\xi$ induced by the foliation. Given such a field $\xi$, a spacelike 
curve $I\ni t\mapsto\gamma(t)$ is called {\it right-directed} if 
$g(\xi,\dot\gamma)>0$ and {\it left-directed} if $g(\xi,\dot\gamma)<0$ 
for one and hence all values of $t$. (A different choice of $\xi$ would 
at most lead to interchanging ``right-directed'' and ``left-directed'' 
since in two spacetime dimensions the set of spacelike vectors at 
each point has two components.) The orientation of spacelike curves 
defined in this way can now be used to specify the two connected 
components of ${\cX}_{M,g}$ in the case of a non-compact Cauchy surface. 
The right component is that 
containing $(\gamma(0),\gamma(1))$ for the endpoints 
$\gamma(0)$ and $\gamma(1)$ of some and hence any right-directed spacelike 
curve $\gamma$. This follows from the previous description in terms of the 
foliation since the spatial component
 is strictly increasing along such a curve. 
\subsection{Appendix to Chapter 2}
{\it Proof of Lemma 2.1 (Sketch) }\\[2pt]
Let $\cO = {\rm int}\,D(G)$ be a regular diamond, $G \subset C$ where
$C$ is an acausal Cauchy-surface, and $p \in \cO^{\perp}$.

Choose a $C^{\infty}$-foliation $F: \bR \times \Sigma \to M$ of $M$ into
smooth, acausal Cauchy surfaces. Then for each $y \in \Sigma$, the
curves $t \mapsto F(t,y)$ are inextendible, future-directed timelike
curves. Therefore, given any acausal Cauchy surface $C_0$, each of
these curves intersects $C_0$ exactly once, at the parameter value $t
= \t_{C_0}(y)$. The function $\t_{C_0}: \Sigma \to \bR$ is a smooth
function and one has $C_0 = \{F(\t_{C_0}(y),y): y \in
\Sigma\}$. Furthermore, the map $\Phi_{C,C_0}: C \to C_0$ induced by
$F(\t_C(y),y) \mapsto F(\t_{C_0}(y),y)$ is a diffeomorphism.

Using the results of \cite{Brown}, one can show that there is an open
neighbourhood $U$ of $\overline{G}$ in $C$ possessing the same
properties (i) and (ii) as $G$, i.e.\ $U$ is the base of a regular
diamond. It is also not difficult to show (cf.\ \cite{Ver2}) that
there exists an acausal Cauchy surface $C_0$ containing $p$ and with the
additional property that
$$ J(\overline{G}) \cap C_0 \subset \Phi_{C,C_0}(U) =: U_0\,.$$
The latter property means there are acausal Cauchy surfaces
$C_0$ passing through $p$ and coming arbitrarily close to
$\overline{G}$.  This entails that $\cO_0 := {\rm int}\,D(U_0)$ contains
$\cO$. Since $\Phi_{C,C_0}$ is a diffeomorphism, $U_0$
satisfies (i) and (ii) with respect to the
Cauchy surface $C_0$.

It remains to show that $U_0\cup \{p\}$ is contained in a
subset $U_1$ of $C_0$ satisfying (i) and (ii)
with respect to the Cauchy surface $C_0$. This is done by connecting
a point in a smooth part of $\partial U_0$ by a smooth curve $\l$ to
$p$ and by attaching to $U_0$ a suitable smooth deformation of a
tubular normal neighbourhood of $\l$. This yields the required set
$U_1$; properties (i) and (ii) follow by construction as does
$$ \cO \cup \{p\} \subset {\rm int}\,D(U_1) =: \cO_1\,. $$
%

\section{Superselection Structure in Curved Spacetimes} 
\subsection{Introduction}

   In this section, we adapt the basic notions and results of the theory of 
superselection sectors to curved spacetime, limiting ourselves  
to globally hyperbolic spacetimes. As we shall see, the basic theory goes 
through smoothly in the case of globally hyperbolic spacetimes with a 
noncompact Cauchy surface and much of it in the case of a compact Cauchy 
surface. The geometry of spacetime fortunately enters the long analysis only 
in establishing a few specific points. We can therefore limit ourselves to 
clarifying these points and otherwise just quoting the consequences.\smallskip

  We let $\cK$ denote the set of regular diamonds in $M$, 
ordered under inclusion. If $M$ is globally hyperbolic with 
a non-compact Cauchy surface, $\cK$ may not be directed 
although it will be in cases of interest. However, when $M$ is 
globally hyperbolic with a compact Cauchy surface, $\cK$ 
will never be directed and we shall meet problems akin to those 
on the circle. The more complicated structures involved have 
been relegated, as far as possible, to the appendix to this 
chapter.\smallskip 

   The set of double cones in $M$ whose causal complement has 
non-empty interior is even less likely to be directed. Both 
sets have in common that they form a base for the topology of $M$ 
and we will consider our nets of observables as being defined over 
$\cK$ with the general philosophy that they can be extended 
to other regions, if necessary. In fact, we will consider a wider 
class of regions in subsequent chapters. Now, the geometry of 
spacetime enters the analysis only through the partially 
ordered set $\cK$ and its relation of causal disjointness, 
introduced below. In view of further applications and despite 
the degree of abstraction involved, we have emphasised the 
relevant properties of $\cK$.\medskip

   The selection criterion for localized charges in Minkowski space 
uses the vacuum representation as a reference. Although there is no 
such preferred representation in curved spacetime, one expects there 
to be a preferred collection of representations satisfying the conditions 
listed in Sec.\ 1.1. In the case of the Klein-Gordon field on a four 
dimensional globally hyperbolic spacetime, we may take the representations 
induced by the pure quasifree Hadamard states\cite{Ver1}. We shall choose 
one of these representations as our reference representation and, whilst 
our sectors will depend on this choice, the superselection structure will not 
since this depends only the net of von Neumann algebras. 
By 4) of Sec.\ 1.1, any two preferred representations generate the same net 
of von Neumann algebras. We will denote our reference representation by 
$\pi^0$ and its Hilbert space by $\cH_0$.\smallskip   
 
  Once the reference representation has been fixed, it is just the causal 
structure of Minkowski space that plays a role in the superselection 
criterion for localized charges. For this reason, it adapts well to curved 
spacetime. The causal structure enters in the form of the relation $\perp$ 
of causal disjointness, defined in Ch.\ 2, and here
 to be considered as a relation 
on the ordered set $\cal{K}$, satisfying
\begin{description}
\item{$a)$} $\cO_1\perp\cO_2\Rightarrow\cO_2\perp\cO_1$.
\item{$b)$} $\cO_1\subset\cO_2$ and $\cO_2\perp\cO_3
\Rightarrow\cO_1\perp\cO_3$. 
\item{$c)$} Given $\cO_1\in\cK$, there exists an $\cO_2\in\cK$ 
such that $\cO_1\perp\cO_2$.
\end{description} 
We write $\cO^\perp:=\{\cO_1\in\cK:\cO_1\perp\cO\}$.
\smallskip 

   As explained above, the geometry of spacetime enters through the partially 
ordered set $\cal{K}$ together with the relation
 $\perp$ of causal disjointness. 
Hence we have to pass from geometric or topological properties of $(M,g)$ to 
properties of $(\cal{K},\perp)$. We will need to know 
whether certain partially 
ordered sets are connected, a notion defined in the appendix. But 
the basic idea is to move from one element $\cO_1$ of $\cal{K}$ to a nearby 
element $\cO_2$, where nearby means that there is a third element
$\cO_3$  containing $\cO_1$  and $\cO_2$. 
A finite series of such moves 
constitutes a path. $\cal{K}$ is connected 
if any two elements can be connected 
by a path. By virtue of Lemma 3A.1, we know that $\cal{K}$ is connected and, 
see Lemma 2.2, that $\cO^\perp$ is connected except when $M$ is two 
dimensional with a non--compact Cauchy surface.\smallskip  

   Lemma 2.2, itself, asserts that the set $\cX_{M,g}$ of pairs of spacelike 
separated points is pathwise connected again unless $M$ is two dimensional 
with a non--compact Cauchy surface. Since pairs of elements of $\cal{K}$ form 
a base for the topology in the product space, 
we can again conclude by Lemma 3A.1 that 
the graph $\cG^\perp$ of the relation $\perp$ is connected, 
$$\cG^\perp=\{\cO_1\times\cO_2:\cO_1\perp\cO_2\}.$$

In the exceptional case, $\cX_{M,g}$ has two pathwise connected components. 
Indeed the causal complement of a point is no longer connected but decomposes 
into a `left' causal complement and a `right' causal complement. 
\smallskip

   These are the basic geometric considerations determining the statistics. 
The remaining condition used in Sec.\ 3.3, the surjectivity of the projection 
from $\cG^\perp_c$, a connected component of $\cG^\perp$, to $\cal{K}$ 
has no geometric relevance seeing that it is automatically satisfied in 
the context of globally hyperbolic spacetimes.
Thus, as will follow from the results of Sec.\ 3.3, in a globally hyperbolic 
spacetime of dimension greater than 2, 
we get a net of symmetric tensor $W^*$--categories, 
$(\cT_t,\varepsilon^c)$, whereas in a 2--dimensional spacetime 
we shall in general get a braided tensor 
$W^*$--category with two different braidings $\varepsilon^\ell$  and 
$\varepsilon^r$  corresponding to the left and right causal complements of 
a double cone. Obviously, $\varepsilon^\ell =\varepsilon^{r*}$, where 
$\varepsilon^*$ is defined by
$$  \varepsilon^*(\rho,\sigma) = \varepsilon(\sigma,\rho)^*.$$

   The next basic step is to establish the properties of charge conjugation. 
The basic tool here is a left inverse. The physical idea behind 
constructing left inverses is that of transferring 
charge to spacelike infinity and a geometric property is obviously involved. 
Expressed as a property of our partially ordered set $\cal{K}$ we need to 
assume the existence of a net $\cO_n$ of elements of $\cal{K}$ such that 
given $\cal{O}\in\cal{K}$ there exists an $n_0$ with $\cO_n\perp\cal{O}$ 
for $n\geq n_0$. We will say that a net $\cO_n$ tends spacelike to 
infinity. Such a net obviously exists whenever $\cal{K}$ is directed but it  
continues to exist for an arbitrary 
globally hyperbolic spacetime with a noncompact 
Cauchy surface. The question of whether one can find a suitable substitute 
for globally hyperbolic spacetimes with compact Cauchy surfaces is still open,
a defect mitigated by the circumstance that a left inverse exists as 
a consequence of the equality of local and global intertwiners, 
postulated in Ch.\ 5.\smallskip 

   In this way, we establish in Sec.\ 3.4, the classification of statistics 
and the existence of charge conjugation for finite statistics 
for the case of a globally hyperbolic spacetime of dimension greater than
two.  
\subsection{The Selection Criterion}
  Our discussion of superselection theory in this and in subsequent sections 
is in terms of a partially ordered 
set $\cal{K}$ together with a binary relation 
$\perp$. The necessary properties will be introduced as needed and there 
will be no specific reference to spacetime. We have adopted this procedure 
for clarity and with future applications in mind. 
Thus the best choice of $\cal{K}$ in a curved spacetime is not altogether 
clear. We have already, for example, thought fit to use regular diamonds in 
place of double cones. On the other hand, we might like to go beyond strictly 
localized charges and work with spacelike cones or to replace causal disjointness by its 
Euclidean counterpart, disjointness, as when working on the circle. In fact, 
we shall need to use results on superselection structure on the circle
 in Ch.\ 4 
and, although these results have been developed previously \cite{FreReS}, 
\cite{GuLo3},
 the formalism presented here includes this case and allows a uniform 
approach to all such problems. We shall also simplify the exposition by making 
use of the freedom to modify the binary relation on $\cal{K}$. Thus 
this degree of abstraction is now called for even if we 
have not been able to derive all results in an adequate generality.\footnote
{Baumg\"artel and Wollenberg\cite{BauWo} treat nets over partially ordered 
sets with a relation of causal disjointness. In their applications to 
superselection structure they assume among other properties that the 
partially ordered set is directed. When the partially ordered set is 
not directed, their notion of representation depends on a choice of enveloping 
quasilocal algebra.} 
\smallskip

   Two nets $\cA$ and $\cB$ of $^*$--subalgebras of 
$\cB(\cH_0)$ over $\cal K$ are said to be {\it relatively local} if 
$$\cA(\cO_1)\subset\cB(\cO_2)',\,\,\text{whenever} 
\,\,\cO_1\perp\cO_2.$$ 
This relation fulfills the analogues of a), b) and c) above.
 Furthermore, there is a maximal net, the {\it dual net} $\cA^d$, which 
is relatively local to $\cA$. It is given by 
$$\cA^d(\cO)=\cap\{\cA(\cO_1)': \cO_1\perp \cO\}.$$ 
Since $\cA^{dd}$ is the largest net local relative to $\cA^d$, 
$\cA\subset\cA^{dd}$. However $\cA\subset\cB$ implies 
$\cB^d\subset\cA^d$, so that $\cA^d=\cA^{ddd}$. A net 
$\cA$ is said to be {\it local} if $\cA\subset\cA^d$ and  
then $\cA^{dd}\subset\cA^d=\cA^{ddd}$ so that $\cA^{dd}$ 
is local, too. We now compute the double dual: 
$$\cA^{dd}(\cO)=\cap_{\cO_1\perp\cO}\cA^d(\cO_1)'
=\cap_{\cO\perp \cO_1}\vee_{\hat{\cO}\perp\cO_1}\cA(\hat{\cO}).$$ 

\noindent
{\bf Definition.} A representation $\pi$ of the net $\cA$ is said 
to satisfy the {\it selection criterion} if 
$$\pi\rest\cO^\perp\simeq\pi^0\rest\cO^\perp,
\quad \cO\in\cal{K}.$$ 

 When $\cal{K}$ is directed this means that for each $\cal{O}$ there is 
a unitary $V_{\cal{O}}$ such that 
$$V_{\cO}\pi(A)=AV_{\cO},\quad A\in\cA(\cO_1),\quad 
\cO_1\in\cO^\perp,$$ 
where, to simplify notation in the sequel, we have omitted the symbol 
$\pi^0$ for the reference representation. We write $T\in(\pi,\pi')$ to 
mean that $T$ intertwines the representations $\pi$ and $\pi'$ and let 
Rep$^\perp\cA$ denote the $W^*$--category whose objects are the 
representations of $\cA$ satisfying the selection criterion and whose 
arrows are the intertwiners between these representations. As far as 
superselection theory goes, the following result allows one to replace 
the original net by its bidual.\smallskip 

\noindent
{\bf The Extension Theorem} {\it If each $\cO^\perp$ is connected, every 
object $\pi$ of ${\rm Rep}^\perp\cA$ admits a unique extension to an 
object of ${\rm Rep}^\perp\cA^{dd}$. Furthermore there is a canonical 
isomorphism of the corresponding $W^*$--categories.}\smallskip 

  This result is proved as Theorem 3A.4 of the Appendix. How to proceed 
when $\cO^\perp$ is not connected is exemplified by the well known 
case of a two dimensional Minkowski space and we will not attempt a general 
analysis here. The theory of superselection 
structure rests on two assumptions. 
The first is a property derived by Borchers 
in Minkowski space as a consequence 
of additivity, locality and the spectrum condition. Here it involves the 
dual net, $\cA^d$.\smallskip

\noindent
{\bf Definition} A net $\cA^d$ satisfies {\it Property B} if given 
$\cal{O}$, $\cO_1$ and $\cO_2$ in $\cal{K}$ such that 
$\cO\perp\cO_2$, and $\cO,\,\cO_2\subset\cO_1$ 
and a projection $E\neq 0$ in $\cA^d(\cO)$, there is an isometry 
$W\in\cA^d(\cO_1)$ with $WW^*=E$.\smallskip

\noindent
{\bf Lemma 3.1} {\it If $\cA^d$ satisfies Property B,
 the set of representations 
satisfying the
 selection criterion is closed under direct sums and (non-trivial) 
subrepresentations. In other words, the $W^*$--category 
${\rm Rep}^\perp\cA$ has direct
 sums and (non--zero) subobjects.}\smallskip

 The proof of this lemma will be omitted as it in no way differs from its 
Minkowski counterpart. The characteristic assumption of superselection theory 
is a duality assumption.\smallskip 

\noindent
{\bf Definition} A net $\cA$ 
is said to satisfy {\it duality} if $\cA=\cA^d$ and {\it essential 
duality} if $\cA^{dd}=\cA^d$.\smallskip

  To simplify notation, we shall suppose here that our net satisfies duality 
but, as a consequence of the Extension Theorem, the results remain valid under 
the weaker assumption of essential duality, whenever each $\cO^\perp$ 
is connected.\smallskip  

  In the Appendix, we have adopted the cohomological approach to 
superselection structure as this provides the most natural expression 
of the selection criterion. In the main text, we shall pursue the 
alternative strategy of working in terms of localized endomorphisms 
rather than $1$--cocycles.

\subsection{Localized Endomorphisms}

       When $\cal{K}$ is directed, the analysis of superselection structure 
rests on the following simple construction: let $\pi$ be a representation 
satisfying the selection criterion, pick a unitary $V_{\cal{O}}$ as above 
and set
$$       \rho(A) := V_{\cO}\pi(A)V_{\cO}^*,\quad   A\in\cA.$$
Obviously $\rho$ is a representation of $\cA$ on $\cH_0$  unitarily 
equivalent to $\pi$ but, in fact, $\rho(\cA)\subset\cA$. To see this, 
pick 
$\cO_1,\,\cO_2\in\cal K$, $\cO_1 \supset\cO$, $\cO_1\perp\cO_2$ 
and $B\in\cA(\cO_2)$ then, writing $V$ for $V_{\cal{O}}$, 
$$\rho(A)B = V\pi(A)V^*B = V\pi(AB)V^* = V\pi(BA)V^* =
 BV\pi(A)V^* = B\rho(A),$$
Hence $\rho(A)\in\cA^d(\cO_1) = \cA(\cO_1 )$, as required. 
Furthermore, $\rho$ is localized in 
$\cal O$, i.e.\
$$\rho(AB)=\rho(A)B,\quad B\in\cA(\cO_1),\quad A\in\cA,\,\,
\cO_1\perp\cO$$
and we refer to $\rho$ as a {\it localized endomorphism}.
 Now if $\rho$ and $\rho'$ are 
localized endomorphisms, an intertwiner $R$ for the corresponding representations 
is automatically in $\cA$. For suppose $\rho$ and $\rho'$ are localized in 
$\cal O$ and $A\in\cA(\cO_1)$, $\cO_1\perp\cal{O}$, then 
$$RA=R\rho(A)=\rho'(A)R=AR$$
so that $R\in\cA^d(\cal{O})=\cA(\cal{O})$.\smallskip

  We can thus write $R\in(\rho,\rho')$ without specifying whether we treat 
$\rho$ as a representation or as an
 endomorphism and, when studying superselection 
sectors, $\text{Rep}^\perp\cA$ may be replaced by the full subcategory 
$\cT_t$ of End$\cA$. 
End$\cA$ is a tensor $C^*$--category and we use the tensor product 
notation. Thus if $S\in(\sigma,\sigma')$, we write $R\otimes S$ to denote the 
intertwiner $R\rho(S)\in (\rho\sigma,\rho'\sigma')$. We characterize $\cT_t$ 
by characterizing the corresponding set $\Delta_t$ of endomorphisms. The representation corresponding to $\rho\in\Delta_t$ satisfies the selection 
criterion precisely when, given $\cal O\in\cal K$, there is an equivalent endomorphism $\sigma$ 
localized in $\cal O$. We then call $\rho$ {\it transportable} since, transporting $\rho$ 
by a suitable unitary $U\in\cA$, it can be localized in any given $\cal{O}\in\cal{K}$. 
$\Delta_t$ is thus the set of transportable localized endomorphisms and $\Delta_t(\cal O)$ shall 
denote the subset of endomorphisms localized in $\cal O$.\smallskip

\noindent
{\bf Lemma 3.2} {\it If $\rho,\,\rho'\in\Delta_t$
 then $\rho\rho'\in\Delta_t$.}\smallskip

\begin{proof}
 As the product of endomorphisms localized in $\cal O$ is again localized in 
$\cal O$, it suffices to observe that if $U\in(\rho,\sigma)$ and $U'\in(\rho',\sigma')$ 
are unitary then $U\otimes U'
\in(\rho\rho',\sigma\sigma')$ is unitary.
\end{proof}

   Thus the unitary equivalence class of 
$\rho\rho'$ depends only on the unitary equivalence classes of $\rho$ and 
$\rho'$ and, regarding charge as the quality distinguishing one sector from 
another, this defines a composition of charges.\smallskip

   When $\cal{K}$ is not directed, this simple scheme must be modified. The 
basic complication is that localized endomorphisms are now not defined on the 
whole net $\cA$. Instead, an endomorphism $\rho$ localized in $\cal{O}$ 
is just defined on the net $\cO_1\mapsto\cA(\cO_1)$ with 
$\cO\subset\cO_1$ and has the property that $\rho(\cA(\cO_1))
\subset\cA(\cO_1)$. As explained in detail in the Appendix, we have 
a net $\cO\mapsto\cT_t(\cal{O})$ of tensor $W^*$--categories, the 
objects of $\cT_t(\cal{O})$ are the transportable endomorphisms localized 
in $\cal{O}$.\smallskip

   It is also shown in the Appendix how a representation $\pi$ satisfying the 
selection criterion gives rise to objects of $\cT_t(a)$, $a\in\Sigma_0$ 
and how an interwiner $T\in(\pi,\pi')$ between two such representations 
leads to arrows $t_a$, $a\in\Sigma_0$, between the corresponding objects of 
$\cT_t(a)$. We can no longer study superselection sectors replacing 
Rep$\cA^\perp$ by $\cT_t(\cal{O})$, more precisely, we have a faithful 
$^*$--functor from Rep$\cA^\perp$ to $\cT_t(\cal{O})$ but cannot assert 
that it is an equivalence of $W^*$--categories. Thus, when $\cal{K}$ is not 
directed, $\cT_t(\cal{O})$ may not give a description of superselection 
sectors. Nevertheless, as we shall see, an analysis of localized endomorphisms 
still provides useful information.\smallskip

   The basic step in this analysis is to investigate the relation between 
causal disjointness and commutation of localized endomorphisms and their 
intertwiners. It is natural to say that an intertwiner $T\in\cT_t(\cal{O})$ 
is localized in $\cal{O}$, but we need a finer notion because we may have 
$T\in(\rho_1,\rho_0)$ where $\rho_i\in\Delta_t(\cO_i)$ with 
$\cO_i\subset\cal{O}$. In this case, we refer to $\cO_1$ as being an 
initial support and $\cO_0$ as being a final support of $T$. 
As explained in the Appendix, we consider the set $\Sigma_1$ of 
$1$--simplices in $\cal{K}$ as a partially ordered set and
 let $\Sigma_1^\perp$ 
denote the subset of $1$--simplices $b$ with $\partial_1b\perp\partial_0b$ 
with the induced order.\smallskip

\noindent
{\bf Lemma 3.3} {\it Let $\Sigma^\perp_{1,c}$ be a connected component of 
$\Sigma^\perp_1$, and suppose that given $\cO_0\in\cal{K}$, there 
is a $b\in\Sigma^\perp_{1,c}$ with $\partial_0b=\cO_0$. 
Let $T_i\in(\rho_i,\rho'_i)$ be arrows in some $\cT_t(\cal{O})$ then 
$$T_0\otimes T_1=T_1\otimes T_0,$$ 
if there are $b,b'\in\Sigma^\perp_{1,c}$ so that 
$\partial_0b$ and $\partial_1b$ 
are initial supports of $T_0$ and $T_1$ and $\partial_0b'$ and $\partial_1b'$ 
are final supports of $T_0$ and $T_1$.}\smallskip

\noindent
\begin{proof}
 We first show that $T_0\rho_0(T_1)=T_1\rho_1(T_0)$. This relation 
is trivial if $T_0$ and $T_1$ are causally disjoint
 in the sense that there is a 
$\hat b\in\Sigma^\perp_1$ such that $\partial_0\hat b$ contains an initial and 
final support of $T_0$ and $\partial_1\hat b$ an initial and final support of 
$T_1$. The idea of the proof is to reduce to this trivial case.
Replace $T_0$ and $T_1$ by $T_2=T_0\circ U_0$ and $T_3=T_1\circ U_1$, where 
$U_0\in(\rho_2,\rho_0)$ and $U_1\in(\rho_3,\rho_1)$ are unitary. Then
$$T_2\otimes T_3=T_0\otimes T_1\circ U_0\otimes U_1,\quad 
T_3\otimes T_2=T_1\otimes T_0\circ U_1\otimes U_0,$$ 
to be understood as valid in some $\cT_t(\hat{\cal{O}})$ for $\hat{\cal{O}}$ 
sufficiently large. Thus if $U_0$ and
 $U_1$ are causally disjoint, the validity 
or not of our relation is unaffected by the passage from $T_0$ , $T_1$ to 
$T_2$, $T_3$. But $b$ and $b'$ lie in a 
connected component $\Sigma^\perp_{1,c}$ 
by hypothesis, so after a finite number of steps we can arrange that the 
initial and final supports of both intertwiners coincide. This is again the 
trivial case so $T_0\rho_0(T_1)=T_1\rho_1(T_0)$, as required.
 It only remains to show that
$$\rho_0\rho_1-\rho_1\rho_0=0.$$ 
The above computations show that the kernel of the left hand side does not 
change if we shift to $\rho_2$ and $\rho_3$. However, by hypothesis, given 
$\cO\supset b'$, we can find $\hat b\in\Sigma^\perp_{1,c}$ with 
$\partial_0\hat b=\cal{O}$ and we
 can take $\rho_3\in\Delta_t(\partial_1\hat b)$,  when 
$$\rho_0\rho_3(A)=\rho_0(A)=\rho_3\rho_0(A),\quad A\in\cA(\cal{O}),$$ 
completing the proof.
\end{proof}

  After this one crucial lemma, the standard results on the existence of a braiding 
follow without further geometric input. Of course the braiding will, in 
general, continue to depend on the choice of connected component.\smallskip

\noindent
{\bf Theorem 3.4} {\it Let $\Sigma^\perp_{1,c}$ be a connected component of 
$\Sigma^\perp_1$. If the projection mapping $b\mapsto\partial_0b$ from 
$\Sigma_{1,c}^\perp$ to $\cal{K}$ 
is surjective then there is a unique intertwiner-valued function 
$(\rho_0,\rho_1)\mapsto\varepsilon^c(\rho_0,\rho_1)\in(\rho_0\rho_1,\rho_1\rho_0)$ 
such that
\begin{description}
\item $a)$  $\varepsilon^c(\rho'_0 ,\rho'_1 )\circ T_1\otimes T_2=T_2\otimes T_1\circ\varepsilon^c
(\rho_0 ,\rho_1 ),\quad  T_i\in(\rho_i ,\rho'_i ),\quad i=0,1,$
\item{$b)$} $\varepsilon^c(\rho_0,\rho_1) =1_{\rho_0\rho_1}$, 
if there is a $b\in\Sigma^\perp_{1,c}$ such that $\rho_i\in\Delta_t(\partial_ib)$ 
$i=0,1$.
\end{description}}\smallskip

\noindent
\begin{proof}
 The uniqueness claim tells us how to go about defining $\varepsilon^c$: given 
$\rho_1 ,\rho_2$  pick $b\in\Sigma^\perp_{1,c}$ and
unitaries $U_i\in(\rho_i,\tau_i)$ where $\tau_i\in\Delta_t(\partial_ib)$ 
and we have no option but to set
$$ \varepsilon^c(\rho_1 ,\rho_2 ) =U^*_2\otimes U^*_1\circ U_1\otimes U_2.$$
By Lemma 3.3, such a choice, however  made,  automatically  satisfies  $b)$.  We  have
$\varepsilon^c(\rho'_1 ,\rho'_2 ) = U'_2{}^*\otimes U'_1{}^*\circ U'_1\otimes U'_2$, where 
$U'_i\in(\rho'_i ,\tau'_i)$ and the product of supports of $\tau'_1$  and
$\tau'_2$ is  contained  in  $X_c$.  Set $S_i=U'_i\circ T_i\circ U_i^*$ then, 
by  Lemma 3.3, $S_1\otimes S_2=S_2\otimes S_1$    and
rearranging this identity gives $a)$ and completes the proof of the
theorem.
\end{proof}

\noindent
{\bf Corollary 3.5} {\it Under the hypothesis of Theorem 3.4
\begin{description}
\item{$a)$}     $\varepsilon^c (\rho_1 \rho_2 ,\rho_3 ) = 
\varepsilon^c (\rho_1 ,\rho_3 )\otimes 1_{\rho_2} \circ 1_{\rho_1}  \otimes\varepsilon^c (\rho_2 ,\rho_3 ),$
\item{$b)$} $\varepsilon^c(\rho_1,\rho_2\rho_3 )=1_{\rho_2}\otimes\varepsilon^c 
(\rho_1,\rho_3)\circ\varepsilon^c (\rho_1,\rho_2)\otimes 1_{\rho_3},$
\smallskip

   If $b\in\Sigma^\perp_{1,c}$ implies 
$\bar b\in \Sigma^\perp_{1,c}$, where $|\bar b|=|b|$, $\partial_0\bar b=\partial_1b$ 
and $\partial_1\bar b=\partial_0b$, then
\item{$c)$} $\varepsilon^c(\rho_2,\rho_1)\circ\varepsilon^c(\rho_1,\rho_2)=
1_{\rho_1\rho_2}$.
\end{description}
}

\noindent
\begin{proof}
 These equalities follow easily from the formula
 $$\varepsilon^c (\rho_1 ,\rho_2 ) = U_2^*\otimes U_1^*\circ U_1\otimes U_2$$
used to define $\varepsilon^c$  in the proof of Theorem 3.4.
\end{proof}

   As a consequence of $a)$ and $b)$ or by direct computation, we also have
$$\varepsilon^c (\rho,\iota) = \varepsilon^c (\iota,\rho) = 1_\rho.$$
In virtue of $a)$ and $b)$, if $\cal{K}$ is directed, the pair 
$(\cT_t,\varepsilon^c )$ is a
 {\it braided} tensor $W^*$--category and when $c)$
holds, too, we get a {\it symmetric} tensor $W^*$--category. In the general 
case we get a net $\cO\mapsto(\cT_t(\cO),\varepsilon^c)$ of 
braided or symmetric tensor $W^*$--categories, where the terminology 
implies that the inclusion $\cT_t(\cO_1)\subset\cT_t(\cO_2)$ 
for $\cO_1\subset\cO_2$ is not only a tensor $^*$--functor 
but also preserves the braiding.\smallskip

   In view of the above results, it is obviously important to be able to 
compute the connected components of $\Sigma^\perp_1$. We first localize  
and try to compute the connected components of 
$$\Sigma^\perp_1(\cO):=\{b\in\Sigma^\perp_1:\,|b|\subset\cO\}$$ 
before trying to compute those of $\Sigma^\perp_1$. Needleess to say, neither 
step can be carried through at this level of generality but
 we shall carry them 
through when $\cal{K}$ is the set of regular diamonds in a globally 
hyperbolic spacetime.\smallskip 

  Note that $\Sigma^\perp_1(\cal{O})$ is closely related to the local graph 
of the relation $\perp$, 
$$\cG^\perp(\cO):=\{\cO_1\times\cO_0:\,\cO_1,\cO_0
\subset\cO,\,\cO_1\perp\cO_0\}.$$ 
There is an obvious order--preserving injection $i:\cG^\perp(\cal{O})
\to\Sigma^\perp_1(\cal{O})$. We simply consider $\cO_i$ as $\partial_ib$ 
and $\cal{O}$ as $|b|$. Conversely, we have an order--preserving surjection 
$s:\Sigma^\perp_1(\cO)\to\cG^\perp(\cal{O})$ mapping $b$ to 
$\partial_1b\times\partial_0b$. $b$ lies in the same component of 
$\Sigma^\perp_1(\cal{O})$ as $i\circ s(b)$. Hence if $s(b)$ and $s(b')$ 
lie in the same component, so do $b$ and $b'$, thus we have computed the 
components of $\Sigma^\perp_1(\cal{O})$ in terms of those of 
$\cG^\perp(\cal{O})$. Now if $\cal{O}$ is a regular diamond in a 
globally hyperbolic spacetime, then $\cal{O}$ itself with the induced 
metric is a globally hyperbolic spacetime with a non--compact Cauchy surface 
and the connected components have been computed in Lemma 2.2.\smallskip 

   For passing from the local to the global computation, the strategy is 
to look for coherent choices of components for the $\Sigma^\perp_1(\cal{O})$, 
i.e.\ we want a component $\Sigma^\perp_{1,c}(\cal{O})$ for each $\cal{O}$ 
such that 
$$\Sigma^\perp_{1,c}(\cO_1)=\Sigma^\perp_{1,c}(\cO_2)\cap\Sigma^\perp_1(\cO_1),
\quad \cO_1\subset\cO_2.$$ 

\noindent
{\bf Lemma 3.6} {\it Given a coherent choice of components $\cO\mapsto
\Sigma^\perp_{1,c}(\cal{O})$, then $\Sigma^\perp_{1,c}~:=~\{b\in\Sigma_1:b\in\Sigma^\perp_{1,c}(|b|)\}$
is a component of $\Sigma^\perp_1$.}\smallskip 

\noindent
\begin{proof}
 $\cal{K}$ being connected, the result will follow from Lemma~3A.3 
once we show that 
$$\Sigma^\perp_{1,c}(\cO)=\Sigma^\perp_{1,c}\cap\Sigma^\perp_1(\cal{O}).$$
But if $b\in\Sigma^\perp_{1,c}(\cal{O})$, $|b|\subset\cal{O}$ and
 since we have 
a coherent choice of components, $b\in\Sigma^\perp_{1,c}(|b|)$ giving an 
inclusion. The reverse inclusion is trivial, completing the proof.
\end{proof}  

   Now when $\cal{K}$ denotes the set of regular diamonds in globally 
hyperbolic spacetime with dimension $\geq 2$, then $\Sigma^\perp_1(\cal{O})$ 
has a single component so that $\Sigma^\perp_1$ is connected by Lemma 3.6. 
It remains to consider the case of a
 globally hyperbolic spacetime of dimension 
two. We know that each $\Sigma^\perp_1(\cal{O})$ now has two components and 
that one passes from one component to the other by reversing the orientation 
of the $1$--simplices. We need a way of specifying a coherent choice of 
components. If the Cauchy surfaces are non--compact, then $\cG^\perp$ 
also has two components and one passes from one component to the other by 
interchanging the two double cones. Hence mapping $b$ to $\partial_1b\times
\partial_0b$ must map the two components of $\Sigma^\perp_1(\cal{O})$ into 
different components of $\cG^\perp$. Denoting the two components of  
$\cG^\perp$ by $\cG^\perp_\ell$ and $\cG^\perp_r$, the inverse 
images under the above map give us a coherent choice of components. Lemma~3.6 
then shows us that $\Sigma^\perp_1$ has precisely two components and that one 
passes from one component to the other by reversing the orientation of 
$1$--simplices.\smallskip 

   On the other hand, in a globally hyperbolic spacetime $(M,g)$ of dimension 
two with compact Cauchy surfaces, we know from the discussion in Sec.\ 3.1 
that $\cal{G}^\perp$ is connected. However, $\Sigma_1^\perp$ continues to 
have two components and we need a different procedure for making a coherent 
choice of local components. To this end, we pick a nowhere vanishing 
timelike vector field and restricting this to a regular diamond $\cal{O}$, 
we have, by the discussion following Lemma 2.2, a coherent way of 
distinguishing the left and right components of the set of spacelike points 
in the regular diamond and hence left and right components of 
$\cal{G}^\perp(\cal{O})$ and $\Sigma_1^\perp(\cal{O})$. Thus by Lemma 3.6, 
$\Sigma_1^\perp$ has two connected components and one passes from one 
component to the other by reversing the orientation of $1$--simplices.

\subsection{The Left Inverse and Charge Transfer}

The classification of statistics makes essential use of left inverses. 
When $\cal{K}$ is directed, we may proceed as follows. 
\smallskip

\noindent
{\bf Definition} A positive linear mapping $\phi$ on 
$\cB(\cH_0)$ is called a {\it left inverse} of a representation $\pi$ 
of $\cA$ on $\cH_0$ if

$$ \phi(A\pi(B)) = \phi(A)B,\quad A\in\cB(\cH_0),\, B\in\cA,
\quad \text{and}\quad \phi(1)= 1.$$

       There  are  some  simple  facts  to 
 be  noted:  first,  a  positive  mapping  is
automatically self-adjoint, $\phi(A^*)= \phi(A)^*$ so that we have 
$\phi(\pi(A)B) = A\phi(B)$, $A,B\in\cA$. Secondly, if $\rho(B)=B$, 
then $\phi(B)=B$. Thus $\phi$ inherits any localization properties of $\pi$. 
In particular, if $\pi$ is localized in $\cal{O}$
$$       \phi(A) = A\quad \text{for}\quad A\in\cA(\cO_2),
\quad \cO_2\perp\cal{O}$$
and, by duality, if $\cO \subset \cO_1$  then 
$\phi(\cA(\cO_1 )) \subset \cA(\cO_1 )$. Consequently 
$\phi$ maps $\cA$ into $\cA$. Furthermore one may show that 
$\phi(A^*A)\geq \phi(A)^*\phi(A)$ and $\|\phi\|\leq 1$.\smallskip

  The complications involved when $\cal{K}$ is not directed are treated in the 
Appendix where the relations with the left inverse of a localized endomorphism 
and the left inverse of a cocycle are also discussed.\smallskip

   Once we have left inverses, we may proceed to the classification of 
statistics. We suppose we have permutation statistics. The basic 
result, stated abstractly, is as follows.\smallskip

\noindent
{\bf Theorem 3.10} {\it Let $\rho$ be an object in a symmetric tensor 
$C^*$--category $(\cT,\varepsilon)$ and $\phi$ a left inverse of 
$\rho$ with $\phi_{\rho,\rho}=\lambda 1_\rho$ for some scalar $\lambda$ 
then $\lambda\in \{0\}\cup\{\pm d^{-1}:d\in\bN\}$ and depends only 
on the equivalence class of $\rho$. The Young tableaux associated with the 
representations of $\bP_n$ on $(\rho^n,\rho^n)$, $n\geq 1$ are all 
Young tableaux:
\begin{description}
\item{a)} whose columns have length $\leq d$, if $\lambda=d^{-1}$ 
(para-Bose statistics of order $d$);
\item{b)} whose rows have length $\leq d$ if $\lambda=-d^{-1}$ 
(para-Fermi statistics of order $d$);
\item{c)} without restriction, if $\lambda=0$ (infinite statistics).
\end{description}
}

   Note that when $\rho$ is irreducible,
 $\phi_{\rho,\rho}(\varepsilon(\rho,\rho))$ 
is automatically a scalar, called the {\it statistics parameter} of $\rho$. 
$d$ is referred to as the {\it statistics dimension} and the sign is the 
{\it statistics phase}, $\kappa_\rho$ and corresponds to the Bose-Fermi 
alternative. In general, we say that $\rho$ has {\it infinite statistics} 
if there is a left inverse $\phi$ with 
$\phi_{\rho,\rho}(\varepsilon(\rho,\rho))=0$. Otherwise $\rho$ is said to 
have finite statistics. Assuming our category $\cal{T}$ has subobjects,
$\rho$ has finite statistics if and only if $\rho$ is a finite direct sum of 
irreducible objects with finite statistics. In the cases where we can have 
braid statistics there is, of course, no correspondingly complete 
classification, not even if we invoke the special setting of a two dimensional
Minkowski space. However, many partial results are known in that case and the 
proofs presumably generalize without essential modification.\smallskip

   As explained in Sec.\ 3.1, to deduce the existence of a left inverse, we 
assume that $\cal{K}$ has an asymptotically causally disjoint net $\cO_n$. 
Thus, given $\cO\in\cal{K}$ there is an $n_0$ with $\cO_n\perp\cal{O}$ 
for $n\geq n_0$.
Under such a hypothesis, every representation $\pi$ satisfying the 
selection criterion can be obtained as a limit of unitary transformations. 
Physically, this would be interpreted as creating charge by transferring it 
from spacelike infinity. We pick unitary intertwiners $U_n\in(\pi_n,\pi)$ 
where $\pi_n$ is localized in $\cO_n$. The corresponding unitary 
transformation $\sigma_{U_n}$, $\sigma_{U_n}(A) :=U_nAU_n^*$, 
may be interpreted 
as an operation which transfers charge from $\cO_n$ to $\cal O$. Now if 
$A\in\cA(\cO_0)$ and $n$ is sufficiently large so that 
$\cO_0\perp\cO_n$ then $\sigma_{U_n}(A)= \pi(A)$ so that, 
as far as $A$ is concerned, we have created a charge in $\cal O$. In the
limit as $n\to\infty$ this holds for all $A\in\cA$ and we have\smallskip

\noindent
{\bf Lemma 3.11}  $\lim_{k\to\infty}\|U_k AU_k^* -\pi(A)\|= 0$,
$A\in\cA$.
\smallskip

   The physical idea is now to create the conjugate charge in $\cal O$ by 
transferring charge to spacelike infinity. More prosaically, we would like
to get a left inverse by replacing $U_k$ by $U_k^*$ and 
taking a limit. This will indeed be the case although the limiting
procedure is more delicate and we cannot work in the 
strong topology (i.e. pointwise
norm topology) for linear mappings on $\cA$.\smallskip

   We consider the space $\cal M$ of bounded linear mappings on 
$\cB(\cH_0)$ equipped with  the  pointwise  $\sigma$--topology,
 i.e.\ a net $\phi_n$ from $\cal M$  
converges  to  $\phi$  if  $\phi_n (A)$
converges to $\phi(A)$ in the  $\sigma$--topology  
for  each  $A\in\cA$.  The  important  fact  for  our
purposes is that the unit ball $\cM_1$  of $\cal M$ 
is compact in this topology,
$$ \cM_1 =\{\phi\in\cM : \|\phi\|\leq 1\}.$$
\noindent
{\bf Lemma 3.12} {\it The net $\sigma_{U_n^*}$ possesses
 at least one limit point 
in $ \cal M$. Every limit point of this net is a left inverse of $\pi$. The 
set of all left inverses of $\rho$ is a nonvoid compact convex subset of 
$\cal M$.}\smallskip

    We omit the proof as it is identical with that already
 given for Minkowski space. 
The existence of an asymptotically causally disjoint  net $\cO_n$ is also 
used in the analysis of left inverses but there are no new
 geometric properties 
involved. 

  Another important aspect of superselection structure which does not 
involve spacetime symmetries is the existence of a complete field net 
with gauge symmetry describing the superselection sectors\cite{DR}. This 
clearly involves no further input of a geometric nature as it is based 
on Corollary 6.2 of \cite{DR1} which refers to a single $C^*$--algebra 
rather than a net of von Neumann algebras. We leave to the reader the task 
of formulating a precise result so as to avoid having to introduce the 
relevant definitions from \cite{DR}.\smallskip

\subsection{Sectors of a Fixed--Point Net} 

  Although we have now succeeded in adapting the main results of 
superselection theory to globally hyperbolic spacetimes with non--compact 
Cauchy surface, there is another important aspect to be discussed. As 
we have seen the Selection Criterion has a natural mathematical 
extension to curved spacetime. In Minkowski space, however, it is 
further justified by there being a simple mechanism producing 
examples of such sectors. Under rather general conditions, it suffices 
to begin with a field net ${\cF}$ in its vacuum 
representation and a group of unitaries, a gauge group, compact in 
the strong operator topology, and inducing automorphisms of the field 
net. Then defining an observable net ${\cA}$ as the fixed--point net: 
${\cA}({\cO}):={\cF}({\cO})^G$, the resulting representation 
decomposes as a direct sum of irreducible representations satisfying the 
selection criterion. The equivalence classes of these representations 
are in 1--1 correspondence with the set $\hat G$ of equivalence classes 
of irreducible, continuous, unitary representations of $G$ and the 
irreducible representation corresponding to $\xi\in\hat G$ 
has multiplicity $d(\xi)$, the dimension of $\xi$. The question is 
whether these results continue to hold in curved spacetime.\smallskip 

  The original result in \cite{DHR1} does not, as it stands, apply to 
curved spacetime as it involves translations and the cluster property. 
However the variant given in \cite{DR} involves only structural elements 
and geometric properties compatible with curved spacetime and therefore can be 
stated here as a result on superselection sectors in curved spacetime. 
In fact, the following result is valid for a directed set $\cal{K}$ 
with a binary relation $\perp$ such that given $\cal{O}\in\cal{K}$, 
there exists $\cO_1,\cO_2\in\cal{K}$ with 
$\cO,\cO_1\subset\cO_2$ and $\cO\perp\cO_1$.
This condition is related to our use of the Borchers Property.\smallskip   
  
\noindent
{\bf Theorem 3.13} {\it Let ${\cF}$ be a field net over $\cal{K}$ acting 
irreducibly on a Hilbert space ${\cal H}$ equipped with a strongly compact 
group $G$ of unitaries inducing automorphisms of the net ${\cF}$. We 
define the observable net ${\cA}$ to be the fixed--point net: 
$${\cA}({\cO}):={\cF}(\cO)^G,\quad \cal{O}\in\cal{K}.$$ 
We assume that the subspace $\cH_0$ of $G$--invariant vectors is 
separable and that ${\cA}$ is represented irreducibly on $\cH_0$,
satisfying duality there and having the Borchers Property. 
Furthermore, ${\cal H}_0$ is supposed to be cyclic for each 
${\cF}({\cal O})$ and ${\cF}({\cal O}_1)$ and 
${\cA}({\cal O}_2)$ to commute whenever 
${\cal O}_1\perp{\cal O}_2$. 
Then ${\cA}'=G''$ and letting $\pi$ denote the defining representation 
of ${\cA}$ on ${\cal H}$  
$$\pi=\sum_\xi d(\xi)\pi_\xi,\quad \xi\in\hat G,$$ 
where the $\pi_\xi$ are inequivalent irreducible representations 
satisfying the selection criterion and $\hat G$ denotes the set 
of equivalence classes of continuous irreducible unitary representations 
of $G$.}\smallskip

   Despite this positive result, we must examine the assumptions 
carefully to see whether they remain reasonable in the context of 
curved spacetime. To test the assumptions we turn to the examples 
of scalar free fields defined using quasifree Hadamard states\cite{Ver1}. 
It is known that duality holds for the Klein-Gordon field on a globally 
hyperbolic spacetime for regular diamonds and that the associated 
von Neumann algebra is the hyperfinite type III$_1$ factor and hence 
satisfies the Borchers property. However, at least in the context 
of Theorem 3.12, this must be regarded as a field net rather than 
an observable net. Furthermore, we actually use duality for the 
modified relation $\tilde\perp$ of causal disjointness to pass 
from cocycles to localized endomorphisms in the next section. 
This strengthened form of duality is equivalent to the original 
form whenever the nets are inner regular, as is the case for the 
Klein--Gordon field. An even stronger form of duality, 
$\hat\perp$--duality, is used in the discussion of left inverses 
in the next section. However, our basic result on regular diamonds, 
Lemma 2.1, shows that it is in fact equivalent to 
$\perp$--duality for additive nets.\smallskip 
   
   As is well known, a geometric property is involved in passing 
from duality for the fields to duality for the observables. We give here 
a variant on the proof of Theorem 4.3 of \cite{R1}, not a priori requiring
each irreducible representation of the gauge group to be realized on  
Hilbert spaces in $\cF$. In view of the $\bZ_2$--graded 
structure of a field net, it is appropriate to define its dual 
net by
$$\cF^d(\cO)=\cap\{\cF^t(\cO_1)': \cO_1\perp \cO\}.$$ 
Here $\cF^t$, the twisted field net, can be defined as the transform 
of $\cF$ under the unitary transformation $2^{-1/2}(1+iV)$, where $V$ 
is the gauge transformation changing the sign of Fermi fields, see e.g.\ 
\cite{DHR1}.

\smallskip 

\noindent
{\bf Theorem 3.14} {\it Let $\cF$ be a field net over $\cal{K}$ 
on a Hilbert space $\cal{H}$ satisfying twisted duality under a compact 
group of unitaries $G$ inducing automorphisms of the net $\cal{F}$. Let  
$\cal{H}_0$, the subspace of $G$--invariant vectors, be cyclic for 
each $\cF(\cal{O})$. Then the fixed--point net $\cA$ satifies 
duality for each $\cal{O}\in\cal{K}$ provided $\cal{O}^\perp$ is connected.}
\smallskip 

\noindent
\begin{proof} Let $E$ denote the projection onto $\cal{H}_0$ then the 
conditional expectation $m$ of $\cF$ onto $\cA$ may either 
be defined by integrating over the action of $G$ or by 
$$m(F)E=EFE,\quad F\in\cF.$$ 
Now 
$$(\cA_E)^d(\cO)=\cap_{\cO_1\perp\cO}(\cA_E(\cO_1)')
=\cap_{\cO_1\perp\cO}(E\cF^t(\cO_1)E\rest\cH_0)',$$ 
Since $E$ is cyclic and separating for each $\cF^t(\cal{O})$ and 
$\cA(\cO)=m(\cF(\cO))$, 
$$(E\cF^t(\cO_1)E\rest\cH_0)'=
(E\cF^t(\cO_1)'E)\rest\cH_0.$$ 
Now using the fact that $E$
 is separating for each $\cF^t(\cal{O}_1)'$ and 
that $\cal{O}^\perp$ is path--connected, we obtain 
$$(\cA_E)^d(\cO)=E\cap_{\cO_1\perp\cO}\cF^t(\cO_1)'E
\rest\cH_0=\cA(\cal{O}),$$ 
since $\cF$ satisfies twisted duality.
\end{proof}

  What is still missing is a result allowing one to pass from the 
Borchers Property for the field net to the corresponding property 
of the observable net. 
 
\subsection{Appendix to Chapter 3}

  In this Appendix, we begin by introducing various notions we shall need 
in connection with the partially ordered set $\cal{K}$. We recall\cite{R} 
that an 
$0$--simplex $a$ of the partially ordered set $\cal{P}$ is just an element 
of $\cal{P}$ and a $1$--simplex $b$ consists of two $0$--simplices denoted 
$\partial_0b$ and $\partial_1b$ contained in a third element $|b|$ of 
$\cal{P}$ called the {\it support} of $b$. More generally, an $n$--simplex  
is an order--preserving map into $\cal{P}$ from the set of subsimplices of the 
standard $n$--simplex, ordered under inclusion. $\Sigma_n(\cal{P})$ or just 
$\Sigma_n$ will denote the partially ordered set of $n$--simplices of 
$\cal{P}$ with the pointwise ordering.\smallskip

  A partially ordered set $\cal{P}$ is {\it connected} if given 
$a,a'\in\Sigma_0(\cal{P})$, there is a path from $a$ to $a'$ in $\cal{P}$, i.e. if there 
exist $b_0,b_1,\dots,b_n\in\Sigma_1(\cal{P})$ with $\partial_0b_0=a$, 
$\partial_1b_n=a'$ and $\partial_0b_i=\partial_1b_{i-1}$, $i=1,2,\dots,n$. 
Obviously, if $\cal{P}$ is not connected, it is a disjoint union of its 
connected components. We will be taking for $\cal{P}$ not only subsets of 
$\cal{K}$ with the induced order but also   
of $\cal{K}\times\cal{K}$ with the product ordering. 
These notions are related to topological
 notions in the following way.\smallskip 

\noindent
{\bf Lemma 3A.1} {\it Let $\cal{P}$ be a
 base for the topology of a space $M$ and 
ordered under inclusion and suppose the elements
 of $\cal{P}$ are open, (non-empty) and 
path--connected. Then an open subset $X$ of $M$ is path--connected if and only 
if $\cP_X$:=$\{\cO\in\cP:\cO\subset X\}$ is connected.}\smallskip  

\noindent
\begin{proof}
 Any two points of $X$ are contained in elements of 
$\cP_X$ so if this is connected and each of its elements are path--connected 
the two points can be joined by a path in $X$. Conversely, given 
$\cO_0,\cO_1\in\cP_X$, there is a path in $X$ beginning in 
$\cO_1$ and ending in $\cO_0$, if $X$ is pathwise connected. Since 
$\cal{P}$ is a base for the topology, it is easy to construct a path in 
$\cP_X$ joining $\cO_1$ and $\cO_0$.
\end{proof}  

  A subset $\cal{S}$ of $\cal{P}$ of the form $\cP_X$ has the property that 
$\cal{O}\in\cal{S}$ and $\cO_1\subset\cal{O}$
 implies $\cO_1\in\cal{S}$. 
Such subsets are referred to as {\it sieves}. If $\cal{P}$ is a base for the 
topology of $M$ then a sieve $\cal{S}$ is a base for the topology of the 
open subset $X_{\cal{S}}:=\cup\{\cal{O}:\cal{O}\in\cal{S}\}$. The 
connected components of a partially ordered set are sieves, the union 
or intersection of sieves is again a sieve. 
\smallskip

\noindent
{\bf Corollary 3A.2} {\it Under the hypotheses of Lemma 3A.1, the connected 
components of $\cal{P}$ are of the form $\cP_X$, where $X$ runs over 
the path--connected components of $M$.}\smallskip 

   We let Open$(M)$ denote the set of open subsets of $M$ ordered 
under inclusion and Sieve$(\cK)$ the set of sieves of $\cK$, then 
defining for a open set $X$ of $M$, $\mu(X)$ to be the set of $\cO\in\cK$ 
contained in $X$, $\mu$ is an injective order-preserving 
map from Open$(M)$ to Sieve$(\cK)$. If we define $\nu(\cS):=X_\cS$, 
then $\nu$ is order-preserving and a left inverse for $\mu$.\smallskip 

  The following result will prove useful in calculating the connected 
components of a partially ordered sets.\smallskip 
 
\noindent
{\bf Lemma 3A.3} {\it Let $i\mapsto\cP_i$ be an order--preserving map 
from a partially ordered set $I$ to the set of sieves of a partially 
ordered set $\cal{P}$ ordered under inclusion. Suppose that 
$\cP=\cup_{i\in I}\cP_i$. Let $\cal{C}\subset\cal{P}$ 
and set $\cC_i:=\cC\cap\cP_i$ then $\cal{C}$ is a union of 
components of $\cal{P}$ if and only if $\cC_i$ is a union of 
components of $\cP_i$ for each $i\in I$. If $I$ is connected and 
$\cC_i$ is either empty or a component of $\cP_i$, $i\in I$, 
then $\cal{C}$ is a component of $\cal{P}$.}\smallskip 

\noindent
\begin{proof}
 If $\cC$ is a union of components and $b\in\Sigma_1(\cP_i)$ 
with $\partial_1b\in C_i$ then $b\in\Sigma_1(\cal{P})$
so $\partial_0b\in \cC\cap\cP_i=\cC_i$ and $\cC_i$ is a union of components. 
Conversely, if each $\cC_i$ is a
 union of components and $b\in\Sigma_1(\cal{P})$ 
with $\partial_1b\in \cC$, then $|b|\in\cP_i$ for some $i$. But 
$\cP_i$ is a sieve so $b\in\Sigma_1(\cP_i)$ and $\partial_1b\in \cC\cap\cP_i$. 
Since $\cC_i$ is a union of components, $\partial_0b\in\cC_i\subset\cal{C}$ 
so $\cal{C}$ is a union of components. Now $\cal{C}$ is a component, if 
any given pair $a\in\cC_i$ and $a'\in\cC_{i'}$ 
can be joined by a path in $\cal{C}$. But $I$ being connected, we may as well  
suppose $i$ and $i'$ have an upper bound $j\in I$.  If $\cC_j$ is 
a component, $a$ and $a'$ can even be joined by a path in $\cC_j$, 
completing the proof of the lemma.
\end{proof}  

   Now an automorphism $g$ of a partially ordered set $\cal{P}$ such that 
given $\cal{O}\in\cal{P}$ there is a $b\in\Sigma_1(\cal{P})$ with 
$\partial_1b\subset\cal{O}$ and $\partial_0b\subset g\cal{O}$ obviously 
leaves each connected component of $\cal{P}$ globally invariant. If 
$G$ is a connected topological group acting continuously on a topological 
space $M$ and $\cal{P}$ is a base for the topology of $M$, then it is easy 
to see that given $\cal{O}\in\cal{K}$ there is a $\cO_1\in\cal{K}$ and 
a neighbourhood $\cal{N}$ of the unit in $G$ such that 
$\cN\cO_1\subset\cal{O}$. It follows that $G$ leaves any 
path--component of $\cal{P}$ globally invariant. Of course, this may also 
be deduced from Corollary 3A.2.\smallskip 

  After these generalities on partially ordered sets, we turn to the theory 
of superselection sectors and need a partially ordered set $\cal{K}$ 
equipped with a binary relation $\perp$ satisfying a), b) and c) of 
Sec.\ 3.1. Note that b) just says that $\cal{O}^\perp$ is a sieve of $\cal{K}$.
There are two derived binary relations $\tilde\perp$ and 
$\hat\perp$ defined by supplementing $\cO_1\perp\cO_2$ by requiring 
that there exists an $\cO_3\in\cal{K}$ such that 
$$\cO_1\perp\cO_3,\,\,\cO_2\perp\cO_3$$ 
or such that
$$\cO_1,\,\,\cO_2\subset\cO_3,$$ 
respectively. These relations automatically satisfy a) and b) but c) 
remains to be checked and will not prove to be a problem in our applications 
to curved spacetime. The operation of passing from $\perp$ to $\tilde\perp$ 
or $\hat\perp$ is idempotent and if $\cal{K}$ is directed, all three 
relations coincide. Furthermore, by Lemma~\ref{hatduality}, the corresponding 
notions of duality coincide for additive nets when $\cK$ is the set of regular diamonds in a 
globally hyperbolic spacetime.\smallskip

  If $\cO_1\perp\cO_2$ and $\cO_3$ has non-trivial causal complement 
in $\cal{O}_2$, i.e.\ if there exists an $\cO_4$ with $\cO_3\perp\cO_4$, 
$\cO_3,\cO_4\subset\cO_2$ then trivially $\cO_1\tilde\perp\cO_3$. 
Now a regular diamond is a union of a sequence of smaller regular 
diamonds with non-trivial causal complement in the original 
regular diamond. Thus when $\cK$ is the set of regular diamonds, 
the difference between the relations $\perp$ and $\tilde\perp$ 
is, in this sense, a boundary effect.\smallskip 

  The difference between $\tilde\perp$ and $\hat\perp$ merely reflects 
the potential difficulty of finding suitably large regular diamonds. If 
we replace the set $\cK$ of regular diamonds by the set $\tilde\cK$ of 
sieves in $\cK$ with non-trivial causal complement, defining the causal 
complement $\cS^\perp$ of a sieve $\cS$ to be the sieve 
$\cS^\perp:=\cap_{\cO\in\cS}\cO^\perp$, then 
$\tilde\perp=\hat\perp$. In fact, if $\cS_1\tilde\perp\cS_2$, then 
$(\cS_1\cup\cS_2)^\perp=\cS_1^\perp\cap\cS_2^\perp\neq\emptyset$ so that 
$\cS_1\tilde\perp\cS_2$.\smallskip

  If $\cK$ is a base of open sets of a topological space $M$ and 
the relation $\perp$ on $\cK$ is induced by a relation $\perp$ on 
Open$(M)$ satisfying a) and b) of Sec.\ 3.1 and which is local in the 
sense that if $X\in$Open$(M)$ and $X\subset\cup_i\cO_i$, then 
$\cO_i\perp\cO$ for all $i$ implies $X\perp\cO$. This condition is obviously 
satisfied by the relation of causal disjointness on a globally hyperbolic 
spacetime. It implies that $\mu(X^\perp)=\mu(X)^\perp$. We also have 
$\nu(\cS)^\perp=\nu(\cS^\perp)$ for any sieve $\cS$ in $\cK$.\smallskip 

\noindent
{\bf Lemma 3.A4} {\sl When restricted to causally closed open sets and sieves, 
the maps $\mu$ and $\nu$ are inverses of one another.}\smallskip 

\noindent
{\bf Proof.} If $\cS$ is a sieve and $X:=\nu(\cS)$, then 
$\mu(X)^\perp=\cS^\perp$. If $\cS$ is causally closed, so is 
$X$ since $\mu$ is injective. On the other hand, if $X$ is causally 
closed and we set $\cS:=\mu(X)$, then 
$S^{\perp\perp}=\mu(X^{\perp\perp})=\mu(X)$ and $\cS$ is causally closed. 
It remains to show that $\cS=\mu\nu(\cS)$ if $\cS$ is 
causally closed. But, in this case, 
$$\cS\subset\mu\nu(\cS)\subset\mu\nu(\cS)^{\perp\perp}
=\cS^{\perp\perp}=\cS,$$ 
completing the proof.\smallskip

   By a representation $\pi$ of a net of von Neumann algebras $\cA$ over 
$\cK$ we mean normal representations $\pi_\cO$ of $\cA(\cO)$ on a 
Hilbert space $\cH_\pi$ such that $\pi_{\cO_1}$ is $\pi_{\cO_2}$ restricted 
to $\cA(\cO_1)$, whenever $\cO_1\subset\cO_2$ in  $\cK$.\smallskip 

  If $G$ is a group of automorphisms of $\cK$ and $(\cA,\alpha)$ is a 
covariant net then a covariant representation is a pair $(\pi,U)$ consisting 
of a representation $\pi$ of $\cA$ and a unitary representation of 
$G$ on $\cH_\pi$ such that $U(g)\pi_\cO(A)=\pi_{g\cO}(\alpha_g(A)U(g)$, 
$A\in\cA(\cO)$, $g\in G$.\smallskip

  We now provide a cohomological interpretation
 of superselection sectors leading to 
a proof of the Extension Theorem of Sec.\ 3.1. To enter into the spirit of 
the cohomological interpretation, we
 regard $\cO^\perp$, $\cO\in\cK$ 
as being a covering of $\cK$, the {\it causal covering}. The selection 
criterion selects those representations that are trivial on the causal cover 
and these representations allow a cohomological description in analogy with 
locally trivial bundles.\smallskip 
 
   For each $a\in\Sigma_0$ we pick a unitary $V_a$ such that 
$$V_a\pi_{\cO}(A)=AV_a,\quad A\in\cA(\cO),\quad \cO\perp a$$ 
and set 
$z(b):=V_{\partial_0b}V_{\partial_1b}^*,\,\,b\in\Sigma_1$. Obviously 
if $\cO\in|b|^\perp$, $z(b)\in\cA(\cO)'$ thus $z(b)\in\cA^d(|b|)$. 
Furthermore,
$$z(\partial_0c)z(\partial_2c)=z(\partial_1c),\quad c\in\Sigma_2$$
so that $z$ is a unitary 1--cocycle with values in the dual net $\cA^d$. 
We consider such 1--cocycles as objects of a category $Z^1(\cA^d)$, 
where an arrow $t$ in this category from $z$ to $z'$ is a
 $t_a\in \cA^d(a)$, 
$a\in\Sigma_0$, such that 
$$t_{\partial_0b}z(b)=z'(b)t_{\partial_1b},\quad b\in\Sigma_1.$$ 
This makes $Z^1(\cA^d)$ into a $W^*$--category. Note that $\|t_a\|$ 
is independent of $a$.\smallskip

   If we were to make a different choice $V'_a$ of unitaries $V_a$, then 
setting $z'(b):= V'_{\partial_0b}V'{}^*_{\partial_1b}$ and $w_a:=V'_aV_a^*$, 
we see that $w_a\in\cA^d(a)$ and
$w_{\partial_0b}z(b)=z'(b)w_{\partial_1b}$. Thus $w\in (z,z')$ is a unitary and the 1--cocycle 
attached to $\pi$ is defined up to unitary equivalence in $Z^1(\cA^d)$. 
More generally, if $T\in (\pi,\pi')$ and $\pi$ and $\pi'$ are trivial on 
the causal cover and $z$ and $z'$ are associated cocycles defined by unitaries 
$V_a$ and $V'_a$, as above, set 
$$t_a:=V'_aTV_a^*,\,\,a\in\Sigma_0.$$ 
Then $t_a\in \cA^d(a)$ and 
$$t_{\partial_0b}z(b)=V'_{\partial_0b}TV_{\partial_0b}^*V_{\partial_0b}
V_{\partial_1b}^*=V'_{\partial_0b}TV_{\partial_1b}^*=V'_{\partial_0b}
V'{}^*_{\partial_1b}V'_{\partial_1b}TV_{\partial_0b}^*=z'(b)t_{\partial_1b},$$ 
so that $t\in (z,z')$. Conversely, if $t\in (z,z')$ then 
$T:=V'{}^*_at_aV_a$ is independent of $a$ so that 
$$T\pi_{\cal O}(A)=\pi'_{\cal O}(A)T,\quad A\in\cA(\cal O),
\quad \cal O\in\cal K.$$ 
and we clearly have a close relation between $Z^1(\cA^d)$ and the 
$W^*$--category Rep$^\perp\cA$ of representations of $\cA$ trivial on 
the causal cover.\smallskip 

   However, any cocycle $z$ arising from such a representation has two special 
properties that may not be shared by a general $1$--cocycle. First, $z$ is 
trivial on $\cal{B}(\cal{H}_0)$, i.e.\ there are unitaries $V_a$, $a\in\Sigma_0$, 
on $\cal{H}_0$ such that $z(b)=V_{\partial_0b}V^*_{\partial_1b}$, $b\in\Sigma_1$.\smallskip 

   If $\cal K$ is directed then $\Sigma_*(\cal K)$ admits 
a contracting homotopy\cite{R}. In this case every $1$--cocycle of $\cA^d$ is trivial 
in $\cB(\cH_0)$. In general, if we consider the graph 
with vertices $\Sigma_0$ and arrows $\Sigma_1$ then the category generated 
by this graph has as arrows the paths in $\cal K$. Thus every $1$--cocycle 
extends to a functor from this category. When $z$ is trivial on $\cB(\cH_0)$ 
then $z(p)$ for a path $p$ depends only on the endpoints $\partial_0p$ and 
$\partial_1p$ of the path. 
Conversely, if $z(p)$ just depends on the endpoints of $p$ and $\cal K$ is 
connected, then $z$ is trivial on $\cB(\cH_0)$. To see this we 
pick a base point $a_0\in\Sigma_0$, then given $a\in\Sigma_0$ a path 
$p_a$ with $\partial_0p_a=a$ and $\partial_1p_a=a_0$ and finally define 
$y(a)=z(p_a)$. $z(p)y(\partial_1p)=y(\partial_0p)$, so we have trivialized 
$z$ in $\cB(\cH_0)$.\smallskip 

Secondly, for any path $p$, $z(p)Az(p)^*=A$ whenever $A\in\cA(\cal{O})$ 
and $\partial_0p$, $\partial_1p\in\cO^\perp$. The full subcategory of 
$Z^1(\cA^d)$ whose objects satisfy these two conditions will be denoted 
by $Z^1_t(\cA^d)$.\smallskip

   The following simple result shows that the second condition is 
automatically satisfied in an important special case.\smallskip 

\noindent
{\bf Lemma 3A.5} {\it If $\cO^\perp$ is connected, then any object $z$ 
of $Z^1(\cA^d)$, trivial on $\cB(\cH_0)$ satisfies 
$$z(p)Az(p)^*=A,\quad \partial_0p\,,\partial_1p\in\cO^\perp,\,\, 
A\in\cA(\cal{O}).$$} 

\noindent
\begin{proof}
 Since $\cal{O}^\perp$ is connected, it suffices to prove the result 
when the path $p$ is a $1$--simplex $b$ with $|b|\in\cO^\perp$. 
But then, $z(b)\in\cA^d(|b|)\subset\cA(\cO)'$.
\end{proof} 
 
   Having discussed these two conditions, we can give our cohomological 
characterization of the selection criterion.\smallskip

\noindent 
{\bf Theorem 3A.6} {\it The $W^*$--categories {\rm Rep}$^\perp\cA$ and 
$Z^1_t(\cA^d)$ are equivalent.}\smallskip

\noindent
\begin{proof} We pick unitaries $V_a^\pi$, $a\in\Sigma_0$, as above, 
for each object $\pi$ of Rep$^\perp\cA$. Given an arrow $T\in (\pi,\pi')$ 
in that category, we define for $b\in\Sigma_1$, $a\in\Sigma_0$
$$F(\pi)(b)=V_{\partial_0b}^\pi V_{\partial_1b}^{\pi *};
\quad F(T)_a:=V_a^{\pi'}TV_a^{\pi*}.$$ 
Then $F$ is a faithful $^*$-functor and our computations above show 
that it is full. Hence, it remains to show that each object $z$ of $Z^1_t(\cA^d)$, 
is equivalent to an object in the image 
of $F$. We show this by constructing a representation $\pi^z$. We pick unitaries 
$V_a$, $a\in\Sigma_0$, on $\cH_0$ such that 
$z(b)=V_{\partial_0b}V_{\partial_1b}^*,\quad b\in\Sigma_1,$ and define
$$\pi^z_{\cal O}(A)=V_a^*AV_a,\quad a\in\cO^\perp,\quad A\in\cA(\cal O).$$
This is well defined since $\cal{K}$ is connected and for any path 
$p$ with $\partial_0p$, $\partial_1p\in\cO^\perp$ we have 
$z(p)\in\cA(\cO)'$. 
Furthermore, the definition respects the net structure since 
$$\pi_{\cO_1}^z(A)=\pi_{\cO_2}^z(A),\quad A\in\cA(\cO_1),
\quad \cO_1\subset\cO_2.$$ 
Hence we get a representation of the net $\cA$, trivial on the covering 
by construction and $V_{\partial_0b}V_{\partial_1b}^*=z(b)$ is an associated 1--cocycle. This completes 
the proof.
\end{proof}

   We now consider the problem of extending
 representations of a net $\cA$, 
trivial on the causal cover, to representations of the bidual net 
$\cA^{dd}$, again trivial on the causal cover.\smallskip 

\noindent 
{\bf Theorem 3A.7} {\it If each $\cal{O}^\perp$ is connected, every object 
$\pi$ of {\rm Rep}$^\perp\cA$ admits a unique extension to an object of 
{\rm Rep}$^\perp\cA^{dd}$. Furthermore there is a canonical isomorphism of 
$W^*$--categories {\rm Rep}$^\perp\cA$
 and {\rm Rep}$^\perp\cA^{dd}$.}\smallskip

\noindent 
\begin{proof}
 Let $V_a$, $a\in\Sigma_0$ be unitaries realizing the equivalence 
of $\pi$ and $\pi^0$ on $a^\perp$. Then $z(b):=V_{\partial_0b}V_{\partial_1b}^*$, 
$b\in\Sigma_1$ is an associated object of $Z^1_t(\cA^d)$. Since 
each $\cal{O}^\perp$ is connected, $z$ is at the same time an object of 
$Z^1(\cA^{ddd})$ by Lemma 3.A.4. 
If we define 
$$\tilde\pi_{\cal O}(A):=V_a^*AV_a,\quad A\in\cA^{dd}(\cO),\quad 
a\in\cO^\perp,$$ 
this gives a well defined element of Rep$^\perp\cA^{dd}$ just as in 
the proof of Theorem 3A.6. Furthermore, $\tilde\pi$ obviously extends $\pi$ 
by the choice of the $V_a$. If we make another choice $V'_a$ of the $V_a$ then 
$V'_aV_a^*\in\cA^d(a)$ so that $\tilde\pi$ remains unchanged and is 
consequently the unique extension of $\pi$ to an object of 
Rep$^\perp\cA^{dd}$. Passing to the extensions does not change 
the intertwiners by Theorem 3A.6.
\end{proof}

  For the further development of superselection theory, we must assume 
duality $\cA=\cA^d$, although essential duality would do whenever 
each $\cO^\perp$ is connected. We shall even need to assume 
$\tilde\perp$--duality, but this coincides with duality in curved 
spacetime whose status is commented on in Sec.\ 4.2.
\smallskip 

  The next goal is to show that sectors have 
a tensor structure. More precisely, we shall show that $Z^1(\cA)$ 
has a canonical structure of a tensor $W^*$--category arising by adjoining 
endomorphisms. If $\cA$ is a 
net of von Neumann algebras, then there is an associated net $\cO\mapsto
\text{End}\cA(\cal{O})$ of tensor $W^*$--categories.
 End$\cA(\cal{O})$ 
has as objects the normal endomorphisms of the net $\cO_1\mapsto
\cA(\cO_1)$, i.e.\ normal endomorphisms $\rho_{\cO_1}$ of 
$\cA(\cO_1)$ compatible with the net structure. An arrow 
$T\in(\rho,\sigma)$ in End$\cA(\cO)$ is a $T\in\cA(\cO)$ 
such that 
$$T\rho(A)=\sigma(A)T,\quad A\in\cA(\cO_1),\,\,\cO
\subset\cO_1.$$ 
The tensor structure is defined on the lines of Sec.\ 3.3 
and the net structure is given  by the obvious restriction mappings.

  The construction of appropriate endomorphisms is just a variant on that 
already used to pass from a $1$--cocycle $z\in Z^1_t(\cA^d)$ to a 
representation $\pi^z$. Given $a\in\Sigma_0$, 
and $A\in\cA(\cal{O})$, $a\subset\cal{O}$ pick a path $p$ with 
$\partial_0p=a$ and $\partial_1p\in\cO^\perp$ and set 
$$y(a)(A):=z(p)Az(p)^*.$$
$y(a)(A)$ is independent of the choice of $p$ since 
$z\in Z^1_t(\cA^d)$. Given $X\in\cA(\cO_1)$ 
with $\cO_1\perp\cal{O}$, $\cO_2$  
with $\cO_2\perp\cal{O}$ and $\cO_2\perp\cO_1$ and choosing 
$\partial_1p=\cO_2$, we see that $y(a)(A)$ and $X$ commute so that 
$y(a)(A)\in\cA(\cal{O})$ by $\tilde\perp$--duality.
Thus $y(a)$ is an object of End$\cA(a)$.\smallskip

But $y(a)$ is not only localized in $a$ in the sense of net
automorphisms but also in the sense of
superselection theory in that $y(a)(A)=A$ 
whenever $A\in\cA(\cO_1)$ where $\cO_1\in a^\perp$ and 
$\cO_1,\,\,a\subset\cal{O}$, since the endpoints of $p$  
lie in $\cO_1^\perp$. We write $\Delta(a)$ to denote the 
objects of End$\cA(a)$ satisfying this second localization 
condition and denote by $\cT(a)$ the corresponding full tensor 
$C^*$--subcategory of End$\cA(a)$. \smallskip 

\noindent
{\bf Lemma 3A.8} {\it Let $p$ be a path with $\partial_1p,\,\partial_0p\subset 
\cal{O}$ then 
$$z(p)y(\partial_1p)(A)=y(\partial_0p)(A)z(p),\quad A\in
\cA(\cal{O}).$$}

\noindent
{\bf Proof.} Given $A\in\cA(\cal{O})$ 
and a path $p$ with $\partial_1p,\,
\partial_0p\subset\cal{O}$, 
pick paths $p',p''$ with $\partial_0p'=
\partial_1p$, $\partial_0p''=\partial_0p$ and $\partial_1p',\partial_1p''
\in\cal{O}^\perp$, then 
$$z(p)y(\partial_1p)(A)=z(p)z(p')Az(p')^*=z(p'')Az(p'')^*z(p)=
y(\partial_0p)(A)z(p),$$ 
as required.\smallskip 

   Furthermore if $t\in(z,\hat z)$, $A\in\cA(\cal{O})$ and $p$ is a path 
with $\partial_0p=a\subset\cal{O}$ and $\partial_1p\subset\cO^\perp$ then 
$$t_ay(a)(A)=t_az(p)Xz(p)^*=\hat z(p)t_{\partial_1p}Az(p)^*=
\hat z(p)A\hat z(p)^*t_a=\hat y(a)(A)t_a.$$ 
In other words $t_a\in(y(a),\hat y(a))$.\smallskip 

   These results admit the following interpretation.\smallskip 

\noindent
{\bf Theorem 3A.9} {\sl Let $\cA$ be a net over $(\cK,\perp)$ 
satisfying $\tilde\perp$--duality. If $z$ is a $1$--cocycle 
of $\cA$ trivial in 
$\cB(\cH_0)$ then $(y,z)$ is a $1$--cocycle in the net 
$\cal{T}$ of tensor $W^*$--categories 
and the map $z\mapsto(y,z)$ together with the identity map on arrows 
is an isomorphism of $Z^1_t(\cA)$ and 
$Z^1_t(\cal{T})$.}\smallskip

   Now, $\cal{T}$ being a net of tensor $W^*$--categories, $Z^1(\cal{T})$ is 
itself a tensor $W^*$--category. Given $1$--cocycles $(y_1,z_1)$ and $(y_2,z_2)$, 
their tensor product is the $1$--cocycle $(y,z)$, where 
$$y(a)=y_1(a)y_2(a),\quad z(b)=z_1(b)y_1(\partial_1b)(z_2(b)).$$ 
If both $(y_1,z_1)$ and $(y_2,z_2)$ are trivial in $\cB(\cH_0)$ 
then so is their tensor product. The tensor product on arrows is defined 
as follows: if $t_i$ maps from $(y_i,z_i)$ to $(y'_i,z'_i)$ for $i=1,2$, 
then the tensor product $t_1\otimes t_2$ is given by 
$$(t_1\otimes t_2)_a=t_{1,a}y_1(a)(t_{2,a}).$$ 
This completes our goal of describing superselection structure in terms 
of a tensor $W^*$--category. Note that we could have used the subnet 
$\cT_t$ in place of $\cal{T}$ defined by requiring an object $\rho$ 
of $\cal{T}(\cal{O})$ to be transportable, i.e.\ there exists a map 
$a\mapsto\rho_a$, where $\rho_a$ is an object of $\cT(a)$ and 
$\rho_a=\rho$ when $a=\cal{O}$ and a map $\Sigma_1\ni b\mapsto u(b)$, 
where $u(b)$ is an arrow from $\rho_{\partial_1b}$ to $\rho_{\partial_0b}$
in $\cT(|b|)$. In fact the tensor $W^*$--categories $Z^1_t(\cal{T})$ and 
$Z^1_t(\cT_t)$ are canonically isomorphic. In Sec.\ 3.3, we show how to get 
a net $(\cT_t,\varepsilon_c)$ of braided tensor $W^*$--categories and 
it is a simple general fact that
 this leads to a braided tensor $W^*$--category, 
$(Z^1_t(\cT_t),\varepsilon_c)$. We need only set 
$$\varepsilon_c(z,z')_a:=\varepsilon(y(a),y'(a)).$$ 
Since this expression obviously acts correctly on 
the arrows evaluated in $a$ and the laws for a braiding 
hold for each $a$, the only point that has to be checked is that 
$\varepsilon(z,z')$ is an arrow from $z\times z'$ to $z'\times z$. 
However, if $b\in\Sigma_1$, $z(b)\in(\rho_{\partial_1b},\rho_{\partial_0b})$ 
in $\cT_t(|b|)$ and similarly for $z'(b)$. Thus 
$$z'(b)\times z(b)\circ\varepsilon(\rho_{\partial_1b},\rho'_{\partial_1b})=
\varepsilon(\rho_{\partial_0b},\rho'_{\partial_0b})\circ z(b)\times z'(b),$$ 
as required.\smallskip 

   Thus the cohomological approach leads to a braided tensor $W^*$--category 
$(Z^1_t(\cT_t)$, $\epsilon_c)$ describing superselection structure 
and in the context of globally hyperbolic spacetimes this is even a 
symmetric tensor $W^*$--category for spacetime dimensions $\geq 2$. It should 
be noted that except when $\cal{K}$ is directed, we have not given a direct 
description of this structure in terms of transportable localized 
endomorphisms. In particular, it not clear that every transportable localized 
endomorphism arises from a $1$--cocycle. Furthermore, if $\rho$ and $\sigma$ 
are in $\Delta_t(\cal{O})$ and $T$ is a bounded operator on the ambient 
Hilbert space, such that 
$$T\rho(A)=\sigma(A)T,\quad A\in\cA(\cO_1),\,\cO\subset\cO_1,$$ 
then $T$ commutes with $\cA(\cO_2)$ for $\cO_2\perp\cal{O}$ 
provided there is a $\cO_1$ with $\cO,\cO_2\subset\cO_1$. 
This means, we would need duality with respect to the modified relation 
$\hat\perp$ to be able to conclude that $T\in\cA(\cal{O})$ and hence 
that $T$ is an arrow from $\rho$ to $\sigma$ in $\cT_t(\cal{O})$. 
Conversely, if $\pi$ and $\pi'$ are representations satisfying the 
selection criterion and restricting to endomorphisms $\rho$ and $\rho'$ 
in $\Delta_t(\cal{O})$ then it is not clear that an
 arrow $T\in(\rho,\rho')$ in 
$\cT_t(\cal{O})$ will at the same time intertwine $\pi$ and $\pi'$.\smallskip 

   These points should be bourne in mind, when, in the main body of the 
text, we avoid the cohomological description and put the emphasis on
transportable localized endomorphisms.\smallskip 

   To proceed with the analysis of statistics, we need to use left inverses 
and we examine, at this point, the notions involved and the relations 
between them. If $\pi$ is a representation of $\cA$ on $\cH_0$ then 
we define a left inverse $\phi$ of $\pi$ to be given by unital positive linear 
mappings $\phi_{\cal{O}}$ on $\cB(\cH_0)$ 
compatible with the net inclusions and satisfying 
$$\phi_{\cO}(A\pi_{\cO}(B))=\phi_{\cO}(A)B,\quad A,B\in\cA(\cal{O}).$$
Note that if $\pi_{\cO}(B)=B$ then $\phi_{\cO}(B)=B$. If $\pi$ is 
localized in $\cal{O}$ in the sense that 
$$\pi_{\cO_1}(A)=A,
\quad \cO\perp\cO_1,\quad A\in\cA(\cO_1),$$ 
then $\phi$ is localized in $\cal{O}$ in the same sense. Furthermore, 
if $\cO_1\hat\perp\cO_2$ and $\cO\subset\cO_2$, then 
$\phi_{\cO_2}(A)B=B\phi_{\cO_2}(A)$ for $A\in\cA(\cO_2)$ 
and $B\in\cA(\cO_1)$. In fact, picking $\cO_3$ with 
$\cO_1,\cO_2\subset\cO_3$ we have 
$$\phi_{\cO_2}(A)B=\phi_{\cO_3}(A)B=\phi_{\cO_3}(A\pi_{\cO_3}(B))$$
$$=\phi_{\cO_3}(A\pi_{\cO_1}(B))=\phi_{\cO_3}(AB).$$ 
Since $A$ and $B$ commute, we interchange them and reverse the steps 
to conclude that $\phi_{\cO_2}(A)$ and $B$ commute. This proves 
the following result.\smallskip 

\noindent
{\bf Lemma 3A.10} {\it If $\phi$ is a left inverse for a representation 
$\pi$ localized in $\cal{O}$ then $\phi$ is localized in $\cal{O}$ and 
if duality holds for the relation $\hat\perp$, $\phi_{\cO_1}\cA
(\cO_1)\subset\cA(\cO_1)$ for $\cO\subset\cO_1$.}\smallskip 

   The restriction of $\pi$ to the net $\cO_1\mapsto\cA(\cO_1)$, 
$\cO_1\supset\cO$ is a localized endomorphism $\rho$ and a object 
of the tensor $W^*$--category End$\cA(\cO)$. The above notion of left 
inverse adapts easily to localized endomorphisms. If $\rho$ is localized 
in $\cO$, a {\it left inverse} of $\rho$ is a family 
$\cO_1\supset\cO\mapsto \phi_{\cO_1}$ of unital positive linear 
mappings on the $\cA(\cO_1)$, compatible with 
the net inclusions and satisfying 
$$\phi_{\cO_1}(A\rho_{\cO_1}(B))=\phi_{\cO_1}(A)B,\quad A,\,B\in\cA(\cO_1).$$ 
Obviously, a left inverse for $\rho$ considered as a representation yields a 
left inverse for the endomorphism $\rho$ on restriction. If $\bar\rho$ is a 
conjugate for $\rho$ then we get a left inverse $\phi$ for $\rho$ by setting 
$$\phi_{\cO_1}(A):= V^*\bar\rho_{\cO_1}(A)V,\quad A\in\cA(\cO_1),\,\,
\cO_1\supset\cO,$$ 
where $V\in(\text{id},\bar\rho\rho)$ is an isometry.\smallskip

   The restriction of $\pi$ to the net $\cO_1\mapsto\cA(\cO_1)$, 
$\cO_1\supset\cO$ is a localized endomorphism $\rho$ and a object 
of the tensor $W^*$--category End$\cA(\cal{O})$. 
We now show that a left inverse $\phi$ for $\rho$ induces a 
left inverse of $\rho$ in the categorical sense
\cite{LoRo}. In other words, we need a set
$$\phi_{\sigma,\tau}:(\rho\sigma,\rho\tau)\to(\sigma,\tau),$$
of linear mappings where $\sigma$, $\tau$ are objects of the category.
These have to be natural in $\sigma$ and $\tau,$ i.e. given
 $S \in (\sigma, \sigma')$ and $T \in
(\tau, \tau')$ we have
$$\phi_{\sigma', \tau'} (1_\rho \otimes T \circ X \circ
1_{\rho} \otimes S^*) = T \circ \phi_{\sigma, \tau} (X) \circ S^*,
\ X \in (\rho \sigma, \rho \tau),$$
and furthermore to satisfy
$$\phi_{\sigma \nu, \tau \nu} (X \otimes 1_\pi) = 
\phi_{\sigma, \tau} (X) 
\otimes 1_\nu, \ X \in (\rho \sigma, \rho \tau)$$
for each object $\nu $. We will require that $\phi$ is 
positive in the sense that 
$\phi_{\sigma, \sigma}$ is positive for each $\sigma$ and normalized in 
the sense that
$$\phi_{\iota, \iota}(1_\rho) = 1_\iota.$$
We say that $\phi$ is faithful if 
$\phi_{\sigma, \sigma}$ is faithful for each object $\sigma$.\par 

  Now, given $T\in(\rho\sigma,\rho\tau)$, we recall that $T\in\cA(\cal{O})$. 
Hence we set 
$$\phi_{\sigma,\tau}(T)=\phi_{\cal{O}}(T)$$ 
and since $\phi_{\cal{O}}(T)\in\cA(\cal{O})$ by Lemma 3A.9, we conclude 
without difficulty that we get a left inverse for $\rho$ in this way.\smallskip 
 
   On the other hand, if we are dealing with a representation satisfying the
selection criterion then we know that, by passing to an associated $1$--cocycle, 
we get a field $a\mapsto y(a)$ of localized endomorphisms under the weaker 
assumption that duality holds for the relation $\tilde\perp$. In this case, 
we would actually like a left inverse for the $1$--cocycle considered as 
an object of the tensor $W^*$--category $Z^1_t(\cA)$. To this end, we 
pick, for each of the associated endomorphisms $y(a)$ a left inverse $\phi_a$ 
and ask whether $a\mapsto \phi_a(t_a)$ is an arrow from $z'$ to $z''$, whenever 
$a\mapsto t_a$ is an arrow from $z\times z'$ to $z\times z''$. Thus 
$t_a\in\cA(a)$ and 
$$(z\times z'')(b)t_{\partial_1b}=t_{\partial_0b}(z\times z')(b).$$ 
It follows that 
$$z''(b)\phi_{\partial_1b}(t_{\partial_1b})=
\phi_{\partial_1b}(y(\partial_1b)(z''(b))t_{\partial_1b})
=\phi_{\partial_1b}(z(b)^*t_{\partial_0b}z(b))z'(b)$$ 
and we deduce the following lemma.\smallskip 

\noindent
{\bf Lemma 3A.11} {\it If $z\in Z^1_t(\cA)$ and $a\mapsto y(a)$ is the 
associated field of endomorphisms. Then a field $a\mapsto \phi_a$ of 
left inverses of the $y(a)$ defines a left inverse for $z$ by the formula 
$$\phi_{z',z''}(t)_a:=\phi_a(t_a)$$ 
provided $\phi_{\partial_0b}=\phi_{\partial_1b}${\rm Ad}$z(b)^*$
 for $b\in\Sigma_1$.}
\smallskip 

  There is no a priori reason to suppose that every left inverse 
for a $1$-cocycle arises from such a field of left inverses. In 
particular a map $t\in(z,z')\mapsto t_a\in(y(a),y'(a))$ might not 
be surjective. We can also not just begin with a left inverse $\phi_a$ for 
$y(a)$ since it is not clear that we get a field of left inverses using 
the cocycle. However, if we assume, as in Sec.\ 3.4, that $\cal{K}$ 
has an asymptotically causally disjoint net $\cO_n$, then we can 
construct left inverses for $1$-cocycles. If $z$ is an object of 
$Z^1_t(\cA)$, we denote by $z(a,n)$, the evaluation of 
$z$ on a path $p$ with $\partial_0b=a$ and $\partial_1b=\cO_n$. 
This is independent of the chosen path. We now define $\phi_a(X)$
to be a Banach--limit over $n$ of $z(a,n)^*Xz(a,n)$. Then 
$\phi_a$ is a positive linear map satisfying 
$$\phi_a(X)A=\phi_a(X\pi^a_{\cO}(A)),\quad A\in\cA(\cO).$$ 
Furthermore, from the cocycle identity we have 
$$\phi_{\partial_0b}=\phi_{\partial_1b}\text{Ad}(z(b)).$$ 
Since each $\phi_a$ defines a left inverse for $y(a)$, we have constructed 
a left inverse for $z$ by Lemma 3A.11.\smallskip 

   One sometimes wishes to consider nets defined over a wider class 
of regions than say just the set of regular diamonds. Thus in Sections 4 
and 5, we are interested in defining the von Neumann algebras of wedge 
regions. Furthermore, another reason for wanting von Neumann algebras 
associated with large rather than small regions is that we can only compose 
endomorphisms if we find a joint localization region for the endomorphisms 
involved. We consider here the task of extending the domain of definition 
of the net in the context of the present formalism where $\cK$ is a partially 
ordered set commenting on the relation with regions of spacetime afterwards. 
Thus instead of a region, we use the notion of a sieve $\cS$, see above, 
and  consider the set $\tilde{\cK}$ of sieves $\cS$ of $\cK$ 
such that neither $\cS$ nor $\cS^\perp$ are the empty set, ordered 
under inclusion. To each such sieve $\cS$, 
we associate the von Neumann algebra $\cA(\cS)$ generated by the 
$\cA(\cO)$ with $\cO\in\cS$ in the defining representation.
\smallskip 

  We now show that a representation $\pi$ of $\cA$ satisfying 
the selection criterion has a natural extension to a representation 
of the net $\cS\mapsto\cA(\cS)$. We pick for each $a\in\Sigma_0$ 
a unitary $V_a$ such that 
$$\pi_{\cO}(A)=V_a^*AV_a,\quad A\in\cA(\cO),\,\,\cO\in a^\perp,$$ 
and then define 
$$\pi_{\cS}(A):=V_a^*AV_a,\quad A\in\cA(\cS),\,\,a\in\cS^\perp.$$ 
Note that this expression is well defined being independent of the choice of 
$a\in\cS^\perp$ since if $a'\in\cS^\perp$ then 
$$V_{a'}V^*_a\in\cap_{\cO\in\cS}\cA(\cO)'=\cA(\cS)'.$$ 
In the same way, we see that $\pi_{\cS}$ is independent of the 
choice of $a\mapsto V_a$. Note, too that we get a representation of the 
extended  net in that if $\cS_1\subset\cS_2$ then $\pi_{\cS_1}$ is 
the restriction of $\pi_{\cS_2}$ to $\cA(\cS_1)$.
Obviously, an intertwiner $T\in(\pi,\p')$ over $\cK$ remains 
and intertwiner over $\tilde\cK$ so that effectively Rep$^\perp\cA$ 
remains unchanged when we extend the net.\smallskip 

  That part of the formalism related to the concept of localized 
endomorphism is however sensitive to exteding the net. Although 
localized endomorphisms do not play the same fundamental role as $1$--cocycles, 
we have found it convenient to use them in developing the theory. 
The problems involved in using them are two: they are not defined on the whole 
net and the natural map $(z,z')\mapsto (y(a),y'(a))$ may not be surjective. 
Extending the net improves matters in that localized endomorphisms 
are then defined on more operators and hence have fewer intertwiners. 
Since localized endomorphisms require subsets satisfying $\hat\perp$--duality, 
we benefit from the equality $\tilde\perp=\hat\perp$ on $\tilde{\cK}$.\smallskip 
 
   Supposing we have as usual a field 
$a\mapsto y(a)$, $a\in\Sigma_0$, of localized endomorphisms derived from a 
$1$--cocycle, then we know that if $\cO\tilde\perp\cO_1$ and 
$a\perp\cO_1$, $y(a)(\cA(\cO))\subset\cA(\cO_1)'$. 
Hence $y(a)(\cA(\cS))\subset\cap_{\cO_1\in\cS^{\tilde\perp}}\cA(\cO_1)'$. 
We conclude that if $\tilde\perp$--duality holds for $\cS$ in the defining 
representation in the sense that 
$$\cA(\cS)=\cA(\cS^{\tilde\perp})',$$ 
then $y(a)$ acts as an endomorphism of $\cA(\cS)$, 
$y(a)(\cA(\cS))\subset\cA(\cS)$. Now if $\cS$ satisfies 
$\tilde\perp$--duality then so does $\cS^{\tilde\perp}$. Furthermore,
$\cA(\cS)=\cA(\cS^\perp)'\supset\cA(\cS^{\perp\perp})$. 
Thus $\cA(\cS)=\cA(\cS^{\perp\perp})$. Hence, we may as 
well restrict attention to causally closed sieves and choose as our index 
set the set $\cal{L}$ of non-trivial causally closed sieves $\cS$ for which 
$\tilde\perp$--duality holds either for $\cS$ or for $\cS^\perp$. This choice 
has the disadvantage of depending 
on the theory under consideration but it allows a smooth treatment of 
endomorphisms. In particular, if $\tilde\perp$ duality holds for $\cS$ 
and $a\in\cS^\perp$ then the endomorphism $y(a)$ associated with a 
$1$--cocycle satisfies $y(a)(\cA(\cS^\perp))\subset\cA(\cS^\perp)$, 
because, as we have seen above, duality holds for $\cS^{\tilde\perp}$ 
and $\cA(\cS^\perp)=\cA(\cS^{\tilde\perp})$.\smallskip 

  We shall be assuming $\tilde\perp$--duality for the elements of $\cK$. 
Thus $\cK\subset\cL$ and $\{\cO^\perp:\cO\in\cK\}\subset\cL$. Thus 
$\cL$ is both coinitial and cofinal in $\tilde\cK$. Let us call 
two localized endomorphisms {\it comparable} if they are both localized 
in a common sieve in $\tilde\cK$ and hence in some element of $\cL$.
In this case, it makes sense to talk about intertwining operators 
between the two localized endomorphisms. If $\rho_i$ is localized 
in $\cS_i$, $i=1,2$, then $\rho_1$ and $\rho_2$ are comparable, if 
and only if $\cS_1\cap\cS_2\neq\emptyset$.\smallskip

  We turn now to the notion of left inverse.    If we consider $\pi$ as 
a representation of the extended net $\cS\mapsto\cA(\cS)$, 
then there is an obvious modification of the notion of left inverse as 
we just need to replace $\cO$ everywhere by $\cS$. Suppose 
$\pi$ is localized in $\cS$ and $\phi$ is a left inverse for $\pi$, 
then given $\cS_1\supset\cS$ and $\cO\in\cS_1^{\hat\perp}$, 
we remark that there is a sieve $\cS_2$ with $\cO\in\cS_2$ and 
$\cS_1\subset\cS_2$. Given $A\in\cA(\cS_1)$ and 
$B\in\cA(\cO)$ we have 
$$\phi_{\cS_1}(A)B=\phi_{\cS_2}(A)B=\phi_{\cS_2}(A\pi_{\cS_2}(B))$$
$$=\phi_{\cS_2}(A\pi_{\cO}(B))=\phi_{\cS_2}(AB).$$ 
Since $A$ and $B$ commute, we interchange them and reverse the steps to 
conclude that $\phi_{\cS_1}(A)$ and $B$ commute. Recalling that 
$\tilde\perp=\hat\perp$ on $\tilde{\cK}$, this proves the 
following result.\smallskip

\noindent
{\bf Lemma 3A.12} {\sl Let $\phi$ be a left inverse for a
representation $\pi$ of the extended net $\cS\mapsto\cA(\cS)$
localized in $\cS$.  Then, if $\cS\subset\cS_1$ and
$\tilde\perp$--duality holds for $\cS_1$,
$\phi_{\cS_1}\cA(\cS_1)\subset\cA(\cS_1)$.}\smallskip


\section{The Conformal Spin and Statistics Relation for Spacetimes
With Bifurcate Killing Horizon}

\setcounter{equation}{0}

In the present chapter, we shall specialize our considerations to the
class of spacetimes with a bifurcate Killing horizon
(bKh), whose definition we now summarize, following
Kay and Wald \cite{KayWald}. The interested reader is strongly
recommended to consult this reference for further details
not spelled out here.
The main purpose here is to show that, from the original theory, we can
construct a family of local algebras localized on the horizon, which
possesses a conformal symmetry. Therefore horizon localized superselection
sectors have a conformal spin and we prove that this coincides with
their statistics phase.

\subsection{Spacetimes with bKh}

A spacetime with a bKh is a triple $(M,g,\tau_t)$ where $(M,g)$ is a
four-dimensional, globally hyperbolic
spacetime, although spacetimes with a bKh generalize to other spacetime
dimensions. $(\tau_t)_{t\in \RR}$  is a non-trivial one-parameter
group of isometries of $(M,g)$, assumed to be $C^{\infty}$, and hence
the flow of a Killing vector field $\xi$ on $M$ for
the metric $g$. We often refer to $(\tau_t)_{t\in \RR}$ as the
Killing flow (of the spacetime with bKh). We shall assume that $(M,g)$
is orientable and that the set $\Sigma\subset M$ of fixed points of
$(\tau_t)_{t\in \RR}$ is a two-dimensional smooth, acausal, orientable,
connected submanifold of $M$. It is worth noting
that $\Sigma$, when compact, automatically lies
in some Cauchy-surface, see \cite{KayWald} for a proof.
\par From this data we can construct the bKh, $\hh$, as follows: at each
point $p \in \Sigma$ we choose a pair of linearly independent,
lightlike, future-oriented vectors $\chi_A(p),\chi_B(p) \in T_pM$,
normal to $\Sigma$. They are unique up to scalars and they may be chosen
so
that $\Sigma \owns p \mapsto \chi_A(p)$ and $\Sigma \owns p \mapsto
\chi_B(p)$
are smooth vector fields along $\Sigma$ since
$M$ and $\Sigma$ are orientable.
Now let $\gamma_{Ap}$ and
$\gamma_{Bp}$ be the maximal geodesics with tangents $\chi_A(p)$ and
$\chi_B(p)$ at $p \in \Sigma$, respectively. Since $(\tau_t)$ leaves
each $p \in \Sigma$ fixed, it  maps each of the curves
$\gamma_{Ap}$ and $\gamma_{Bp}$ into itself. Moreover,
$\gamma_{Ap}$ and $\gamma_{Ap'}$ do not intersect for $p \neq p'$, and
the same holds with $B$ in place of $A$.
Now one defines sets $\hh_A$ and $\hh_B$ to be the lightlike
hypersurfaces in $M$ formed by the $\gamma_{Ap}$ and
$\gamma_{Bp}$, respectively, as $p$ ranges over $\Sigma$. Then $\hh :=
\hh_A \cup \hh_B$ is the bKh, and one distinguishes the following subsets:
\begin{eqnarray}
\hh^R_A &:=& (\hh_A \backslash \Sigma) \cap J^+(\Sigma)\,, \quad
\hh_A^L\ := \ (\hh_A \backslash \Sigma) \cap J^-(\Sigma)\,,
\nonumber \\
\hh_B^R &:=& (\hh_B \backslash \Sigma) \cap J^-(\Sigma)\,, \quad
\hh_B^L\ :=\ (\hh_B \backslash \Sigma) \cap J^+(\Sigma)\,. \nonumber
\end{eqnarray}
The Killing vector field $\xi$ is conventionally assumed to be
future oriented on $\hh_A^R$. The bKh divides the spacetime $M$
locally into four disjoint parts, $F:= J^+(\Sigma)$, $P :=
J^-(\Sigma)$, $R:= (J^-(\hh_A^R) \backslash \hh_A^R) \cap
(J^+(\hh_B^R) \backslash \hh_B^R)$ and $L:= (J^-(\hh_B^L)\backslash
\hh^L_B)\cap (J^+(\hh_A^L)\backslash \hh_A^L)$, the future, past,
right and left parts of the bKh, respectively.

To give a rather simple illustration, consider $(M,g)$ as
Minkowski spacetime (of dimension 4).
Then choose an inertial coordinate system and define
$\Sigma$ as the two-dimensional hyperplane $\{(x^0,x^1,x^2,x^3)\in
\RR^4: x^0 = x^1 = 0\}$. There is a smooth, one-parameter group
$\tau_t = \Lambda_t$, $t \in \RR$, of pure Lorentz transformations
leaving $\Sigma$ fixed; they are defined by
\begin{equation}
\Lambda_t(x_0,x_1,x_2,x_3):=
(\cosh (t) x_0 + \sinh (t)x_1,\sinh (t)x_0 + \cosh (t) x_1,x_2,x_3)\,.
\end{equation}
 Then $\hh = \hh_A \cup \hh_B$ is a bKh,
where $\hh_A =\{(u,u,x_2,x_3): u \in \bR,\ (x_2,x_3) \in \bR^2\}$ and
 $\hh_B=\{(v,-v,x_2,x_3) : v \in \bR,\ (x_2,x_3) \in \bR^2\}$.
  Here, the regions $R$ and
$L$ correspond to the usual ``right wedge'' and ``left wedge'' regions
in Minkowski spacetime. Other important examples of spacetimes with a
bKh include e.g.\  deSitter and Schwarzschild-Kruskal spacetimes,
 as well as the
Schwarzschild-deSitter spacetimes and (certain regions of the)
Kerr-Newman spacetimes. (The latter have at least two bKhs with
different surfaces gravities, see below. This leads
\cite{KayWald} to conclude that there are no regular, Killing-flow
invariant states of the free scalar field on such spacetimes.)

Let us now look at how the Killing flow acts on the bKh in greater detail.
Each of the geodesic generators $\gamma_{Ap}$ of the
$\hh_A$-part of the bKh is defined on some interval $I_p$. We may
choose an affine parametrization of $\gamma_{Ap}$, with affine
parameter $U$, such that $\gamma_{Ap}(U=0) = p$ and
 $\left.\frac{d}{dU}\gamma_{Ap}\right|_{U=0} = \chi_A(p)$ for all $p
 \in \Sigma$. This parametrizes all the geodesics and,
 since the vector field $\chi_A(p)$ depends
 smoothly on $p \in \Sigma$, by assumption, the affine parametrization of
 the curves $\gamma_{Ap}$ depends smoothly on $p \in \Sigma$. Since
 $\gamma_{Ap}$ is left invariant under the Killing flow, $I_p$ must be
invariant under a (non-trivial)
smooth representation of the additive group $\RR$ (with $0$ as the only
fixed point), and thus $I_p = \RR$. A similar
result holds for the domains of the geodesic generators
$\gamma_{Bp}$ of $\hh_B$. Therefore, each point $q \in \hh_A$ is
uniquely determined by the pair $(U,p)$, where
$q = \gamma_{Ap}(U)$. Hence we have
a diffeomorphism $\psi_A: \hh_A \to \RR\times \Sigma$
assigning to $q \in \hh_A$ the pair $(U,p) \in \RR \times \Sigma$ with
$q = \gamma_{Ap}(U)$. \footnote{Notice that $\psi_A$ depends on the
  choice of the vector field $\Sigma \owns p \mapsto \chi_A(p)$ along
  $\Sigma$. It may be rescaled at each point:
 $\tilde{\chi}_A(p) = \phi(p)\chi_A(p)$, with $\phi:\Sigma \to \RR$
a smooth, strictly positive function, would
  serve just as well when constructing $\hh_A$. A similar
  remark applies to the $\hh_B$-horizon.}
As explained below, certain choices of
$\chi_A$ and $\chi_B$ turn out to be particularly useful for our purposes
and lead to the following relation (cf.\ \cite{KayWald},
 see also \cite{SumVer}):
\begin{equation}
 \tau_t \lcrc \psi_A{}^{-1}(U,p) = \psi_A{}^{-1}({\rm e}^{\kappa
 t}U,p)\,, \quad t,U\in\RR,\ p \in \Sigma\,,
\end{equation}
where the number $\kappa >0$, called the surface gravity, is
an invariant of the bKh under consideration. (For the
Schwarzschild-Kruskal spacetime of a black hole with mass $m_{\rm bh} >
0$,
$\kappa$ is proportional to $m_{\rm bh}$. The reader is referred to
\cite{KayWald},\cite{WaldI} for more information about
 the notion of surface gravity.)
Constructing a diffeomorphism $\psi_B :\hh_B \to \RR
 \times \Sigma$, similarly, where $\psi_B(q) = (V,q)$ iff $q =
\gamma_{Bp}(V)$,
 the affine geodesic parameter being now denoted by $V$,
 one can show that
\begin{equation}
 \tau_t \lcrc \psi_B{}^{-1}(V,p) = \psi_B{}^{-1}({\rm e}^{-\kappa
 t}V,p)\,, \quad t,V\in\RR,\ p \in \Sigma\,,
\end{equation}
with the same $\kappa > 0$ as in the previous equation.
\\[10pt]
There are a few other geometric actions on $\hh_A$ and $\hh_B$,
induced by identifying these parts of the bKh with
$\RR\times \Sigma$ via the maps $\psi_A$ and $\psi_B$. First, there
are the affine translations
\begin{eqnarray}
 \ell_a \lcrc \psi_A{}^{-1}(U,p) &:=& \psi_A{}^{-1}(U + a,p) \,,\\
 \ell_a \lcrc \psi_B{}^{-1}(V,p) & :=& \psi_B{}^{-1}(V + a,p)\,, \quad
 a,U,V \in \RR,\ p \in \Sigma\,.
\end{eqnarray}
In contrast to the dilations on $\hh_A$ and $\hh_B$, induced
by restricting the Killing flow to the bKh, the translations
will not, in general, extend to isometries of the full
spacetime. Another action is the (affine) reflection,
\footnote{The definitions of
  $\ell_a$ and $\iota$ involve $\psi_A$ (or $\psi_B$) so these quantities,
  cf.\ the previous footnote, depend on the scaling freedom when
  choosing $\psi_A$ (or $\psi_B$).}
\begin{eqnarray}
 \iota \lcrc \psi_A{}^{-1}(U,p) & := & \psi_A{}^{-1}(-U,p)\,, \\
 \iota \lcrc \psi_B{}^{-1}(V,p) & :=& \psi_B{}^{-1}(-V,p)\,, \quad U,V
 \in \RR,\ p \in \Sigma\,.
\end{eqnarray}
Again, $\iota$ need not extend
to an isometry of the full spacetime to the bKh. However,
Kay and Wald \cite{KayWald} have shown that, if the spacetime with bKh
is analytic, there is a neighbourhood $N$ of $\hh$ and an
orientation and chronology-reversing
isometry $\j$ of $N$ (``horizon reflection'') commuting with the action of
$(\tau_t)$ which reflects the affine parameter of
geodesics passing orthogonally through $\Sigma$.
\\[10pt]
In the next step, we shall specify some families of regions
analogous in some respects to the ``shifted wedges'' in
Minkowski spacetime. With their help, we can then formulate a
version of geometric modular action for quantum field theories on
spacetimes with a bKh in the operator-algebraic framework.
To begin with, we note (cf.\ \cite{KayWald}) that the parts $F$, $P$,
$R$ and $L$ of a spacetime with bKh (see above) satisfy
\begin{eqnarray}
& &  F \cap P = \Sigma\,, \quad F \cap R = \emptyset\,,\quad P \cap R =
\emptyset\,, \\
& & F \cap L = \emptyset\,, \quad\,\,  P \cap L = \emptyset\,. \nonumber
\end{eqnarray}
Thus, as we have already seen from the example above, $R$ and $L$ may
be viewed as playing the role of the right and left wedge regions in
Minkowski spacetime. If $\tilde{M} := F \cup P \cup L \cup R$,
then $\tilde{M}$, $L$ and $R$, with the
appropriate restrictions of $g$ as Lorentzian metric, are globally
hyperbolic spacetimes. It may, however, happen that $\tilde{M} \neq
M$, see \cite{KayWald} for examples. As we shall later assume that
 $M =\tilde{M}$, this possibility need
not concern us. One can see from (2.3) that
the regions $F$, $P$, $R$ and $L$ are invariant under the Killing flow
$(\tau_t)$. This implies that $\tilde{M}$ is also invariant under
$(\tau_t)$.
For open intervals $(a,b)$ with $a <b$ and $a,b \in \RR \cup \{\pm
\infty\}$, we now define
\begin{equation}
 \hh_A(a,b) := \{\psi_A{}^{-1}(U,p): a < U < b,\ p \in \Sigma\}\,;
\end{equation}
with an analogous definition of $\hh_B(a,b)$. Notice that with this
notation,
\begin{equation}
\hh_A^R = \hh_A(0,\infty)\,, \quad \ \  \hh_A^L = \hh_A(-\infty,0)\,.
\end{equation}
The ``shifted right wedge'' can then be defined as
\begin{equation}
R_a := R\ \backslash\ {\rm cl}\, J^-(\hh_A(-\infty,a))
\end{equation}
for $a >0$, where cl means ``closure''.
\begin{Lemma}
\begin{equation}
     \tau_t(R_a) = R_{{\rm e}^{\kappa t} \cdot a} \quad {\rm for \
     all}\ \ \  t \in \RR,\ a \ge 0\,.
\end{equation}
\end{Lemma}
\begin{proof} Since $(\tau_t)$ is a group of isometries
leaving $R$ invariant,
\begin{eqnarray}
 \tau_t(R_a) & = & \tau_t \left( R\ \backslash\ {\rm
 cl}\,J^-(\hh_A(-\infty,a))\right) \\
 & = & \tau_t(R)\ \backslash\ \tau_t({\rm cl}\,J^-(\hh_A(-\infty,a)))
 \nonumber\\
& = & R\ \backslash\ {\rm cl}\,J^-(\hh_a(-\infty,{\rm e}^{\kappa t}\cdot
 a)) \nonumber \\
& = & R_{{\rm e}^{\kappa t}\cdot a}\,. \nonumber
\end{eqnarray}
\end{proof}
Similarly, setting
\begin{equation}
L_{-a} := L\ \backslash\ {\rm cl}\,J^+(\hh_A(-\infty,-a))
\end{equation}
for $a > 0$ (!), we find as before that
\begin{equation}
\tau_t(L_{-a}) = L_{-{\rm e}^{-\kappa t}\cdot a} \,, \quad t \in \RR,\
a>0\,.
\end{equation}

In this section, a non-void open $\cO \subset M$ is called a {\it
diamond} if it is of the form $\cO = {\rm int}\,D(G)$ where $G$ is an
open subset of a Cauchy surface $C$ (not necessarily acausal) such
that $\partial G$ is continuous and $\cO^{\perp}$ non-void; moreover
$\cO$ or $\cO^{\perp}$ is required to be connected.

Below we study nets of von Neumann algebras indexed by the
diamond regions in a given spacetime with bKh. Hence we would like the
regions $R_a$ and $L_{-a}$ to be diamonds. Our task is thus to
verify this if $\chi_A$ and $\chi_B$ are chosen suitably.
By assumption, there is an acausal Cauchy surface $C$
passing through $\S$. Let $C_1$ be another acausal Cauchy surface
lying strictly in the future of $C$, i.e.\ $C_1 \subset {\rm int}\,J^+(C) =
J^+(C)\backslash C$. Then we suppose that $\chi_A$ has been chosen
such that each point $q \in \hh_A \cap C_1$ has affine parameter $U
=1$, which means that $q = \psi^{-1}(1,p)$ for some $p \in \S$. Clearly
such a choice is always possible (it amounts to a suitable
choice of the smooth rescaling function $\phi : \S \to \bR$). Under
the Killing flow $\tau_t$  we get a family
$C_{{\rm e}^{\k t}}:= \tau_t(C_1)$, $t \in \bR$, of acausal
Cauchy surfaces (not necessarily forming a foliation) having the
property that each $q \in \hh_A \cap C_{{\rm e}^{\k t}}$ is
represented as $q = \psi_A^{-1}({\rm e}^{\k t},p)$ with suitable $p
\in \S$. Obviously, a similar construction can be carried out with a
Cauchy surface $C_{-1}$ lying strictly in the past of $C$ and
leads to family of acausal Cauchy surfaces $C_{-{\rm e}^{\k t}} =
\tau_t(C_{-1})$. (Moreover, similar constructions can be made for
$\chi_B$, $\hh_B$.) As we first chose $C_1$ and then
adjusted $\chi_A$ to give all points of $C_1 \cap \hh_A$
affine parameter $U = 1$ it is not obvious that
we can choose $C_{-1}$ to give all points of $C_{-1} \cap \hh_A$
affine parameter $U = -1$.
It would suffice if there were a global isometry of M acting as
a horizon-reflection symmetry $\j$
since then one may simply choose $C_{-1} =
\j(C_1)$. The existence of such an isometry will be required later, but
not for the next lemma, where an arbitrary pair of Cauchy surfaces
$C_1$ and $C_{-1}$ with the indicated properties is assumed given, and
the corresponding vector fields $\chi_A$ and
$\chi_A^{(-)}$ assumed chosen so that each point on $C_1\cap \hh_A$ has
affine parameter $U = 1$ with respect to $\chi_A$ and each point on
$C_{-1} \cap \hh_A$ affine parameter $U = -1$ with respect to
$\chi_A^{(-)}$.
\begin{Lemma}
 If $M = \tilde{M}$, then $R^{\perp} = L$, $L^{\perp} = R$ and $R,L$
 and $R_a,L_{-a}$, $a > 0$, are diamonds.
\end{Lemma}
\begin{proof}
 By assumption, we have $M = F \cup P \cup R \cup L$, and $F \cup P =
 J(\S)$. Since $\S$ is part of a Cauchy surface, it follows that
 $\S^{\perp} = {\rm int}\, D(C \backslash \S)$. Hence $R \cup L = {\rm
 int}\, D(C\backslash \S)$. Now define $C_R := C \cap R$, $C_L := C
 \cap L$. Then $C_R \cap C_L = \emptyset$ since $L \cap R = \emptyset$
 (see \cite{KayWald}), and $C_L \cup C_R = C\backslash \S$. Therefore we
 obtain ${\rm int}\, D(C\backslash \S) = {\rm int}\,D(C_R \cup C_L) =
 {\rm int}\,D(C_L) \cup {\rm int}\,D(C_R)$ where the last equality is
 a consequence of the fact that $C_L$ and $C_R$ are disjoint open
 subsets of a Cauchy surface. The boundary of $C_L$ and $C_R$ is in
 both cases the smooth manifold $\S$. Hence $L = {\rm
 int}\,D(C_L)$ and $R = {\rm int}\,D(C_R)$ are  diamonds, and
 since $C_L$ and $C_R$ are disjoint and their union yields $C$ up to
 the common boundary $\S$ of $C_L$ and $C_R$, this entails $R^{\perp}
 = L$ and $L^{\perp} = R$.

Now we define the following sets: $\S_a := C_a \cap \hh_A$, $C_{aR} :=
C_a \cap R$, $C_{aL} := C_a \cap L$, $C_{aF} = C_a \cap F$. One can
see that $C_a \cap P = \emptyset$, for there would otherwise be causal curves
joining pairs of points on $C_a$ and this is excluded. It follows that
$C_a = C_{aL} \cup C_{aR} \cup C_{aF}$ is the union of three disjoint
parts, and ${\rm int}\,D(C_{aR}) = (C_{aF} \cup C_{aL})^{\perp}$. The
common boundary of $C_{aR}$ and $C_{aL} \cup C_{aF}$ is the smooth
manifold $\S_a$, implying that ${\rm int}\,D(C_{aR})$ is a
diamond. Moreover, it is obvious that $\hh_A(a,\infty) \subset
J^+(\S_a)$, $\hh_A(-\infty,a) \subset J^-(\S_a)$, and by standard
arguments it follows that $J^+(\S_a) = {\rm cl}\,J^+(\hh_A(a,\infty))$
and $J^-(\S_a) = {\rm cl}\,J^-(\hh_A(-\infty,a))$. Let us check that
$R_a = {\rm int}\,D(C_{aR})$. First we notice that ${\rm
  int}\,D(C_{aR})\subset R$ is fairly obvious ($R$ is causally closed,
i.e.\ $R^{\perp}\,{}^{\perp} = R$, and $C_{aR}$ is an acausal
hypersurface in $R$), and so is ${\rm int}\, D(C_{aR}) = (C_{aF} \cup
C_{aL})^{\perp} \subset \S_a^{\perp} = M \backslash J(\S_a)$,
implying ${\rm int}\,D(C_{aR}) \subset R_a$.
 To show
the reverse inclusion it is sufficient to prove that $R_a \cap {\rm
  cl}\, J(C_{aF} \cup C_{aL}) = \emptyset$. We have ${\rm
  cl}\,J(C_{aF} \cup C_{aL}) = {\rm cl}\,J(C_{aF}) \cup {\rm
  cl}\,J(C_{aL})$ and $C_{aL} \subset L$ and $R =
L^{\perp}$ imply that $R \cap {\rm cl}\,J(C_{aL}) =
\emptyset$. Now consider an arbitrary past-directed causal curve
$\gamma$ starting at some point on $C_{aF}$. For $\gamma$
to meet $R_a$, it must intersect $\hh_A$. However, any
intersection of $\gamma$ with $\hh_A$ must be contained in
$\hh_A(-\infty,a]$ since $\gamma$ is past-directed and we have seen
that $\hh_A(a,\infty) \subset J^+(\S_a) \subset J^+(C_{aF})$. Thus,
since only the part of $\gamma$ lying in the causal past of its
intersection with $\hh_A$ can enter $R$, $\gamma$ never meets $R_a = R
\backslash {\rm cl}\,J^-(\hh_A(-\infty,a))$, showing that ${\rm
  cl}\,J(C_{aF}) \cap R_a = \emptyset$. Therefore $R_a = {\rm int}\,
D(C_{aR})$ is a diamond. An analogous argument works
for $L_{-a}$.
\end{proof}

\subsection{Conformal Spin-Statistics Relation}\label{4.2}

Our aim in this  subsection will be to show that
the net $\cO \mapsto \cA(\cO)$ on a spacetime with bKh induces a net of von
Neumann algebras $(a,b) \mapsto \cC(a,b)$, indexed by the open intervals
$(a,b)$ of the real line and allowing an extension to a conformally
covariant theory on the circle $S^1$.  Moreover, we shall see that
this net is to be viewed as containing precisely the observables
localized arbitrarily closely to the $\hh_A$-horizon.  (A
similar construction works for the
$\hh_B$-horizon).
 The variant of Wiesbrock's results on modular inclusion \cite{Wies1}
 which is needed to show this may be familiar to experts, but for the
 reader's convenience we present the arguments in an appendix to this
 chapter (Sec.\ 4.3).

 Earlier results \cite{GuLo3,GLWi1} on
the spin-statistics connection for conformally covariant theories on
$S^1$ then apply, yielding a conformal spin-statistics theorem
for the subnets of the initial theory consisting of
observables concentrated on the parts $\hh_A$ and $\hh_B$ of the horizon.

We begin with a
spacetime with a bKh, $(M,g,\t_t,\S,\hh)$, where we assume henceforth that $M =
\tilde{M}$ (cf.\ Sec.\ 4.1). Furthermore,
 we assume given a net $\cK \owns \cO \mapsto
\cA(\cO)$ assigning to each member $\cO$ in the collection $\cK$ of
regions in $M$ a von Neumann algebra $\cA(\cO)$ on a Hilbert space
$\cH_{\cA}$.  
   For convenience, we shall work not with $\cal{K}$, the collection of 
regular diamonds ordered under inclusion, but extend the domain of our 
observable net $\cA$ in the canonical way 
to include a larger collection $\cL$ of open subsets of our spacetime. 
As discussed in the appendix to 
Sec.\ 3, this choice does not change the superselection structure 
in that each representation satisfying the selection criterion based 
on $\cal{K}$ extends uniquely to a representation satisfying the 
selection criterion based on $\cL$, the intertwining operators 
thereby remaining unchanged. Again as discussed in the Appendix to 
Sec.\ 3, the formalism changes only in so far as the localized 
endomorphisms are now defined on larger algebras and this proves to be 
an advantage. We choose $\cL$ to be the 
set of non-empty causally closed subsets $\cS$ of $M$ with 
non-empty causal complements such that for the given 
net $\cA$ $\tilde\perp$--duality holds either for $\cS$ or for $\cS^\perp$. 
By virtue of Lemma 3A.4, this is the same as the partially ordered set $\cL$ 
defined in the Appendix to Sec.~3 in terms of sieves. We recall, too, 
that if $\cA$ is additive, or even inner regular, as a net over $\cK$, 
then $\tilde\perp$--duality coincides with $\perp$--duality. 
\smallskip

Indeed, even though we assume duality for all diamonds, such
assumption is actually used only for two kinds of regions, the
translated wedges $L_a$ and $R_a$, and some tubular neighborhoods of
the horizon intervals $\hh_A(a,b)$ or $\hh_B(a,b)$, which are in turn
tubular neighborhoods in $\hh_A$ or $\hh_B$ of a suitable
 translation of $\S$. We observe
that the obstructions to duality are usually homological in nature,
and that is why duality is generally assumed to hold for regular diamonds. 
On the other hand the surface $\S$, even though not
necessarily homologically trivial, is often relatively trivial,
meaning that $k$-cycles in $\S$ which are trivial in $M$ are trivial
in $\S$ too.
\smallskip

In the following we shall consider the subnet of $\cO \mapsto
\cA(\cO)$ generated by the observables located arbitrarily closely to
the (half) horizon $\hh_A$.  Let us adopt the setting of Lemma 4.2 and
start with a given acausal Cauchy surface $C$ containing $\S$ and
choose an acausal Cauchy surface $C_1$ lying strictly in the future of
$C$ and the vector field $\chi_A$ so that each point on $C_1 \cap
\hh_A$ has affine parameter $U=1$. Then we define for $ 0 < a < b <
\infty$,
\begin{equation}
\cB^R_A(a,b) := \bigcap_{\cO}\{\cA(\cO)\,: \cO\supset\overline{\hh_A(a,b)}\}
\end{equation}
where the intersection is taken over diamonds $\cO$.  Likewise, one
may also assume that another acausal Cauchy surface $C_{-1}$, lying
strictly in the past of $C$, has been selected and that another
(possibly identical) copy $\chi_A^{(-)}$ of $\chi_A$ has been chosen
to give each point of $C_{-1} \cap \hh_A$ an affine parameter $U =
-1$. Correspondingly, we set for $-\infty < -b < -a < 0\,$;

\begin{equation}
 \cB_A^L(-b,-a) : = \bigcap_{\cO}\{\cA(\cO)
\,:\cO \supset\overline{\hh_A(-b,-a)}\}.
\end{equation}
Finally, with these assumptions, one may also define
\begin{equation}
\cB_A(a',b') := \bigcap_{\cO}\{\cA(\cO):\cO\supset \overline{\hh_A(a',b')}\},
\end{equation}
for $-\infty < a' < b' < \infty$. Substituting $B$ for $A$ in the
above, algebras $\cB^{R/L}_B(a,b)$, $\cB_B(a',b')$ can be defined and
all results formulated in the sequel for the algebras $\cB_A$ hold
with obvious modifications for the algebras $\cB_B$ too.
\begin{Lemma} Suppose that the net $\cO \mapsto \cA(\cO)$ satisfies
the following assumptions:
\begin{itemize}
\item{{\rm (I)}} {\it Irreducibility:}
$\vee_{\cO\in\cK}\cA(\cO)=B(\cH).$
\item{{\rm (II)}} {\it Additivity:}
 $\cO\subset \cup_{i\in I}\cO_i$,\quad $\cO_i,\,\cO \in \cK$ \quad
$\imply$ \quad $\cA(\cO)\subset\vee_{i \in I}\cA(\cO_i)$.
\item{{\rm (III)}} {\it Haag duality:}
 $\cA(\cO^{\perp})=\cA(\cO)'$, \quad $\cO \in \cK$ \qquad (implying locality).
\end{itemize}
 Then
\begin{eqnarray}
 \cB^R_A(a,b) &=&
 \cA(R_a) \cap \cA(R_b)'\, \\
\cB^L_A(-b,-a) &=&
 \cA(L_{-a}) \cap \cA(L_{-b})'\, \\
\cB_A(-a',b) &=&
 \cA(L_{-a'})' \cap \cA(R_b)'
\end{eqnarray}
for all $0 < a < b < \infty$, $-a'<0$.
\end{Lemma}
\begin{proof} We shall only give the proof of the first equality,
  since the remaining cases are completely analogous, requiring some
 largely obvious notational changes.

 We recall that $C_a = \tau_{\ln a/\k}(C_1)$ for any $a >
 0$, and also the notation $\S_a = C_a \cap \hh_A$, $C_{aR} = C_a \cap
 R$, $C_{aF} = C_a \cap F$
 and $C_{aL} = C_a \cap L$ used in the proof of Lemma 4.2. Then we
 define the subsets $\widetilde{L}_a := (R_a)^{\perp} = {\rm
 int}\,D(C_{aL} \cup C_{aF})$, $F_a := J^+(\S_a)$ and $P_a :=
 J^-(\S_a)$, and analogous sets with $a$ replaced by $b$.
Next, we define $C^{\vee} := C_{aL} \cup C_{aF} \cup
 \overline{\hh_A(a,b)} \cup C_{aR}$, and aim at demonstrating that
 this set is a Cauchy surface.
 It is fairly obvious that $C^{\vee}$ is achronal, i.e.\  $C^{\vee}
 \cap {\rm int}\,J^{\pm}(C^{\vee}) = \emptyset$. It is also not difficult to
 check that $M = \widetilde{L}_a \cup R_b \cup F_a \cup P_b$ where the sets
 forming the union are pairwise disjoint except for the
 intersection $F_a \cap P_b = \overline{\hh_A(a,b)}$. Now let $\gamma$
 be an arbitrary endpointless causal curve in $M$. If $\gamma$ enters
 $\widetilde{L}_a$ or $R_b$, it must intersect $\overline{C_{aL} \cup
 C_{aF}}$ or $\overline{C_{bR}}$, hence $C^{\vee}$. Suppose that
 $\gamma$ enters $F_a$. Since $F_a$ is past-compact, $\gamma$ must
 intersect one of the regions $R_b$, $\widetilde{L}_a$ or $P_b$, as
 $\gamma$ would otherwise have a past-endpoint. On the other hand, a
 causal curve without endpoint intersecting $F_a$ can only meet $P_b$ if
it intersects $\overline{\hh_A(a,b)}$, too. Hence, if $\gamma$
 enters $F_a$, it must also intersect $C^{\vee}$. Using the same
 argument with obvious modifications for the case that $\gamma$ enters
 $P_b$, one arrives at the same conclusion. This shows that
 every causal curve without endpoints in $M$ intersects $C^{\vee}$,
 implying $M = D(C^{\vee})$, and therefore $C^{\vee}$ is a
 Cauchy surface.

Now we note that ${\rm int}\,D(U) \supset \overline{\hh_A(a,b)}$
for each open neighbourhood $U$ of
 $\overline{\hh_A(a,b)}$ in $C^{\vee}$ since
  $J(\overline{\hh_A(a,b)}) = P_b \cup F_a$ has empty
 intersection with ${\rm cl}(C^{\vee} \backslash U)$. Thus
 $\overline{\hh_A(a,b)}$ is an intersection of  diamonds.
 Moreover, whenever $\cO \supset \overline{\hh_A(a,b)}$ is any
 diamond, it is obvious that we can find some open subset $U$ of $C^{\vee}$
 with piecewise smooth boundary $\overline{\hh_A(a,b)} \subset
U \subset \cO \cap C^{\vee}$, implying $\overline{\hh_A(a,b)} \subset
 {\rm int}\,D(U) \subset \cO$. Hence, to establish the lemma, it
 suffices to consider  diamonds of
 the form $\cO = {\rm int}\,D(U)$. Obviously, the causal
 complement $\cO^{\perp}$ of each such $\cO$ may be written as
 $\cO^{\perp} = \cO_R^{\perp} \cup \cO_L^{\perp}$
where $\cO^{\perp}_R
 = \cO^{\perp} \cap R_a = {\rm int}\,D(C_{aR} \backslash \overline{U})$ and
 $\cO_L^{\perp} = \cO^{\perp} \cap \widetilde{L}_a
 = {\rm int}\,D((C_{aL} \cup
 C_{aF})\backslash \overline{U})$ are both  diamonds. Notice that the
 union of $\cO_L^{\perp}$ and $\cO_R^{\perp}$ over all $\cO = {\rm int}\,D(U)$
yield $\widetilde{L}_a$ and
 $R_b$, respectively. Consequently we have
\begin{eqnarray*}
 \bigcap_{\cO} \cA(\cO)  & = & (\, \bigvee_{\cO} \cA(\cO)'\,)'
                     \   = \  (\, \bigvee_{\cO}
 \cA(\cO^{\perp})\,)'\\
                         & = & (\, \bigvee_{\cO} \cA(\cO_L^{\perp}
 \cup \cO_R^{\perp})\,)'
\   = \  (\, \bigvee_{\cO} \cA(\cO_L^{\perp}) \vee
 \cA(\cO_R^{\perp})\,)' \\
 & = & (\,\cA(\widetilde{L}_a) \vee \cA(R_b)\,)'
\   = \  (\,\cA(R_a)' \vee \cA(R_b)\,)'\\
& = & \cA(R_a) \cap \cA(R_b)'\,,
\end{eqnarray*}
where the second equality follows from Haag duality, the third has been
justified above,
the fourth and fifth equalities use additivity and the last but one
again follows from Haag duality.
\end{proof}

The formulation of the subsequent result necessitates introducing
further assumptions and related notation.

We shall write $\cB_A^R(a,\infty) := \bigvee_{b > a}\cB_A^R(a,b)$, and
define the other horizon-algebras associated with unbounded intervals
in a similar manner by additivity.
Let $\O \in \cH$ be a unit vector vector, then we denote by
$\cH^R_A(\O) := \overline{\cB^R_A(0,\infty)\O}$, $\cH^L_A(\O):=
\overline{\cB^L_A(-\infty,0)\O}$ and $\cH_A(\O):=
\overline{\cB_A(-\infty,\infty)\O}$ the Hilbert subspaces generated by
applying the various algebras of observables concentrated on the
$\hh_A$-horizon on that vector.
We say that $(\cB^R(0,\infty),\O)$ is a standard pair if $\O$ is
separating for $\cB^R(0,\infty)$. It is by definition cyclic with
respect to the Hilbert subspace $\cH^R_A(\O)$. The modular objects
(with respect to $\cH^R_A(\O)$) of such a standard pair will be
denoted by $J_{R,\O}$, $\D_{R,\O}$. The like objects for $L$ in place
of $R$ are defined similarly.

In the following, we shall focus attention on the next two
assumptions:
\begin{itemize}
\item[(IV)] {\it Geometric modular group on the horizon: } There is a
  unit vector $\O\in \cH$ so that\\[2pt]
(i) \quad $(\cB^R_A(0,\infty),\O)$ is a standard pair, and
\begin{equation}
 \D^{it}_{R,\O}\,\cB^R_A(a,\infty)\,\D^{-it}_{R,\O} = \cB^R_A({\rm
   e}^{-2\p t/\k}a,\infty)\,,
\end{equation}
(ii) \quad $(\cB^L_A(0,\infty),\O)$ is a standard pair, and
\begin{equation}
 \D^{it}_{L,\O}\,\cB^L_A(-\infty,-a)\,\D^{-it}_{L,\O} = \cB^L_A(-\infty,-{\rm
   e}^{2\p t/\k}a)\,,
\end{equation}
for all $a > 0$, $t \in \bR$, where $\k > 0$ is the surface gravity
of the bKh.
\item[(V)] {\it Geometric modular conjugation on the horizon: }
For the $\O$ as in (IV), we have $\cH_A(\O) = \cH_A^R(\O) =
\cH_A^L(\O)$ and moreover
\begin{equation}
                J_{R,\O}\,\cB^R_A(a,\infty) J_{R,\O} =
                \cB_A^L(-\infty,-a)\,, \quad a \ge 0\,.
\end{equation}
\end{itemize}
Let us now assume that the net $\cO \mapsto \cA(\cO)$ satisfies
assumptions (I--IV). Thus we see that $(\cB_A^R(1,\infty) \subset
\cB_A^R(0,\infty),\O)$
is a +hsm inclusion and $(\cB_A^L(-\infty,-1) \subset
\cB_A^L(-\infty,0),\O)$
 is a --hsm
inclusion. Then the results of \cite{Wies1,ArZs1} yield two
continuous unitary groups $U^{R/L}(a)$, $a \in \bR$, having
positive/negative spectrum and satisfying the following relations
for $a > 0$:\\[6pt]
\begin{tabular}{rclrcl}
 $\D^{-it}_RU^R(a)\D^{it}_R$&= &$  U^R({\rm e}^{2\pi t}a)\,,\ \ \ $ &
$\D^{it}_LU^L(a)\D^{-it}_L $& =&$  U^R({\rm e}^{2\pi t}a)\,,$\\[4pt]
 $J_RU^R(a)J_R$ & = & $U^R(-a)$\,, &$J_LU^L(a)J_L$ & = & $U^L(-a)$\,,\\[4pt]
$U^R(a)\cB^R_A(0,\infty)U^R(-a)$ &=&$\cB_A^R(a,\infty)\,,$ & & & \\[4pt]
  $U^L(a)\cB_A^L(-\infty,0)U^L(-a)$ &=&
$\cB^L_A(-\infty,-a)$ & & &
 \end{tabular}
\\[10pt]
where we have dropped the index $\O$ on the modular objects to
simplify notation. Without further assumptions,
$U^R$ and $U^L$ are unrelated and so are the
nets $\cB_A^R$ and $\cB_A^L$. However, if we suppose that (V)
holds, too, then it follows from the way these unitaries are
constructed (cf.\ \cite{Wies1}), that $J_RU^R(a)J_R = U^L(a)$,
$a \in \bR$.
 Therefore we obtain the following:
\begin{Cor}
Under assumptions {\rm (I--IV)}  the nets of horizon-algebras indexed
by the intervals of the half real lines,
\begin{align*}
 (a,b) \ & \mapsto\ \cB_A^R(a,b)\,,\quad& 0< a < b< \infty\,, \\
 (-b,-a)\ & \mapsto\ \cB_A^L(-b,-a)\,,\quad& -\infty < -b < -a < 0\,,
\end{align*}
extend to local conformal nets $I \mapsto
\cM^R(I)$ and $I \mapsto \cM^L(I)$ of von Neumann algebras on $S^1$ on the
Hilbert spaces $\cH_0^R = \overline{\cB_A^R(a,b)\O}$ and
$\cH _0^L = \overline{\cB_0^L(-b,-a)\O}$, respectively (where the
$0<a<b<\infty$ are arbitrary).
\\[6pt]
 If {\rm (V)} is assumed, too, then the net $(a',b') \mapsto
 \cB_A(a',b')$ on the full real line
extends to a local conformal net $I \mapsto \cM(I)$ on $\cH_A(\O)$.
\end{Cor}
\begin{proof}
The first part is a variant of Wiesbrocks's result
\cite{Wies1,Wies4}, cf.\ also \cite{GLWi1}.
 We supply the relevant argument as
Proposition 4A.2 in Sec.\ 4.3.

If assumption (V) is added so that $J_R$ intertwines $U^R$ and $U^L$,
the adjoint action of $U^R(a)$  on the net $\cB_A$ is geometrically
correct, i.e.\ $U^R(a)\cB_A(a',b')U^R(-a) =
\cB_A(a'+a,b'+a)$, $a \in \bR$, $a' < b'$. Thus the net $\cB_A$
together with its dilation and translation symmetries coincides with
both $\cC^R$ and $\cC^L$ (derived from the nets $\cB_A^R$ and $\cB_A^L$
 as in Proposition 4A.2) and their respective translation and
dilation symmetries. Thus the corresponding extensions to conformally
 covariant theories coincide.
\end{proof}

 Condition (IV) may be viewed as a weak form of  the Hawking-Unruh
effect: an observer moving with
the Killing flow of the bKh registers a thermal ensemble in the
``vacuum'' state (see \cite{Sew,Wald2}). The term ``vacuum'' here means
a state invariant under the space-time isometries and
fulfilling additional stability conditions, in fact (IV) and (V) may
be viewed as a weak form of
such conditions, namely applying to the subsystem of
observables concentrated on the horizon. As the
group of affine translations along the geodesic generators of the
horizon has positive generator derived from the modular
inclusion of horizon-algebras, $\O$ can be justly interpreted
as a vacuum vector for the horizon-algebras
 (cf.\ the principle of geometric modular
action \cite{BuSu1} or modular covariance \cite{BGLo2}). Clearly, if
$\O$ induces a KMS-state for the Killing flow at the Hawking temperature
on $\cA(R)$, then (IV,i) follows by Lemma 4.3. Likewise, if $\O$
induces a KMS-state for the Killing flow at negative Hawking
temperature on
$\cA(L)$, then (IV,ii) follows by Lemma 4.3.

The motivation for Condition (V) is that, a horizon (or wedge)
reflection symmetry should be implemented
in a ``vacuum'' representation by the modular conjugations $J_R$,
in analogy with the
Bisognano-Wichmann result for quantum fields in Minkowski space
\cite{BiWi,Bor1,Bor2,Sew}. Our condition
is actually a bit weaker in that $J_R$ need not implement a
point-transformation of the underlying spacetime manifold. However,
Condition (V) implicitly imposes a relation between the horizon
segments $\hh_A(-\infty,-a)$ and $\hh_A(a,\infty)$.

 We finally comment on whether these assumptions
 are realistic. For the free scalar field, conditions (I,II) hold generally
 in representations induced by quasifree Hadamard states (for
 $\cO \in \cK$ based on relatively compact subsets of Cauchy surfaces,
 and, in more special cases, even when the base is unbounded),
 see \cite{Ver1}. The Hartle-Hawking state, i.e.\ the
 candidate for the ``vacuum'' state of the free scalar field on the
 Schwarzschild-Kruskal spacetime, should also satisfy all the
 assumptions  \cite{Kay,KayWald} ((III) has not been checked
in the generality formulated here, but a version of
 (III) sufficient to imply
 the spin-statistics theorem in the sequel does hold).
 As is known from the Bisognano-Wichmann result \cite{BiWi}, the
 assumptions are fulfilled for local von Neumann algebras generated by
 (finite-component) Wightman fields in Minkowski spacetime ((III) then
 holds for wedge-regions and this suffices to establish the spin-statistics
 relations $\cite{GuLo2,Kuc}$). Results of Borchers \cite{Bor1,Bor2}
 yield (III--V) generally for algebraic quantum field theories in
 two spacetime dimensions. With additional conditions these generalize
to higher dimensions\cite{Bor3,Bor4,Wies5}.

\bigskip

Now we can formulate the conformal spin and statistics theorem.  Our
aim is to define the spin of a sector as the conformal spin on the
horizon. To this end we need to restrict to considering sectors that are {\it
horizon localizable}, namely having a representative which acts
trivially on the algebras $\cB_A(a,b)'$ for some $a,b\in\bR$ (or the
same for the $B$ horizon). However this is not sufficient in general
because the sector on the horizon may not be covariant. As shown in
\cite{GLWi1} covariance of localized endomorphisms with finite
statistics is automatic when the net is strongly additive, which is
always the case for the dual net. Unfortunately extending a sector on
a conformal net to a sector on the dual net may produce soliton
sectors.  Therefore we shall only consider those sectors which are not
only horizon localizable, but also {\it dual localizable}, namely
which give rise to a localized sector on the dual net of the horizon
conformal net.  Clearly if we have a dual localizable sector on the
net $\cO \mapsto \cA(\cO)$ satisfying assumptions (I--V) with non-zero
statistical parameter $\l$, we obtain a covariant sector on the dual
net on the horizon with the same statistical parameter, since this is
determined by the intertwiners. The following theorem is now a simple
consequence of the conformal spin and statistics theorem in \cite{GuLo3}.

\begin{Thm} Let $\cO\mapsto\cA(\cO)$ be a theory on a spacetime with bKh
satisfying assumptions {\rm (I--V)} and $\r$ a dual localizable sector
with finite statistics. Then $\r$ gives rise to a covariant sector on
the dual net on the horizon, therefore a conformal spin $s_\r$ is
defined, and the conformal spin and statistics relation holds, namely
$s_\r=\k_\r$.
\end{Thm}

\newpage\noindent
\bigskip\noindent{\bf Remarks Concluding
  Section 4.2 }\quad\\[6pt]
The idea of passing from a quantum field theory initially formulated over
a four-dimensional spacetime to observables concentrated on a
lightlike hypersurface (i.e.\ pieces of a bKh) is not a new one and
once was popular in quantum field theory under the keyword ``infinite
momentum frame''. \cite{LeuKlSt,Rohr} are just two
references in this direction. This matter is studied for the first time
in the operator algebraic framework in \cite{Dri}. One
motivation is that symmetries may be enhanced by restricting to a subtheory
concentrated on a lightlike hypersurface, a particularly
attractive possibility for quantum field theory in curved spacetimes where
symmetries of the underlying four-dimensional
spacetime are rather limited.
As proved in this section, for bKh spacetimes restricting to the horizon
does indeed give conformally covariant nets.

Sewell \cite{Sew}
was the first to observe that this allows one to formulate a
Bisognano-Wichmann theorem relating to the Hawking effect
for quantum fields on blackhole spacetimes, in the setting of
a Wightman field theory
(see e.g.\ \cite{Wald2} for further discussion). In this context,
two papers rigorously establish related results
for free field models \cite{Kay,DimKay}. Kay and Wald
\cite{KayWald} realized that such results may be generalized to
spacetimes with a bKh and obtained strong theorems for free fields in
this setting. An operator-algebraic version of aspects of Sewell's
work appears in \cite{SumVer} where the nets $\cB_A(a,b)$ are used. 

We ought to mention that in general it is not very clear how ``big''
the algebras $\cB_A(a,b)$ (or $\cB_B(a,b)$) are in the original
algebras $\cA(\cO)$.

If $\O$ is cyclic for $\cB_A(a,b)$ then it is resonable to expect that
sectors are horizon localizable (on the $A$-horizon). Moreover in this
case the conformal net on the horizon is strongly additive by
definition, therefore it coincides with its dual net
(cf. \cite{GLWi1}), and then horizon-localizability and dual
localizability are equivalent.

It is known that $\O$ is cyclic for $\cB_A(a,b)$ when free fields on
the $n$-dimensional Minkowski space are considered, $n\ne2$. We give
here a simple argument based on \cite{BGLo3}.

By a ``free field'' on Minkowski space we here mean a local net $\cA$
of von Neumann algebras indexed by regions of Minkowski space which
can be constructed by second quantization from a net $K$ of real
vector spaces in a complex Hilbert space $H$, plus the usual
assumptions of Poincar\'e covariance, positive energy, and in
particular the Bisognano-Wichmann property and irreducibility:
$\cap_{W}\cA(W)=\bC I$.

Working in the first quantization space $H$ from now on, we first
observe that irreducibility means $\cap_{W}K(W)=\{0\}$ and, by the
Bisognano-Wichmann property, this is equivalent to there being no
fixed vectors for the action of the Poincar\'e group on $H$.

Then, by a Theorem of Mackey (cf.\ e.g.\ \cite{Zimm}, Proposition
2.3.5), the absence of invariant vectors for the whole Poincar\'e
group is equivalent, when $n\ne2$, to the absence of invariant vectors
for any given translation, hence the spectrum of the generator of any
light-like translation is strictly positive, i.e. zero is not an
eigenvalue.

Now, given two wedges $W_1$, $W_2$, the cyclicity of
the vacuum vector $\Omega$ for $\cA(W_1) \cap \cA(W_2)$ is equivalent to
 $(K(W_1)\cap K(W_2)) +
i(K(W_1)\cap K(W_2))$ being dense in $H$, this being in turn
equivalent to having
$\{v\in{\rm dom}(s_{W_1})\cap {\rm dom}(s_{W_2}):s_{W_1}v=s_{W_2}v\}$ 
dense in $H$, where $s_{W_j}$ denotes the ``first quantized'' Tomita
operator defined by $s_{W_j}(\chi + i\phi) := \chi - i\phi$,
$\chi,\phi \in K(W_j)$.
 When $W_1=\{(t,x):x_1>|x_0|\}$ and $W_2$ is a translation of
the causal complement of $W_1$, $W_2=\{(t,x):x_1-c<-|t-c|\}$, $c>0$,
the situation met when considering the vector space associated with
the interval $(0,c)$ on the $A$-horizon, this is in turn equivalent,
again by the Bisognano-Wichmann property, to the density of the space
 \begin{equation}\label{cyc}
\{v\in {\rm dom}(\d_1^{1/2}T(c)\d_1^{1/2}):T(c)\d_1^{1/2}T(c)\d_1^{1/2}v=v\},
 \end{equation} 
where $a\to T(a)$ denotes the representation of the light-like
translations along the $A$-horizon and $\d_1$ denotes the 
``first quantized'' modular
operator for the space $K(W_1)$. This property clearly depends only on
the restriction of the representation of the Poincar\'e group to the
subgroup $\bP_1$ generated by boosts and light-like translations
with strictly positive generator
(relative to the wedge $W_1$).  As the logarithm of the generator of
translations  and the generator of the boosts give rise to (and are
determined by) a representation of the CCR in one dimension, the
strictly positive energy representations of $\bP_1$ have a simple
structure: they are always a multiple of the unique irreducible
representation. Therefore the density of the space in eqn.
(\ref{cyc}) holds either always or never, and hence can be checked in
the irreducible case. But this is the case of the current algebra on
the circle, where cyclicity holds by conformal covariance.

Of course, the vector $\O$ is not expected to be
cyclic in general for the algebra generated by the $\cB_A(a,b)$, and
it might even happen that $\cB_A(a,b)$ contains only multiples of the
identity. Field nets giving rise to non-trivial superselection
sectors of the observable net localizable on the horizon can easily
be constructed just by requiring the vacuum to be cyclic for the
horizon field algebras. However it is not clear, in general, how strong
the requirement of dual localizability is.
\subsection{Appendix to Chapter 4}
For the benefit of the non-expert reader, we present in detail in this
Appendix  the arguments leading from
the results in \cite{Wies1,Wies4,GLWi1} to Corollary 4.5.
 To begin with, we
state a result about modular inclusions needed in
the following.
\\[6pt]
{\bf Lemma 4A.1. } {\it
Let $(\cN\subset \cM,\O)$ be a pair of von Neumann algebras with a
unit vector $\O$ cyclic and separating for $\cM$ and such
that $\Delta^{it}\cN\Delta^{-it} \subset \cN$ for all $- t \ge 0$
(or $t \ge 0$),
where $\Delta^{it}$, $t \in \bR$, is the modular group of $\cM,\O$.
Then $\cM=\vee_{t\in\bR}\D^{it}\cN\D^{-it}$ if and only if $\O$ is
cyclic for $\D^{it}\cN\D^{-it}$ for some (hence for any)
$t\in\bR$.}
\begin{proof}
If $\O$ is cyclic for $\D^{it}\cN\D^{-it}$ for a given $t$, then it is
cyclic for $\vee_{t\in\bR}\D^{it}\cN\D^{-it}$, too. However this von~Neumann
algebra is invariant under the modular group of $\cM$,  and hence coincides
with $\cM$ by Takesaki's theorem. Conversely, let $\x$ be orthogonal to
$\D^{it}\cN\D^{-it}\O$. Then for any
$x\in \D^{it}\cN\D^{-it}$ we have $x\O\in {\rm dom}(\D^{1/2})$,
 hence the function
 $z\mapsto (\D^{iz}x\O,\xi)$ is analytic on the strip $-i/2<\Im z<0$ and
continuous on
the boundary. But as we have a $+$hsm inclusion, it vanishes
for negative real $z$
and hence everywhere.  Thus $\x$ is orthogonal to
$\vee_{t\in\bR}\D^{it}\cN\D^{-it}\O=\cM\O$, completing the proof.
\end{proof} \smallskip
{\bf Proposition 4A.2. } {\it
Let $(\cN\subset \cM,\O)$ be a a pair of von Neumann algebras with a
unit vector $\O$ which is cyclic and separating for $\cM$ and such
that $\vee_{t\in\bR}\Delta^{it}\cN\Delta^{-it} =\cM$ and
$\Delta^{it}\cN\Delta^{-it} \subset \cN$ for all $- t \ge 0$,
where $\Delta^{it}$, $t \in \bR$, is the modular group of $\cM,\O$.
Then, setting
\begin{align}
\cH_0
&=\ov{(\cN\cap(\D^{-i}\cN\D^{i})')\O}\\
\cC(a,b)
&=(\D^{-i\frac{\log a}{2\pi}}\cN\D^{i\frac{\log a}{2\pi}})\cap
(\D^{-i\frac{\log b}{2\pi}}\cN\D^{i\frac{\log b}{2\pi}})'\rest {\cH_0},
\quad 0<a<b,
\end{align}
the family $(a,b) \mapsto \cC(a,b)$ extends to a local conformal net
of von~Neumann algebras acting on the Hilbert space $\cH_0$.}

\begin{proof} Set
\begin{equation*}
\cN_a=\D^{-i\frac{\log a}{2\pi}}\cN\D^{i\frac{\log a}{2\pi}},\quad a>0
\end{equation*}

By the previous lemma $\O$ is cyclic for $N_a$, $a>0$, therefore we may 
apply a result of Wiesbrock and
Araki-Zsido (\cite{Wies1,ArZs1}) to
the $+$hsm inclusion $(\cN \subset \cM,\O)$ and get
a one parameter group of unitaries
$U(a)$ on $\cH$ with positive generator satisfying
\begin{align*}
\D^{-it}U(a)\D^{it}&=U(e^{2\pi t}a)\\
JU(a)J&=U(-a)
\end{align*}
Hence we have
$$
\cN_a=U(a) \cM U(a)^*\,,\qquad a\geq0\,,
$$
and this equation is used to define $\cN_a$ for negative $a$.

We now set
\begin{align*}
\cC(a,b)&=\cN_a\cap \cN_b'\rest {\cH_0}\,, \quad&-\infty<a<b <+\infty\\
\cC(-\infty,b)&=\vee_{a<b}\cC(a,b)\,, \quad&-\infty<b<+\infty\\
\cC(a,+\infty)&=\vee_{b>a}\cC(a,b)\,, \quad&-\infty<a<+\infty
\end{align*}
and the definition of $\cC(a,b)$ clearly agrees with (4.26) when
$0 < a < b < \infty$. Furthermore,
 $$
\cH_0=\ov{\cN\cap \cN'_{e^{2\pi}}\O}=\ov{\cC(1,e^{2\pi})\O}.
 $$
Moreover, the operators $J,\D$ restricted to $\cH_0$ give the modular
conjugation and operator of
$(\cC(0,\infty),\O)$. Similarly, using the results of \cite{Wies1,ArZs1}
anew, the restriction of $U(a)$ to $\cH_0$
(again denoted by $U(a)$) coincides with the unitary group
derived from the
+hsm inclusion $(\cC(1,\infty) \subset \cC(0,\infty),\O)$.
Now a standard Reeh-Schlieder argument, based on the positivity of the
generator of $U(a)$, shows that $\ov{\cC(-\infty,b)\O}$ is independent of
$b$, while the ``modular'' Reeh-Schlieder argument in
Lemma 4A.1
shows that $\ov{\cC(a,b)\O}$ is independent of $a\in(-\infty,b)$.
Thus the inclusion $(\cC(1,\infty) \subset \cC(0,\infty),\O)$ is standard.
We have proved that $\cH_0=\ov{\cC(a,b)\O}$ for any $-\infty\leq
a<b\leq+\infty$, and that $\cC$ gives a translation-dilation covariant net
of von~Neumann algebras on $\cH_0$. Then we get a conformally covariant
net by a result of Wiesbrock (\cite{Wies4}, see also \cite{GLWi1}).
\end{proof}

\section{The Spin and Statistics Relation for Spacetimes with Rotation
Symmetry}

\setcounter{equation}{0}

In this section, we present a proof of the spin and statistics
relation for superselection sectors on a globally hyperbolic spacetime with
some rotational symmetry.

The main assumption here is the existence of a suitable family of regions,
called
wedges, each being equipped with a reflection mapping it to its causal
complement and of a net of von~Neumann algebras with a common cyclic vector
whose modular conjugations implement the said reflections, in the
spirit of \cite{BDFS1} and \cite{Long3}.

Moreover we assume the existence of rotational spacetime symmetries,
rotating a wedge to its causal complement and belonging
to the commutator of the spacetime symmetry group. As we shall see, our
geometric assumptions are satisfied in many interesting spacetimes and
form the geometric basis for the rotational spin and statistics theorem,
explained in more detail below.

\subsection{Geometric Assumptions}

A spacetime with rotation and reflection
symmetry is a quadruple $(M,\cW,\Gp,\j)$, where $M$ is a globally
hyperbolic spacetime, $\cW$ is a family of open subregions called
wedges, $\Gp$ is a Lie group of proper (i.e.\ orientation preserving)
transformations of $M$ and $\j$
is a map from $\cW$ to the antichronous (i.e.\ time reversing) reflections
in $\Gp$; we write it as $W \mapsto \j_W$. We denote the
 orthochronous subgroup of $\Gp$ by $\Gpo$ and
the identity component of $\Gp$ by $\Gpnot$. The universal covering of
$\Gpnot$ is denoted by $\Gcov$. The $\bZ_2$ action implemented by any
$\j_W$ on $\Gpnot$ lifts to an action on $\Gcov$. The quadruple has to
satisfy the following properties:

\begin{itemize}
 \item[(a)]
$\j$ leaves $\cW$ globally invariant and verifies $\j_W(W)=W^{\perp}$ and
$\j_{gW}=g\j_W g^{-1}$, $W\in\cW$, $g\in\Gp$.
 \item[(b)] There is $W\in\cW$ and an element $\h$ in the Lie algebra of
$\Gpnot$ such that
 \\
 $(1)$ $\exp (2\pi \h)$ is the identity in $\Gpnot$,
 \\
 $(2)$ $\j_W\exp(t\h)\j_W=\exp(-t\h)$,
 \\
 $(3)$ $\exp(\pi \h)W=W^{\perp}$,
 \\
 $(4)$ $\cap_{0\leq t\leq\pi/2}\exp(t\h)W$ is non--empty.
 \item[(c)] $\h$ belongs to the commutator of the Lie algebra of $\Gpnot$.
\end{itemize}

\begin{rem}\label{rem:ortho} Two wedges $W$, $\Wtilde$ are called
{\it orthogonal} if $\j_W\Wtilde=\Wtilde$ and $\j_{\Wtilde}W=W$. It is
easy to see that $W$ and $\exp(\frac\pi2\h)W$ are orthogonal. Indeed,
making use of assumptions (b~$2$) and (b~$3$), we get

 \begin{align*}
\j_W\exp(\frac\pi2\h)W&=\exp(-\frac\pi2\h)\j_W W\cr
        &=\exp(-\frac\pi2\h)W^{\perp}=\exp(-\frac\pi2\h)\exp(\pi\h)W\cr
        &=\exp(\frac\pi2\h)W.
 \end{align*}
The second equation is proved analogously.
\end{rem}

We shall also consider spaces where property (c) is replaced by
the following property: \\[8pt]
(c$'$) There exists a wedge $\Wtilde$, orthogonal to $W$, such that
$\j_{\Wtilde}$ commutes with $\exp(t\h)$.
\\[8pt]
 \begin{rem}\label{rem:rotgeom} (i) Assumption (a) has to be seen as a
part of the definition of a wedge. The first part of property (a) says
that any wedge is $\Gp$--equivalent to its causal complement, hence a
wedge is in some sense ``a half" of $M$ or, more precisely, is the causal
completion of ``a half'' of a Cauchy surface.  The second part means
that $\j_W$ commutes with the stabilizer of $W$ and, when $\Gp$ acts
transitively on $\cW$, says that $\j$ is determined by its value on
one wedge.
 \\
 (ii) Properties (b)  describe the rotation symmetry.
In view of property (b~$1$) we call the group elements $\exp(t\h)$
rotations. Property (c) ensures that all characters of
$\Gcov$ are trivial on the cycle $\{\exp (t\h), t\in[0,2\pi]\}$, since
the latter belongs to the commutator subgroup of $\Gcov$ where all
characters are trivial. As we shall see, this makes the spin well
defined.
 \\
 (iii) The element $\j_W$, seen as an automorphism of the Lie algebra of
$\Gpnot$, has eigenvalues $1$ and $-1$ and by (a) the eigenspace corresponding
to $1$ consists of generators of transformations preserving $W$.
Therefore (b~$2$) essentially says that not all rotations
preserve $W$. More precisely, $W$ may be rotated to its spacelike
complement by (b~$3$).
 \\
 (iv) Property (b~$3$) mainly expresses the fact that $2\pi$ is the
minimal period of the one-parameter group $\exp(t\h)$.
 \\
 (v) Property (b) is stated for one wedge $W$, but then holds for
any wedge in the family $\cW_0:=\{gW : g\in\Gp\}$. We are of course
interested in the case where the cycle $\{\exp (t\h), t\in[0,2\pi]\}$
is not homotopy trivial and hence gives rise to a non-trivial notion of
spin. However
this is not needed for the proof of the spin and statistics theorem
 nor do we require that the $\exp(2\pi \h)$ generate the homotopy
group of $\Gpnot$.
 \\
 (vi) Property (c$'$) implies that $W$,
$e^{\pi/2\h}W$ and $\Wtilde$ are mutually orthogonal. It also implies
that $r_W:=\j_W\j_{\Wtilde}$ is an involution in $\Gpo$ and that
$\exp(2t\h)=[\exp(t\h),r_W]$, where the square brackets here denote the
multiplicative commutator. As a consequence, the rotations $\exp(t\h)$
belong to the commutator subgroup of $\Gpo$. In this sense (c$'$)
is a weak form of (c). $\Gpo$ and $\Gpnot$ do not always
coincide. Of course $r_W\in\Gpo$, but we do not require that $r_W$
belongs to $\Gpnot$.
 \\
(vii) Property $(b)$ fixes the the generator $\h$ up to a sign, indeed (b~$1)$
fixes the generator up to an integer, (b~$3)$ implies this integer to be odd,
and (b~$5)$ requires this integer to be 1 or $-1$.
 When the spacetime is two-dimensional, i.e.\ when the Cauchy
surface is 1-dimensional, the orientation fixes a direction on any
spacelike curve (from left to right). In this case we choose the sign in such a
way that the element $\h$ generates a rotation in the prescribed
direction. \end{rem}

\bigskip\noindent{\bf Assumptions (a), (b), (c) and (c$'$) in some
spacetimes}\quad\\[6pt]
 In the case of the $n$-dimensional Minkowski
spacetime $M^n$, a wedge is any $\Gp$--transform of the region
$W=\{|x_0|<x_1\}$ if $n>2$, and of the region $\{x>0\}$ if $n=1$.
Taking $\j_W$ to be the reflection w.r.t. the edge of the wedge, the
map $\j$ turns out to be uniquely defined by property (a).

When $\Gp$ is the proper Poincar\'e group and $n\geq3$, property (b)
holds with $W$ as above and $\h$ as the generator of rotations in
the $(x_0,x_1)$-plane. Indeed the proper orthochronous Poincar\'e group
is perfect, hence property (c) is obviously satisfied.  If
$n\geq4$ then (c$'$) is satisfied too, with $\Wtilde=\{|x_0|<x_2\}$.

When $G_+$ is the proper conformal group, properties (b) and (c) are satisfied
for any $n\geq1$, $\h$ being the generator of a suitable
group of (conformal) rotations. Property (c$'$) is satisfied when
$n\geq3$, $W$ being as before, $\h$ being the generator of rotations in
the $(x_0,x_1)$-plane and $\Wtilde$ a double cone with spherical
basis centred on the origin.

Since the $n$-dimensional de\,Sitter spacetime $D^n$ may be defined as
the hyperboloid $x_0^2+1=|{\mathbf x}|^2$ in $M^{n+1}$, the wedges can
be defined as the intersection of this hyperboloid with the wedges in
$M^{n+1}$ whose edge contains the origin. Then properties (b), (c) or (c$'$)
hold for $D^n$ if and only if properties (b), (c) or (c$'$)
hold for $M^{n+1}$ (with Poincar\'e symmetry), respectively.

Note that the Cauchy surface of $D^n$ is compact and the same
 is true for $M^n$
with conformal symmetry, since in this case the quantum field theories
actually live on (a covering of) the Dirac-Weyl compactification of
$M^n$ (cf.\ \cite{BGLo1}).

Whenever the spin makes sense in the above examples, i.e.\ whenever
(c) or (c$'$) holds, the group $\Gpo$
has no non-trivial finite dimensional representations,
a much stronger requirement than (c) or (c$'$).
In this case
the spin and statisitcs relation may be proved as in \cite{Long3}.

Moreover, in these examples, modular covariance makes sense,
i.e.\ there is a natural definition of the geometric action of
$\Delta^{it}$, furthermore, the Bisognano-Wichmann property has been proved
for Wightman fields (\cite{BiWi,BrMo}), wedges separate
spacelike points and every double cone is an
intersection of wedges. Therefore geometric modular
conjugation follows from modular covariance (as in \cite{GuLo2},
cf.\ \cite{BDFS1}) and modular covariant free fields may be
constructed canonically as in \cite{BGLo3} by second quantizing
(anti-)unitary representations of $\Gp$.

We now describe a class of spacetimes where these additional
features do not hold, namely where the group admits one-dimensional
representations and the wedges do not separate points. Nevertheless,
these cases are still covered by the
spin and statistics theorem we are going to present below.

\newpage\noindent
\bigskip\noindent{\bf Spherically symmetric black holes}\quad\\[6pt]
We call spherically symmetric black holes those spacetimes $(K,g_K)$
whose structure is very similar to the Schwarzschild-Kruskal
spacetime, i.e.\ they are isometric to $X \times S^n$, $X$ being the
set of points $(x_0,x_1) \in \bR^2$ with $x_0^2 - x_1^2 < \mu^2$, $\mu
\in \bR \cup \{\infty\}$, with the metric\footnote{It is customary
to write coordinate indices as upper indices, but our deviating from
this convention is unlikely to cause confusion.}
\begin{equation}
ds_K^2 = a(x_0^2 - x_1^2)(dx_0^2 -dx_1^2) - b(x_0^2 -
x_1^2)d^2\sigma\,,
\end{equation}
where $d^2\sigma$ is the usual Riemannian metric on the sphere $S^n$
and $a$ and $b$ are smooth, strictly positive functions.  Then the
hypersurface $x_0=0$ is a Cauchy surface and $(K,g_K)$ is globally
hyperbolic.  The structure of such spacetimes is in some respects
similar to that of Minkowski spacetime. For instance, if points in $X
\times S^n$ are represented as $(x_0,x_1,\sigma)$, then one may define
a one-parametric group of isometries $\Lambda_t$, $t \in \bR$, by
replacing the pair $(x_2,x_3) \in \bR^2$ by $\sigma \in S^n$ in
definition (4.2) and then define $\Sigma$ and $\hh_A$ and
$\hh_B$, correspondingly.  Hence $(K,g_K)$ has the structure of a spacetime
with bKh,
where the Killing flow is $\tau_t = \Lambda_t$, $t \in \bR$. Moreover,
there is an horizon reflection ${\bf j}(x_0,x_1,\sigma) =
(-x_0,-x_1,\sigma)$ which is a PT symmetry, i.e.\ an orientation and
chronology-reversing isometry.

Let us investigate further the isometries of such spacetimes.
To simplify the matter a bit, we assume that $(K,g_K)$ does not admit
translations in the $X$-part of $K = X \times S^n$ as
symmetries. (This is not really a restriction; our findings
can be modified by taking the semidirect product of the translational
symmetry group $T_X$ with the non-translational symmetry group $G$
in the presence of such symmetries. For our treatment of the connection
between rotational spin and statistics, translational symmetries are
irrelevant.)
Since $(K,g_K)$ is orientable and time-orientable, we consider the
groups $\Gp$ and $\Gpo$ of proper (i.e.\ orientation preserving) and proper
orthochronous (i.e.\ time-orientation preserving) isometries, respectively.
In the following, we describe the proper orthochronous subgroup $\Gpo$.

The form of the metric tensor $g_K$ and the assumed triviality of $T_K$
imply that
all isometries leave $\S$ globally fixed and that an element of $\Gpo$
acting trivially on $\S$ has to be an element of the Killing flow.
Conversely, orientation preserving isometries of $\S$, i.e. elements of
$SO(n+1)$,
naturally give rise to symmetries in $\Gpo$. Indeed,
$\hat{R}(x_0,x_1,\sigma)=(x_0,x_1,R\sigma)$, $R\in SO(n+1)$,
gives an isometry of $(K,g_K)$.
To extend orientation reversing isometries of $\S$ to orientation
preserving isometries of $K$, we obviously need a different
procedure.
 To this end we note that each orientation reversing isometry of $\S
 \equiv S^n$ can be written as a product of a rotation
 in $SO(n+1)$ and an equatorial reflection $r_Q$,
 where $Q$ denotes the $S^{n-1}$ equator of fixed points of
 such a reflection. More precisely, $r_Q$ reflects points
 on $S^n$ about $Q$ along the great circles orthogonal to the equator
 $Q$. In other words, $r_Q$ acts as a reflection of the normal geodesic
 spray of $Q$ in $S^n$. Note that such equatorial reflections generate
 the action of $O(n+1)$ on $S^n$.
 In fact, if an equator $Q_1$ is inclined at angle $\phi$
 to an equator $Q_2$, then $r_{Q_1}r_{Q_2}$ is a rotation
 by $2\phi$ about the axis defined by the intersection of $Q_1$ and $Q_2$.

Now choosing a normalized, timelike,
future-oriented, rotation-invariant normal vector field $\xi_0$ along $\S$
there is a unique normalized, spacelike, outward-oriented,
rotation invariant normal vector field $\xi_1$ along $\S$
 such that $\xi_0+\xi_1$
is parallel to the vector field $\chi_A$. It is therefore equivalent to
choosing an orthonormal frame on the $X$--component of
$K= X \times S^n$. Moreover the Killing flow acts transitively on the set
of possible such choices.

 An equatorial reflection $r_Q$ extends to an orientation and
 chronology-preserving isometry $\hat{r}_{Q,0} \in \Gpo$ by setting
$\hat{r}_{Q,0} := (x_0,-x_1,r_Q\sigma)$ and we define
$\hat{r}_{Q,t} := \Lambda_t \hat{r}_{Q,0}\Lambda_t^{-1}$, where
 $\Lambda_t$, $t \in \bR$, is the Killing flow. Each $\hat{r}_{Q,t}
 \in \Gpo$ is an involution. On the other hand, by the above
 observation, each involution in $\Gpo$ restricting to some $r_Q$
 on $\S$ must be of the form $\hat{r}_{Q,t}$ for some $t \in \bR$.
Clearly,  $\hat{r}_{Q,t}$ determines a unique normalized, spacelike,
outward-oriented, rotation invariant normal vector field $\xi_1$ along $\S$
which is anti-invariant under $\hat{r}_{Q,t}$.

Thus $\Gpo$ is generated by the Killing flow, the (extensions of the)
orientation preserving isometries of $\S$ and the reflections
$\hat{r}_{Q,t}$ so that $\Gpo\equiv(\bR \times
SO(n+1))\times_\s\bZ_2$, where $\sigma$ denotes the conjugation by
$\hat{r}_{Q,0}$, for some given equator $Q$.  Consequently
$\Gpnot\equiv(\bR \times SO(n+1))$ and $\Gp$, being generated by
$\Gpo$ and the horizon reflection $\j$, is isomorphic to $(\bR\times
SO(n+1))\times_\s\bZ_2\times\bZ_2$. The following lemma obviously
holds.

\begin{Lemma}\label{Lemma:comm} On a spherically symmetric black hole,
the commutator subalgebra of the Lie algebra of the identity component
$\Gpnot$ of the group of proper isometries is isomorphic to $so(n+1)$.
The commutator subgroup of $\Gpnot$ is isomorphic to $SO(n+1)$.
\end{Lemma}

We now show that the reflection symmetries $\hat{r}_{Q,t}$ are
naturally associated with wedge-like subregions of $K$. Indeed, given
a normalized, spacelike, outward-oriented, rotation-invariant normal
vector field $\xi_1$ along $\S$, its (two-sided, maximally extended)
geodesic spray gives a geodesic-foliated Cauchy surface containing
$\S$, and it is easy to see that all such Cauchy surfaces arise in
this way. Therefore, given a reflection $\hat{r}_{Q,t}$ and an open
hemisphere $E$ in $\S \equiv S^n$ with $\partial E= Q$, we may
consider the open causal completion $W(E,t)$ of the part of the Cauchy
surface generated by the spacelike vectors determined by
$\hat{r}_{Q,t}$ and based on $E$.  Put differently, defining
$\hat{E}_0 := \{(0,x_1,\sigma) : x_1 \in \bR\,, \ \sigma \in E\}$ and
$\hat{E}_t := \Lambda_t \hat{E}_0\Lambda^{-1}_t$, $t \in \bR$, then
$W(E,t) = {\rm int}\,D(\hat{E}_t)$ where $D(\hat{E}_t)$ is the domain
of dependence of $\hat{E}_t$.  We also mention that the edge of the
wedge $W(E,t)$ is the spacelike cylinder generated by the geodesic
spray of the vectors of $\xi_1$ based on $\partial E$, i.e.\ the set
$\Lambda_t\{(0,x_1,\sigma): x_1 \in \bR\,,\ \sigma \in \partial E \}$.
Hence each $W(E,t)$ is a diamond.  The set of such wedge--regions will
be denoted by $\cW_0$. The following proposition immediately follows.

\begin{Prop}\label{Prop:wedges} 
{$(i)$} $W(E,t)^{\perp}
=\hat{r}_{\partial E,t}W(E,t)=W(E',t)$, where
$E'$ denotes the interior of the complement of $E$.
\item{$(ii)$} $\hat{R}\ W(E,t)=W(R\, E,t)$, for any $R\in SO(n+1)$.
\item{$(iii)$} $\L_s W(E,t)=W(E,s+t)$.
\item{$(iv)$} The group $\Gpo$ acts transitively on the family $\cW$ of
 wedges $W(E,s)$.
\item{$(v)$} The group $\Gpo$ is generated by the reflections $\hat{r}_{Q,t}$.
\end{Prop}

Now we show that these spacetimes fit in the scheme proposed at the
beginning of this section. Let us define $\cW$ as
$\cW_0\cup\{R\}\cup\{L\}$, $\j_R=\j_L$ as the horizon reflection $\j$ and
$\j_{W(E,t)}=\j_R\ \hat{r}_{\partial E,t}$.

\begin{Prop} If $n\geq2$ then properties (a), (b), (c) and (c$'$) hold.
 If $n=1$ then  properties (a), (b) and (c$'$) hold.
\end{Prop}

\begin{proof} Proposition~\ref{Prop:wedges} immediately gives (a). Then let
$W=W(E,t)$ and choose $\h\in so(n+1)$ as an eigenvector with eigenvalue $-1$ of
$\j_{W(E,t)}$, normalized in such a way that $\exp(\th\h)$ is a rotation
through an angle $\th$. Then property (b) is obviously
satisfied and choosing $\Wtilde=R$ we get property (c$'$). When $n\geq2$,
(c) follows by Lemma~\ref{Lemma:comm}.\end{proof}


\subsection{Quantum Field Theories on Spacetimes with Rotation Symmetry}

Now we consider a net $\cO \mapsto\cA(\cO)$ of von~Neumann algebras
indexed by elements $\cO \in \cK \cup \cW$ where $\cK$ is the set of
regular diamonds and $\cW$ is a set of wedges with the properties
discussed in the previous section; this net describes the observables
of a local quantum theory on $M$. 
  We require irreducibility, additivity
and Haag duality as in assumptions (I-III) of Sect.\ \ref{4.2} and, moreover,

\begin{itemize}
\item[(VI)] {\it Reeh-Schlieder property:} There exists a unit vector
        $\O$ (vacuum) cyclic for the von~Neumann algebras associated
        with all wedge regions.
\item[(VII)] {\it Geometric modular conjugation:}
         $$J_W\cA(\cO)J_W=\cA(\j_W\cO),$$ where $\cO$ is any regular
         diamond and $J_W$ denotes the modular conjugation associated
         with the algebra $\cA(W)$ and the vector $\O$, cf.\ Sect.\ \ref{4.2}.
\item[(VIII)] {\it Covariance:} There exists a unitary representation $U$ of
         the group $\Gpo$ such that $U(g)\O=\O$ for any $g\in\Gpo$,
         $U(g)\cA(\cO)U(g)^*=\cA(g\cO)$ for any $g\in\Gpo$ and any
         regular diamond $\cO$ and $J_WU(g)J_W=U(\j_W g\j_W)$ for
         any wedge $W$.
\end{itemize}

Let us note that, under the previous hypotheses, the representation $U$
extends to an (anti)-unitary representation of $\Gp$ with a geometric
action on the net verifying $U(\j_W)=J_W$.


\begin{Prop}\label{hatduality} Under the above assumptions, the net
satisfies duality for the relation $\hat\perp$, namely
$$
\cA(\cO)=\cap_{\cO_1\hat\perp\cO}\cA(\cO_1)'
$$
where (cf. Appendix to Section 3) $\cO_1\hat\perp\cO$ if $\cO_1\perp\cO$
and $\exists\cO_2\in\cK$ $:$ $\cO_1$, $\cO\subset\cO_2$.
\end{Prop}

\begin{proof} Let $\cO_1\perp\cO$. By Lemma \ref{RainerLemma},
for any $x\in\cO_1$ there exist $\cO_x$,
$\tilde\cO_x\in\cK$ such that $\cO\perp\cO_x$ and $\cO$,
$\cO_x\subset\tilde\cO_x$, in particular $\cO\hat\perp\cO_x$. Then
$$
\cA(\cO)\subseteq
\cap_{\cO_1\perp\cO}\cap_{x\in\cO_1}\cA(\cO_x)'=
\cap_{\cO_1\perp\cO}\cA(\cO_1)'=\cA(\cO)
$$
where the first equality follows by additivity and the second by duality.
\end{proof}

\begin{itemize}
\item[(IX)] {\it equivalence of local and global intertwiners}:
         Given a representation $\pi$ satisfying the selection criterion
         and localized in a wedge $W$, let $\rho_W$ denote the
         associated endomorphism of  $\cA(W)$, then
         $$
         (\pi,\pi)=(\r_W,\r_W).
         $$
\end{itemize}

\begin{rem}\label{assumption7}
$(i)$ This assumption implies factoriality for the algebras associated
with wedge regions, that irreducibility of representations coincides with
irreducibility on a wedge and that the equality
$(\pi,\pi')=(\r_W,\r'_W)$ holds for pair of representations (see
\cite{GuLo3}).
 \\ 
$(ii)$ Assumption (IX) has been shown to follow from dilation
invariance \cite{Robe1}, and it is conjectured that it already follows
from the existence of a non-trivial scaling limit. We give an explicit
proof of its validity for Minkowski space of any dimension in the
Appendix to this section.
  \\ 
$(iii)$ If we assume $\Gpo$ to be continuously represented by
automorphisms $\a_g$, $\Gp$ to be generated by $\{\j_W, W\in\cW\}$ and
${\rm Ad}\, J_{W_1}J_{W_2}=\a_g$, with $g=\j_{W_1}\j_{W_2}$, we get
covariance (VIII). Moreover we obtain algebraic covariance for any
sector with finite statistics, namely $\r\simeq\a_g\r\a_g^{-1}$,
$g\in\Gpo$. By an argument of M\"uger \cite{Mueg1}, this implies that
any sector is covariant w.r.t. a continuous representation of a
central extension of $\Gpo$. When the wedges separate spacelike
points, i.e.\ regular diamonds are intersections of wedges, geometric
modular conjugation (VII) also follows (cf. \cite{BDFS1}).
\end{rem}
${}$\\[6pt]
\noindent{\bf Spin and Statistics under property
(c)}
\\
\begin{Thm}\label{PCT} Let $\pi$ be a representation satisfying the 
selection criterion and localized in $\cO\subset W$. Suppose the 
associated endomorphism $\rho_W$ of the von Neumann algebra of the wedge $W$ 
has finite index. Let $j$ be the antilinear morphism implemented by the 
modular conjugation of $(\cA(W),\Omega)$. Then $j\cdot \pi\cdot j$
 is a conjugate  of $\pi$ and $\pi$ has finite statistics.
\end{Thm}
\begin{rem}\label{index} To inclusions of von Neumann algebras one can assign
an invariant, a positive number called the index (cf.\ \cite{Long2} and
refs.\ cited there). The index of the endomorphism $\rho_W$ is that
assigned to the
inclusion $\rho_W(\cA(W)) \subset \cA(W)$. For discussion of 
the relation between
the statistical dimension of a superselection sector in quantum field
theory in Minkowski spacetime and the index of its associated
localized endomorphisms, the reader is again referred to \cite{Long2}.
\end{rem}
\begin{proof} Pick a representation $\pi'$ equivalent to $\pi$ and localized 
in $W^\perp$. Then arguing as in \cite{GuLo1}, we see that $j\pi j$ and $\pi'$ 
yield conjugate endomorphisms of the von Neumann algebra of the wedge
$W^{\perp}$.  
The next step is to deduce from Assumption IX that $j\pi j$ and $\pi$ are 
conjugate representations. This circumstance is obscured by the fact that 
the product even of localized representations is defined only up to 
equivalence. For this reason, we use cocycles from $Z^1_t(\cA)$ 
instead of representations, recalling Theorem 3A.5. We have a faithful 
tensor $^*$-functor $Z^1_t(\cA)\to\cT(a)$ taking a cocycle 
$z$ into the associated endomorphism $y(a)$ in $a$ and an arrow $t$ 
into $t_a$. If $a\subset W$, then there is a tensor $^*$-functor from 
$\cT(a)$ into the category of endomorphisms of the von Neumann 
algebra of the wedge $W$, mapping an object $\rho$ onto its restriction 
to the algebra of the wedge $\rho_W$ and acting as the identity on arrows. 
Assumption IX means that the composition of these functors is even full. 
Thus if $y(a)_W$ and $\bar y(a)_W$ are the images of $z$ and $\bar z$ and 
are conjugates, 
$z$ and $\bar z$ are conjugates. If $z$ is a cocycle associated with $\pi'$ 
and $\bar z$ is a cocycle associated with $j\pi j$, then the 
endomorphisms of $\cA(W^{\perp})$
 obtained by restriction are conjugates and so 
are the equivalent endomorphisms $y(a)_W$ and $\bar y(a)_W$. Hence 
$z$ has a left inverse and finite 
statistics.
\end{proof} 
By assumption, $J_W$ implements a spacetime reflection consisting
of a time reversing (since $J_W$ is anti-unitary) and a space 
reversing transformation since, preserving the overall orientation, 
it has to reverse
the orientation of any globally invariant Cauchy surface. Therefore
the previous theorem is indeed a PCT theorem.

In the following we choose a rotationally symmetric spacetime $(M,\cW,\Gp,\bf{j})$
satisfying properties (a), (b) and (c), a local net $\cO\mapsto\cA(\cO)$ verifying
the above assumptions and an irreducible, $\Gcov$-covariant, superselection
sector with finite statistics. 

   If $\pi$ is a representation obeying the selection criterion with 
finite statistics, as above, let $\rho$ be a localized endomorphism 
defined using an associated cocycle. The standard left inverse for 
the cocycle gives us a left inverse $\phi$ for $\rho$, cf.\ Lemma 3A.10. 
When the statistics operator $\varepsilon(\rho,\rho)$ is uniquely defined, 
namely when the space-time dimension is greater than or equal to 3,
$\phi_{\rho,\rho}(\varepsilon(\rho,\rho))$ is an intertwiner between 
$\pi$ and itself. Therefore, when $\pi$ is irreducible, it is a 
complex number, cf. Sec.\ 3.4.\par
   When the dimension of a Cauchy surface is one, there are two choices 
for the statistics and correspondingly two choices for the statistics 
parameter. In this case, we choose the statistics operator $\varepsilon$ 
associated with the connected component of $\cG^\perp$ where the 
$1$-simplices have the chosen orientation (cf.\ Remark 5.2 (vii)).\par 
   Let us note that, by Assumption IX, a left inverse exists even when 
a Cauchy surface is compact.\par 
   The preceding theorem shows that the statistics phase is well defined. 
In fact, the same is true for the spin, as the following proposition shows.\smallskip 

\begin{Prop}\label{Prop:defspin}  Let $\pi$ be a representative of the given sector 
and $(\pi,U_\pi)$ a covariant representation. Then:
\begin{description}
\item{(i)} The quantity $s:=U_\pi(\exp(2\pi\h))$ is a complex number 
of modulus one depending only on the equivalence class of $\pi$ and not on the 
representation $U_\pi$. It is called the spin of the sector. 
\item{(ii)} Given $U_\pi$, let $\nu:={\rm Ad}V\cdot\pi$ be an 
equivalent representation, then $(\nu,U_\nu)$ is a covariant 
representation, where 
$$U_\nu:=VU_\pi V^*$$ 
does not depend on the intertwiner $V$.
\end{description}
\end{Prop}
\begin{proof} (i) Since $\pi$ is irreducible, $U_\pi$ is fixed up to 
a one-dimensional representation. By Assumption (c), one-dimensional 
representations are trivial on $\exp(t\h)$, hence 
$U_\pi(\exp(2\pi\h))$ does not depend on the chosen representation. 
Since $\exp(2\pi\h)$ is the identity element in $G_0$, the corresponding 
element in $\tilde G$ is a central element, so $U_\pi(\exp(2\pi\h)$ 
is a scalar by irreducibility. Equation 5.2 shows that $s$ does not depend 
on the representative $\rho$.
(ii) is obvious.\end{proof} 

  A priori $s$ depends on the Lie algebra element $\h$. However, this 
possibility is ruled out a posteriori by the spin and statistics relation. 
In the following, we fix the assignment $\pi\mapsto U_\pi$ for any 
representative $\pi$, as described in the above proposition.\par 
  Now we may state the main theorem of this section. The proof will 
require some lemmas.\smallskip 

 \begin{Thm}\label{Thm:sns} Let us consider a local net 
$\cal{O}\mapsto \cA(\cal{O})$ on a rotationally symmetric spacetime 
$(M,\cW,G_+,\bf{j})$, satisfying the above assumptions (I-III),
(VI-IX), and an 
irreducible $\tilde G$-covariant superselection sector with finite 
statistics on such a net. Then the spin of the sector agrees with its 
statistics phase.\end{Thm} 

   Let $\pi$ be a representative of a sector with finite statistics, let 
$\cO$ be contained in a wedge $W$ and let $\rho$ be an  
object of End$\cA(\cO)$ associated with $\pi$ and set 
$$\bar\rho:=j\cdot\rho\cdot j,$$ 
where $j$ is the modular antilinear morphism associated with $\cA(W)$ 
and $\Omega$. $\bar\rho$ is an object of End$\cA({\bf j}_W\cO)$. 
Let $V$ denote the 
Araki-Connes-Haagerup standard implementation (cf.\ e.g.\ \cite{GuLo3}) 
of the restriction of $\rho$ to $\cA(W)$.\smallskip 

\begin{Lemma}\label{UniIso} (cf.\ Lemma 3.1 of \cite{GuLo3}) Let $\tilde W$ be a wedge 
orthogonal to $W$. Let $\rho_{\tilde W}$ and 
$\bar\rho_{\tilde W}$ denote the restrictions of $\rho$ and $\bar\rho$ to 
$\cA(\tilde W)$, then $({\rm id},\rho_{\tilde W}\bar\rho_{\tilde W})$ 
is one dimensional and 
$V\in({\rm id},\rho_{\tilde W}\bar\rho_{\tilde W})\cap\cA(\tilde{\cal{O}})$, 
where $\tilde{\cO}$ is any element of $\cL$ \footnote{Recall that
  $\tilde{\cO}$ is in $\cL$ if $\tilde{\perp}$-duality holds either
  for $\tilde{\cO}$ or for $\tilde{\cO}^{\perp}$, cf.\ the discussion
  at beginning of Sect.\ 4.2.}
 containing $\cO$ and 
${\bf j}_W\cal{O}$ with $\tilde\cO\subset\tilde W$.\end{Lemma} 
 
\begin{proof} We remark that the existence of conjugates for finite 
statistics depends on Assumption IX and was discussed in the proof 
of \ref{Thm:sns}. Since we are dealing with a sector, Assumption IX 
implies that $(\text{id},\rho_{\tilde W}\bar\rho_{\tilde W})$ is 
one dimensional and contained in $\cA(\tilde{\cal{O}})$. In fact, 
let $z$ yield $\rho$ in $\cO$, i.e. $y(a)=\rho$ for $a=\cO$, then 
the cocycle $\bar z$, defined by 
$$\bar z(b)=j(z(\j_W b)),\quad b\in\Sigma_1,$$ 
yields $j\cdot\rho\cdot j$ in $\bar a=\j_W\cO$. Let $\hat b\in\Sigma_1$ 
be defined by $\partial_o\hat b=a$, $\partial_1\hat b=\bar a$ and 
$|\hat b|=\tilde\cO$. A simple computation shows that 
$$y(a)(\bar z(\hat b))VA=y(a)\bar y(a)(A)y(a)(\bar z(\hat b))V,
\quad a\in\cA(\tilde W).$$ 
Thus by Assumption IX, $y(a)(\bar z(\hat b))V\in\cA(\cO)$. But 
$\bar z(\hat b)\in\cA(|\hat b|)=\cA(\tilde\cO)$. Hence 
$V\in\cA(\tilde\cO)$ as claimed. Obviously, 
an isometry $V$ in $(\text{id},\rho_{\tilde W}\bar\rho_{\tilde W})$ 
will implement $\rho_W$. Now a simple computation shows that  
$j(V)\in(\text{id},\bar\rho_{\j_W\tilde W}\rho_{\j_W\tilde W})$. 
But $\j_W\tilde W=\tilde W$ since $W$ and $\tilde W$ are orthogonal. 
Hence, we may suppose that $V=j(V)$ and differs at most by a sign from 
the standard implementation of the restriction of $\rho$ to $\cA(W)$.
\end{proof}  

  Let $\pi$ be a representative of a sector with finite statistics and 
let $z$ be an associated cocycle. 
Let $\cO$ be a diamond contained in the intersection 
of two wedges $W_1$ and $W_2$ and $\rho$ the object of End$\cA(\cO)$ 
associated with $z$. Write $j_i$ for the modular antilinear  
morphism associated with $\cA(W_i)$ and $\bar\rho_i$ for 
$j_i\cdot\rho\cdot j_i$, $i=1,\,2$.\smallskip 

\begin{Lemma} Let $\rho$, $\bar\rho_i$ and $W_i$, $i=1,\,2$, be as 
above and suppose there exists a $g\in \tilde G$ with $W_2=gW_1$. The 
following identity between representations of the net 
$\cO_1\mapsto\cA(\cO_1)$, $\cO_1\supset\cal{O}$, holds: 
$$\pi\bar\rho_1={\rm Ad}U_\pi(\j_1g\j_1g^{-1})\pi\bar\rho_2
{\rm Ad}U(\j_1g\j_1g^{-1})^*,$$ 
where $g\mapsto \j_1g\j_1$ denotes by abuse of notation the 
action of $\j_1$ lifted to $\tilde G$ and $\j_1:=\j_{W_1}$.
\end{Lemma}

\begin{proof} We have $J_2=U(g)J_1U(g)^*$, hence $J_1J_2=U(\j_1g\j_1g^{-1})$ 
and $j_1j_2=\text{Ad}U(\j_1g\j_1g^{-1})$, therefore 
$$\bar\rho_1={\rm Ad}U(\j_1g\j_1g^{-1})\bar\rho_2{\rm Ad}U(\j_1g
\j_1g^{-1})^*.$$ 
Thus by covariance
$$\rho\bar\rho_1=\rho{\rm Ad}U(\j_1g\j_1g^{-1})\bar\rho_2{\rm Ad}
U(\j_1g\j_1g^{-1})^*
$$
$$={\rm Ad}U_\pi(\j_1g\j_1g^{-1})\rho\bar\rho_2{\rm Ad}
U(\j_1g\j_1g^{-1})^*.$$ 
\end{proof}

\begin{Lemma} Let $\rho$, $W_1$ and $W_2$ and $g$ be as in the previous lemma. 
Then there is a (unique) complex number $c(\rho,W_1,g)$ of modulus one 
such that
\begin{equation}\label{crho} 
U_\pi(\j_1g\j_1g^{-1})V_2U(\j_1g\j_1g^{-1})^*=c(\rho,W_1,g)V_1.
\end{equation} 
\end{Lemma}

\begin{proof} By Lemma~\ref{UniIso}, $V_1\in(\text{id},\rho_{\tilde W_1}\bar\rho_{\tilde W_1})$. 
Furthermore, by the previous lemma,\linebreak 
$U_\pi(\j_1g\j_1g^{-1})V_2U(\j_1g\j_1g^{-1})^*$ 
belongs to the same one dimensional space of intertwiners. 
\end{proof}

 \begin{Lemma}\label{Lemma:Ginvariance}Let $\rho$ and $\sigma$ be two endomorphisms 
associated with a given sector as above, the first localized in $W_1\cap W_2$, 
$W_2=gW_1$, the second in $hW_1\cap hW_2$, $g,\,h\in\tilde G$. Then 
$c(\rho,W_1,g)=c(\sigma,hW_1,hgh^{-1})$.
\end{Lemma}

\begin{proof}We first observe that if $\sigma=\text{Ad}W^*\rho$ for some 
unitary $W\in\cA(W_1\cap W_2)$, then $V_i^\rho=W^*J_iW^*J_iV_i^\sigma$ 
and this implies that $c(\sigma,W_1,g)=c(\rho,W_1,g)$. Then we note that 
$c(\rho,W_1,g)=c(\alpha_h^{-1}\rho\alpha_h,hW_1,hgh^{-1})$, where 
$\alpha_h=\text{Ad}U(h)$, because $U(h)$ establishes an isomorphism 
between the original structure and the structure transformed by $h$. 
Since $\alpha_h^{-1}\rho\alpha_h$ and $\sigma$ are associated with the 
same sector and both localized in $hW_1\cap hW_2$ and $hgh^{-1}hW_1=hW_2$, 
the result now follows.
\end{proof} 

  The previous lemma shows that for the given sector there 
is a well defined function $c(W,g)$ satisfying 
$$c(W,g)=c(hW,hgh^{-1})$$
whenever $W\cap gW\neq\emptyset$.
\begin{Lemma}\label{Lemma:group} Let $W\in\cW$.
 Then the function $g\mapsto c(W,g)$ 
is a local group character, namely for any $g,h\in\tilde G$ such that 
$W\cap gW\cap ghW\neq\emptyset$, we have 
$$ c(W,g)c(W,h) = c(W,gh)\,.$$
\end{Lemma} 
\begin{proof}Choose associated endomorphisms localized in $W\cap gW\cap ghW$ 
and denote the involutions associated with $W$ and $gW$ by $\j_1$ and 
$\j_2$, respectively. Then from the definition 
of $c$ for the pairs $(W,g)$ and $(gW,hgW)$ and the equality 
$$(\j_1g\j_1g^{-1})(\j_2h\j_2h^{-1})=\j_1g\j_1g^{-1}
(g\j_1g^{-1})h(g\j_1g^{-1})h^{-1}=\j_1hg\j_1(hg)^{-1}\,$$
one obtains the relation 
$$c(W,g)c(gW,h)=c(W,gh)$$
which means that the function $c$ is a local groupoid character.
Then, making use of Lemma~\ref{Lemma:Ginvariance} we get 
$$c(W,g)c(W,h)=c(W,ghg^{-1})=c(W,(ghg^{-1})g)=c(W,gh).$$ 
\end{proof}

  In Proposition~\ref{Prop:defspin}, we only used properties (b\,1), (b\,2). The rest of 
the argument makes essential use of further properties, more precisely, 
(b\,2) and (b\,3) are used in the following proposition, whilst (b\,3) and 
(b\,4), or rather, the orthogonality of Remark~\ref{rem:ortho}, are used to 
conclude the proof of  Theorem~\ref{Thm:sns}.\smallskip 

 \begin{Prop}\label{Prop:c=1} Under the given assumptions, we have
$c(W,\exp{\frac{\pi}2 \h})=1$.
\end{Prop}

\begin{proof} Since $g\mapsto c(W,g)$ is a local representation, it is
locally trivial on the commutator of $\Gcov$, hence, by assumption
(c), there exists $\eps>0$ such that $c(W,\exp{t \h})=1$ for
$|t|\leq\eps$. Because of assumption (b~$4$) the result follows 
by applying Lemma~\ref{Lemma:group} sufficiently often.
\end{proof}
 
\begin{Lemma}\label{Lemma:stat} Let $\r$ be an endomorphism associated with the sector 
and localized in $\cO\subset W_1\cap W_2$, where $W_1$ and $W_2$ are orthogonal 
wedges, $W_2:=\exp(\frac\p2\h)W_1$ (cf. Remark~\ref{rem:ortho}). Let
the standard implementations of its restriction to 
the algebras $\cA(W_1)$, $\cA(W_2)$  be denoted by $V_1$, $V_2$ 
as before. Then the statistics parameter $\lambda_\rho$ can be 
written as $\lambda_\rho=V_1^*V_2^*V_1V_2$. 
\end{Lemma}

\begin{proof} As in \cite{GuLo3}, Lemma 3.5,
 we first show $\l_\r=\r(V_1^*)V_1$; 
indeed if
$\r'$ is localized in $W_1\cap W^\perp_2$ and $u$ is a unitary in
$(\r,\r')$ in End$\cA(W_1)$, then $u\in\cA(W_1)$. Since 
$(W_1\cup W_2)^\perp\neq\emptyset$, $u^*A=u^*\rho'(A)=\rho(A)u^*$, 
for $A\in\cA(W_2)$. But $V_1\in\cA(W_2)$ by Lemma 5.12. Thus 
$\r(V_1^*)V_1 = u^*V_1^*uV_1$. Now $\bar\rho_1:=j_1\cdot\rho\cdot j_1$ 
is localized in $W^\perp_1\cap W_2$ and, again since 
$(W_1\cup W_2)^\perp\neq\emptyset$, $\rho$, $\rho'$ and $\bar\rho_1$ are 
comparable and $\hat\rho_1(u)=u$. Thus 
$V_1^*uV_1=\phi(u)$, where $\phi$ is the left inverse of $\rho$. 
Hence $\rho(V_1^*)V_1= u^*\phi(u) = \phi (\varepsilon(\r,\r)) =
\l_\r$. Now $V_2\in\cA(W_1)$ and implements $\r$ on $\cA(W_2)$. 
$\bar\rho_2:=j_2\cdot\rho\cdot j_2$ is localized in $W_1\cap W^\perp_2$ and 
since $(W_1\cup W_2)^\perp\neq\emptyset$, $\bar\rho_2(V_1)=V_1$ 
so we have 
\begin{equation}\label{eq:stat}
V_1^*V_2^*V_1V_2=V_1^*\phi(V_1)=\phi(\r(V_1^*)V_1)=\phi(\l_\r)=
\l_\r.
\end{equation}
\end{proof}

  Before proceeding to the proof of Theorem~\ref{Thm:sns}, we prove a 
result about orthogonal wedges. 

\begin{Lemma} Given two orthogonal wedges $W_1$, $W_2$ with 
reflections $\j_1$ and $\j_2$, there is a region $\cO$ with non--empty 
causal complement which is invariant
under $\j_1$ and $\j_2$.
\end{Lemma}

\begin{proof} Take $\cO_1$ and $\cO_2$ orthogonal to each other and contained
in $W_1\cap W_2$, and set $\cO=\cO_1\cup \j_1\cO_1\cup \j_2\cO_1\cup
\j_1\j_2\cO_1$. Clearly $\cO$ is causally disjoint from $\cO_2$ and 
invariant under $\j_1$ and $\j_2$.
\end{proof}

\noindent {\it Proof of Theorem~\ref{Thm:sns}}. We follow the reasoning
 of \cite{GuLo3}.  Consider the two orthogonal
 wedges $W_1$, and $W_2=\exp(\frac\p2\h)W_1$ as in the preceding lemma and 
 choose a representative endomorphism localized in a regular diamond 
 $\cO\in W_1\cap W_2$ and chosen such that there is an $\tilde\cO$ containing 
$\cO$, $\j_1\cO$, $\j_2\cO$ and $\j_1\j_2\cO$. Then 
$\rho\bar\rho_1j_2\rho\bar\rho_1j_2=\rho\bar\rho_2j_1\rho\bar\rho_2j_1$ 
and are objects of End$\cA(\tilde\cO)$. $V_1J_2V_1J_2$ and $V_2J_1V_2J_1$ 
intertwine from the identity to this object in End$\cA(\tilde\cO)$. Thus
 $\b_\r:=(V_1J_2V_1J_2)^*V_2J_1V_2J_1 $ is a scalar and we first show that 
it belongs to
 $(0,1]$, as in Lemma~3.4 in \cite{GuLo3}, observing that
 \begin{equation}\label{beta2}
\b_\r=V_1^*U(\exp{\pi\h})V_1^*V_2U(\exp{\pi\h})V_2.
\end{equation}
 Then, by Equation \ref{crho} with $g=\exp{\frac\pi2\h}$ and 
 its adjoint and using the equation
 $c(W,\exp{\frac{\pi}2 \h})=1$, proved in 
 Proposition~\ref{Prop:c=1}, we get
 \begin{equation}\label{spin2}
 V_2^*V_1= s_\r U(\exp{\pi\h})V_1^*V_2U(\exp{\pi\h}).
 \end{equation}
 Inserting this equation into the expression for the statistics 
 parameter given by Lemma~\ref{Lemma:stat} and comparing with 
 Equation~\ref{beta2} we obtain
 $$\l_\r=V_1^*V_2^*V_1V_2=s_\r
V_1^*U(\exp{\pi\h})V_1^*V_2U(\exp{\pi\h})V_2=
s_\r\b_\r$$
and the result follows easily.
\hfill {\Large $\Box$}\\[6pt]

We conclude this subsection showing that the Spin and Statistics relation makes
sense and is true for reducible covariant representations, too. Clearly the
result follows from the irreducible case once we can show that the irreducible
subrepresentations are still covariant.

 \begin{Prop}\label{Prop:reducible} Let $\pi$ be a representation
   satisfying the 
selection criterion and with finite statistics
and covariant under the group $\Gcov$. Then there exists a covariant
representation $(\pi,U_\pi)$, where $U_\pi$ acts trivially on $\pi(\cA)'$.  
$U_\pi$ is unique up to a one
dimensional representation and any other choice of $U_\pi$ may be
written as a product of $U_\pi$ and a representation $U^o_\pi$ contained in
$\r(\cA)'$. In particular, each irreducible component of $\r$ is
$\Gcov$-covariant.
\end{Prop}

\begin{proof}Since $\pi$ has finite statistics, $\pi(\cA)'$ and  hence the
centre of $\pi(\cA)$ are finite dimensional. 
Therefore if $(\pi,\widetilde U_\pi)$ 
is a covariant representation, $\widetilde U_\pi$ is trivial on such
centre. Then, since $\widetilde U_\pi$ implements automorphisms of
$\pi(\cA)$, it implements an action of $\Gcov$ by automorphisms of
$\r(\cA)'$, preserving any factorial component. Thus this
action is implemented by a unitary representation $U^o$ in $\pi(\cA)'$. Then
$g\in\Gcov\to \widetilde U_r(g)U^o(g)^*$ is a representation of $\Gcov$
acting trivially on $\pi(\cA)'$.  Clearly such a representation
decomposes into representations of the irreducible components of
$\pi$, so these are $\Gcov$-covariant.
\end{proof}

\begin{rem}The given proof of the spin and statistics relation does
not rely on the continuity of the representations $U$ or $U_\pi$.  Even
Proposition~\ref{Prop:reducible} does not require continuity, because it
relies on the fact that a connected Lie group acts trivially on a finite
set and this is true without assuming continuity. 
\end{rem}

\bigskip\noindent{\bf Spin and Statistics under property (c$'$)}\quad
\\[6pt]
Now we give a proof of the Spin and Statisitics Theorem for
rotationally symmetric spacetimes satisfying $(c')$ rather than
$(c)$, such as the 3-dimensional Schwarzschild-Kruskal spacetime,
for example.

Recall that in this case there is an involution
$r_W:=\j_W\j_{W_0}\in\Gpo$ anticommuting with $\h$
(cf. Remark~\ref{rem:rotgeom} (vi)).

Let us denote by the subgroup of $\Gpo$ generated by $\Gpnot$
and $r_W$ by $\Gd$. If $r_W$ does not belong to $\Gpnot$, $\Gd$ is isomorphic
to $\Gpnot\times_\s\bZ_2$, where $\s={\rm Ad} r_W$. In the same way we can
consider the group $\Gdcov\equiv\Gcov\times_\s\bZ_2$, where, by an abuse of
notation, $\s$ lifted to $\Gcov$ is still denoted by $\s$. We shall also
denote the corresponding element in $\Gdcov$ by $r_W$. Clearly the
covering map extends to an epimorphism from $\Gdcov$ to $\Gd$. We want
to show that any $\Gcov$-covariant sector with finite statistics
is $\Gdcov$-covariant, too.

 \begin{Prop}\label{G1cov} Let $\pi$ be an irreducible representation of $\cA$
obeying the selection criterion, with finite statistics, and covariant under the
group $\Gcov$. Then it is covariant under $\Gdcov$.
 \end{Prop}

\begin{proof} Of course we may restrict to the case
$r_W\notin\Gpnot$. Since $r\equiv r_W$ is the product of
$\j_W$ with $\j_{W_0}$, such reflections are implemented by the
corresponding modular involutions $J$, $J_0$, and $j\r j$ is
equivalent to $j_0\r j_0$, both being conjugate endomorphisms of $\r$,
there exists a unitary $U_r$ intertwining $\r$ and
$\alpha(r)\r\alpha(r)$.  Since $r^2=1$, $U_r^2$ implements the trivial
action on $\cA$, hence, $\r$ being irreducibile, $U_r^2$ is a
constant and we may choose $U_r$ selfadjoint. Then $U_rU_\r(rgr)U_r$
is another representation of $\Gcov$ realizing the covariance of
$\r$. Since $\r$ is irreducible, we get
$U_rU_\r(g)U_r=\chi(g)U_\r(rgr)$, where $\chi(g)$ is a character of
$\Gcov$. Applying this relation twice, we get
$U_\r(g)=\chi(g)\chi(rgr)U_\r(g)$, namely $\chi(g)\chi(rgr)=1$.  Now
observe that, since $\chi$ is a Lie group representation, it is the
exponential of a Lie algebra morphism $\kappa$ from the Lie algebra of
$\Gpnot$ to $\bR$. Since $\Gcov$ is simply connected, $\kappa/2$
exponentiates to a character, which we denote by $\sqrt\chi$, whose
square gives $\chi$, and we get
 $$
U_r\sqrt{\chi}(g^{-1}U_\r(g))U_r=\sqrt{\chi}(rg^{-1}r)U_\r(rgr),
 $$
 namely $\sqrt{\chi}(g^{-1})U_\r(g)$ and $U_r$ yield the
required representation of $\Gdcov$.
 \end{proof}

 \begin{Thm} Let $(M,\Gp,\j,\cW)$ be a rotationally symmetric
spacetime satisfying properties (a) and (b) and (c$'$), $(\cA,U,\O)$ a
covariant net verifying the mentioned axioms (I-III), (VI-IX),
and let $\r$ be a $\Gcov$-covariant sector with finite statistics.
Then the spin and statistics relation holds.  \end{Thm}

\begin{proof}  By property (c$'$),
$s_\r$ does not depend on $U_\r$, as observed in
Remark~\ref{rem:rotgeom} (vi).  Concerning the relation between spin
and statistics, we may define a function $c(W,g)$, $g\in\Gdcov$,
as in the proof of Theorem~\ref{Thm:sns}, which is indeed a local group
representation namely, if $g,h\in\Gdcov$ verify $W\cap gW\cap
ghW\ne\emptyset$, we have $c(W,g)c(W,h)=c(W,gh)$.
 \\
Setting $\tilde r:=\exp(\frac\pi2\h)r_W\exp(-\frac\pi2\h)$, we get
$$
\tilde rW=\exp(\frac\pi2\h)r_W\exp(-\frac\pi2\h)W=\exp(\pi \h)r_W W=W
$$
and $\tilde r\exp(t\h)\tilde
r=\exp(-t\h)$.  Hence, for sufficiently small $t$,
 \begin{align*}
c(W,\exp(2t\h))&=c(W,\tilde{r})c(W,\tilde{r}\exp(t\h))c(W,\exp(t\h))\cr
&=c(W,\tilde{r}\exp(t\h)\tilde{r})c(W,\exp(t\h))\cr
&=c(W,\exp(-t\h))c(W,\exp(t\h)=1.
 \end{align*}
 The proof now continues as in Theorem~\ref{Thm:sns}.
 \end{proof}

As before, the Spin and Statistics relation for reducible representations
follows as soon as we prove that the irreducible representations are $\tilde
G_1$-covariant, and this is a consequence of Proposition~\ref{Prop:reducible}
and Proposition \ref{G1cov}.

 \begin{rem} Generally speaking, asking
for an irreducible endomorphism to be covariant corresponds to
asking for a projective representation of the group $\Gpnot$,
namely a representation of a central extension of $\Gpnot$ by some
subgroup of $\bT^1$ implementing the action of $\Gpnot$ on
$\r(\cA)$. This means that there may be an extension at the Lie
algebra level, not only a covering.  However, we are not aware of any
physical example where non-trivial Lie algebra central extensions
exist (for the Poincar\'e group on the two-dimensional Minkowski
space, such non-trivial extensions exist, but are incompatible with
the positive energy requirement).  As a consequence, we have only treated
the case of the universal covering.
\end{rem}

\def\A_\z{\A_{\infty}}
\def\bP{\bf P}
\subsection{Appendix.  Equivalence
between local and global intertwiners in Minkowski spacetime}

In this appendix we prove that Assumption IX concerning the
equivalence of local and global intertwiners holds for sectors
localized in a wedge region of a Minkowski space of arbitrary
dimension. The argument is a straightforward adaptation of 
that given in \cite{GuLo3} for a conformal net on $S^1$.

 In the following, $\cA$ is a net of von Neumann algebras on the
 $d+1$-dimensional Minkowski spacetime. We assume Poincar\'e
 covariance with positive energy and uniqueness of the vacuum,
 additivity and Haag duality
 $$
 \cA(\cO)=\cA(\cO')'
 $$
 if $\cO$ is either a double cone or a wedge region.

If $\r,\s$ are  endomorphisms of  $\cA$ localized in 
the wedge region $W$,  we consider their intertwiner space
$(\r_W,\s_W):=\{T\in \cA(W):\s(A)T=T\r(A), \, \forall A\in \cA(W)\}$.
By duality we always have $(\r,\s)\subset(\r_W,\s_W)$.

\smallskip
{\bf Theorem A5.1} {\sl Let $W$ be a wedge region and $\r$, $\s$ be
endomorphisms with
finite dimension localized in a double cone $\cO\subset W$. Then
 $$
(\r_W,\s_W)=(\r,\s) .
 $$
Namely, if
$T\in(\r_W,\s_W)$ then $T$ intertwines the representations
$\r$ and $\s$ of $\cA$.}
\smallskip

In the following $\r$ denotes an endomorphism with finite
dimension of the quasi-local
observable C$^*$-algebra $\cA$ localized in a double cone
$\cO$ contained in the wedge
$W$. We may assume that $W=\{x\in{\mathbb R}^{d+1}: -x_1>|x_0|\}$.

We shall denote by ${\mathbb R}^2$ the 2-dimensional $x_0 - x_1$-plane
and by $\bP$ the corresponding 2-dimensional Poincar\'e group,
namely the semidirect product of the 2-dimensional translations
$\{T(x)\}_{x\in\mathbb R^2}$
and boosts $\{\L(s)\}_{s\in\mathbb R}$ associated to $W$:  each $g\in\bP$
can be written uniquely
as a product $g=T(x)\L(s)$.

 The endomorphism $\r$ restricts  to an
endomorphism  of  the C$^*$-algebra associated with $W+x$
and then extends to the von Neumann algebra $\cA(W+x)$, for
 $x_1 >0$, hence giving rise to an
endomorphism the $C^*$-algebra
$\cA_\infty$, the norm closure of
$\cup_{x\in{\mathbb R}^2}\cA(W+x)$. We will still use $\r$ 
to denote this endomorphism.
Since $\bP$ is simply connected, there is a unitary representation 
 $U_\r$  of $\bP$ expressing the covariance
of $\r$ with respect to $\bP$
\begin{equation}\label{(2.2)}
\b_g(A)=U_\r( g)AU_\r(g)^*=z_\r(g)U(g)AU(g)^*z_\r(g)^*,
\quad A\in \cA_\infty,\,g\in \bP.
\end{equation}
As the cocycle $z_\r$ is a local operator by Haag duality
(this is the essential point about the 2-dimensional $x_0 - x_1$-net
inherited from the higher dimensional original net) $\b$ 
is an action of $\bP$ by automorphisms of $\cA_\infty$.

We consider now the semigroup $\bP_0$, the semidirect product 
of the boosts $\L(s)$ with the
positive translations, where we say that $T(x)$ is positive if
 $x\in{\mathbb R}^2$ with $x_1>|x_0|$.
$\bP_0$ is an amenable semigroup and we need an invariant mean $m$
constructed as
follows: first we average (with an invariant mean) over positive
translations and then over boosts.
 Observe that  $f\to\int_{\bP_0}f(g)dm(g)$  gives an invariant mean
on all $\bP$ vanishing on  $f$ if, for any given $s\in{\mathbb R}$, the map
$x\in{\mathbb R}^2\to f(T(x)\L(s))$  vanishes on a right wedge.

Then we associate to $m$ the completely positive map $\Phi$
of $\cA_\infty$ to $\cB(\cH)$ given by
 $$
 \Phi(A):=\int_{\bP_0} z_\r(g)^*
Az_\r(g)dm(g), \quad A\in \cA_\infty .
 $$
\smallskip
{\bf Lemma A5.2\ } {\sl $\Phi$ is a  left inverse
of $\r$ on $\cA_{\infty}$. Moreover $\Phi$ is locally normal,
i.e. has  normal restriction to $\cA(W+x)$, $x\in{\mathbb R}^2$, and
$\bP$-invariant, namely
 $$
\Phi=\a_g^{-1}\Phi\b_g,\quad g\in\bP. $$
We have set $\a_g\equiv {\rm Ad}U(g)$.}
 \smallskip

\begin{proof} Let $A$ belong to $\cA(W+x)$, $x\in{\mathbb R}^2$. By
formula \ref{(2.2)}
 $$
\Phi(\r(A))=\int_{\bP_0} \a_g(\r(\a_{g^{-1}}(A)))dm(g)=A
 $$
because of the above property of $m$ since the integrand is
constantly equal to $A$ on the set ${g\in{\bP_0}:g^{-1}W\cap
\cO=\emptyset}$.
 Then the localization of $\r$ and Haag duality imply that
the range of $\Phi$ is contained in $\cA_\infty$.

Setting $E=\r\cdot\Phi$ gives a conditional expectation
of $\cA_\infty$ onto the range of $\r$ that restricts to a
conditional expectation $E_x$ of $\cA(W+x)$ onto
$\r(\cA(W+x))$ if $W+x\supset \cO$. Since
$\r_{W+x}$ is assumed to have finite index,
$E_{x}$ is automatically normal \cite{Long2}.
Therefore $\Phi\rest{\cA(W+x)}=
\r_{W+x}^{-1}E_{x}$ is
normal for $x=(0,x_1)$ with $x_1>0$, hence for any $x$.

Concerning the $\bP$-invariance of $\Phi$ we have, making use of
the cocycle condition,

\begin{equation*}\begin{split}
\a_g^{-1}\Phi\b_g(A)
&=\a_g^{-1}( \int_{\bP_0} z_{\r}(h)^*\b_g(A)z_\r(h)dm(h))\\
&=\a_g^{-1}(\int_{\bP_0}z_\r(h)^*z_\r(g)
  \a_g(A)z_\r(g)^*z_\r(h)dm(h))\\
&=\int_{\bP_0} z_\r(hg^{-1})^*A z_\r(hg^{-1})dm(h)=\Phi(A)
\end{split}\end{equation*}
\end{proof}
\smallskip
{\bf Corollary A5.3\ }{\sl $\f = \omega\Phi$ is a locally
normal $\b$-invariant state  on $\cA_\infty$, where $\omega =
(\ \cdot\ \Omega, \Omega)$. }\smallskip
\begin{proof}
 We have
$\f\b_g=\omega\Phi\b_g=\omega\a_g\Phi=\omega\Phi
=\f$ and $\f$ is locally normal because both
$\omega$ and $\Phi$ are locally normal.
\end{proof}

Let $\{\p_\f,\x_\f,\cH_\f\}$ be the GNS triple
associated with the above state $\f$ and $V$ be the unitary
representation of $\bP$ on $\cH_\f$ given by $V_g
A\x_\f=\b_g(A)\x_\f$ for $A\in\cA_{\infty}$. Notice that
$V$ is strongly continuous because $\f$ is locally normal.

\smallskip\noindent
 {\bf Lemma A5.4\ }{\sl If $\r$ is irreducible then
 $$
\f(x)
=\int_{\bP_0}\b_g(x)dm(g),\quad
x\in\cA_\infty
 $$}
 \smallskip
\begin{proof}
If $A\in\cA(W+x)$ and $B\in\cA_\infty$ is localized in a
double cone, the commutator function
${\mathbb R}^2 \owns x \mapsto[\b_{T(x)\L(s)}(A),\r(B)]
=\b_{T(x)\L(s)}([A,\r(\a_{T(x)\L(s)}^{-1}(B)])$
vanishes on a right wedge, hence
$[\int_{\bP_0}\b_g(A),\r(B)dm(g)]=
\int_{\bP_0}[\b_g(A),\r(B)]dm(g)=0$.

 Since $\r$ is locally normal, $\int_{\bP_0}\b_{g}(A)dm(g)$
commutes with every $\r(\cA(W+x))$, thus with
$\r(\cA_\infty)$;
 but $\r$ being irreducible, it is therefore a scalar equal to its
vacuum expectation value:
 $$
\int_{\bP_0}\b_g(A)dm(g)= \int_{\bP_0}\omega(\b_g(A))dm(g)
=\int_{\bP_0}\omega(z_g^*Az_g)dm(g)=\omega\Phi(A)=\f(A),
$$
as $\omega$ is normal and $\a$-invariant.
\end{proof}
\smallskip
{\bf Corollary A5.5\ } {\sl If $\r$ is irreducible, the two-parameter
unitary translation group $V(T(x))$ satisfies the spectrum condition.}
\smallskip
\begin{proof}
One may repeat the proof of Corollary 2.7 of \cite{GuLo3} for each of the
one-parameter light-like
unitary translation groups.
 \end{proof}
\smallskip
 {\bf Corollary A5.6\ }{\sl If $\r$ is irreducible, $\f$ is faithful
on $\cup\r(\cA(W+x))$.}
\smallskip
 \begin{proof}
$\cA_\infty$ is a simple $C^*$-algebra since it is the inductive limit
of type III factors (that are simple $C^*$-algebras). Therefore
$\p_\f$ is one-to-one and the statement will follow if we show
that $\x_\f$ if cyclic for
$\cB_x\equiv\r(\cA(W+x))',\, x_1 >0$. To this end we may use a
classical Reeh-Schlieder argument. If $\psi\in\cH$ is orthogonal to
$\cB_x\x_\f$, and $x-y\in W$, then for all
$A\in\cB_{y}$ we have $(A\x_\f,V(T(x))\psi)=0$ for $x$ in a
neighborhood of $0$, thus for all $x\in\bR^2$ by the spectrum condition
shown by Corollary A5.5.
 Hence, setting $\a_x\equiv\a_{T(x)}$ and $\b_x\equiv\b_{T(x)}$,
$\psi$ is orthogonal to $(\cup_x \b_x(\cB_{y}))\x_\f$, thus
$\psi=0$ because $\cup_x \b_x(\cB_y)$ is irreducible since
\begin{equation*}\begin{split}
(\bigcup_x\b_x(\cB_{y}))'
&=\bigcap_x \b_x(\r(\cA(W+y))=
\bigcap_x\r(\a_x(\cA(W+y)))\\
 & = \r(\bigcap_x
\a_x(\cA(W+y))) =\bigcap_x \cA(W+x)= {\mathbb C}
\end{split}\end{equation*}
 by the local normality of $\r$.
 \end{proof}
\smallskip
 {\bf Proposition A5.7} {\sl $(\r_{W+x},\r_{W+x})$ does not
depend on the wedge  $W+x\supset \cO$.}
\smallskip
\begin{proof}
 We begin with the case where $\r$ is irreducible and assume
for convenience that $\bar \cO\subset W$. Notice then that
 $(\r_{W},\r_{W})$ is finite-dimensional
and, by covariance, globally $\b_g$-invariant with $g$ in the
subgroup of boosts because these transformations preserve
$W$.  Therefore
$(\r_{W},\r_{W})\x_\f$ is a
finite-dimensional subspace of $\cH_\f$ globally invariant for
$V(\L(s))$, $s\in{\mathbb R}$. By Proposition B.3 of \cite{GuLo3}
we thus have $V(T(x))A\x_\f=A\x_\f$ for every element $A\in
(\r_{W},\r_{W})$, thus
$\b_{T(x)}(A)=A$ because $\x_\f$ is separating.
 It follows that if $A\in(\r_{W},\r_{W})$ and
$B\in \cA(W)$
 $$
[A,\r(\a_g(B))]=\b_g([\b_g^{-1}(A),\r(B)])=\b_g([A,\r(B)])=0
 $$
namely
 $$
A\in (\r_W,\r_W)\Rightarrow A\in
(\r,\r)=\mathbb C\ .
 $$
Since the converse implication is obvious by wedge duality we have
the equality of the two intertwiner spaces.

Now if $\r$ is any endomorphism with finite index,  $(\r,\r)$ is
finite-dimensional because $(\r,\r)\subset (\r_W,\r_W)$ and $\r$ decomposes
into a direct sum of irreducible endomorphisms of $\cA_\infty$ which are
covariant, therefore the preceding analysis shows
that $(\r_W,\r_W)=(\r,\r)$ in this case, too.
 Since $(\r,\r)$ is translation invariant, we get
$(\r_{W+x} ,\r_{W+x})=(\r,\r)$
whenever $\cO\subset W+x$ and, since $x$ was arbitrary,
the result follows.
 \end{proof}

\noindent 
{\it Proof of Theorem A5.1.\ } The case $\s=\r$ follows immediately by
Proposition A5.6: if $T\in (\r_W,\r_W)$ then $T$ also belongs to
$(\r_{\tilde W},\r_{\tilde W})$ for any wedge $\tilde W\supset
W$ hence by additivity $T$ is a self-intertwiner of $\r$ on
the whole algebra $\cA$.

To handle the general case, consider  a  direct sum endomorphism
$\eta:=\r\oplus\s$ localized in $W$, then
$$
{\rm dim}(\eta_W,\eta_W)={\rm dim}(\r_W,\r_W)+{\rm
dim}(\s_W,\s_W) +2{\rm dim}(\r_W,\s_W)
$$
while
$$
{\rm dim}(\eta,\eta)={\rm dim}(\r,\r)+{\rm
dim}(\s,\s) +2{\rm dim}(\r,\s)
$$
therefore ${\rm dim}(\r_W,\s_W)={\rm dim}(\r,\s)$ and since we
always have $(\r,\s)\subset (\r_W,\s_W)$ these two intertwiner
spaces coincide.
\hfill {\Large $\Box$}

${}$\\[14pt]
{\large\bf Acknowledgments}
\\[6pt]
 R.V.\  has been in part supported 
through the Operator Algebras Network funded by the EU under contract 
CHRX-CT94-0566. R.V.\ also wishes to thank
all the members of the operator algebra group at the Dipartimento di
Matematica, Universit\`a di Roma ``Tor Vergata'', for their kind
hospitality in 1996.

Three of the authors (D.G., J.R., R.V.) would like to thank the Erwin
Schr\"odinger Institute, Vienna, as well as the Organizers of the
Workshop on Quantum Field Theory in September 1997, 
 D.\ Buchholz and J.\ Yngvason, for
the opportunity of participating the workshop. The excellent working 
conditions provided a basis for discussions relevant to the present paper. 

We would also like to thank K.-H.\ Rehren for pointing out a gap in an
earlier version of Section 4.


\begin{thebibliography}{[22]}

\bibitem{ArZs1} H. Araki, L. Zsido, ``Extension of the structure
theorem of Borchers and its application to half-sided modular inclusions''
               manuscript, preliminary version (1995), to appear
\bibitem{BauWo} H. Baumg\"artel, M. Wollenberg, Causal nets of
  operator algebras, Akademie Verlag, Berlin, 1992
\bibitem{BeemEh} J.K. Beem, P.E. Ehrlich, Global Lorentzian
  geometry, Marcel Dekker, New York, 1981
\bibitem{BiWi} J.J. Bisognano, E.H. Wichmann, ``On the duality
  condition for quantum fields'', J. Math. Phys. {\bf 17}, 303 (1976)
\bibitem{Bor1} H.-J. Borchers, ``The CPT-theorem in two-dimensional
  theories of local observables'', Commun. Math. Phys. {\bf 143}, 315
  (1992)
\bibitem{Bor2} H.-J. Borchers, ``On modular inclusion and spectrum
  condition'', Lett. Math. Phys. {\bf 27}, 311 (1993)
\bibitem{Bor3} H.-J. Borchers, ``When does Lorentz invariance imply
  wedge duality?'', Lett. Math. Phys. {\bf 35}, 39 (1995)
\bibitem{Bor4} H.-J. Borchers, ``Half-sided modular inclusions and the
  construction of the Poincar\'e Group'', Commun. Math. Phys. {\bf
  179}, 703 (1996)
\bibitem{BoBu} H.-J. Borchers, D. Buchholz,
``Global properties of vacuum states in de Sitter space'',
  Ann. Inst. H. Poincar\'e {\bf 70}, 23 (1999)
\bibitem{BrMo}  J. Bros, H. Epstein, U. Moschella,
``Analyticity properties and thermal effects for general quantum field 
theory on de Sitter space-time'',
 Commun. Math. Phys. {\bf 196}, 535 (1998)
\bibitem{Brown} M. Brown, ``Locally flat imbeddings of topological
  manifolds'', Annals of Math. {\bf 75}, 331 (1962)
\bibitem{BrFre} R. Brunetti, K. Fredenhagen, ``Interacting quantum
  fields in curved space: Renormalizability of $\varphi^4$'',
in the Proceedings of the Conference ``Operator algebras and quantum field
theory'' held in Rome, July 1996, S. Doplicher, R. Longo, J. Roberts, 
L. Zsido eds, International Press, 1997;\\
---, ``Microlocal analysis and interacting quantum field theories:
Renormalization on physical backgrounds'', preprint math-ph/9903028
\bibitem{BFK} R. Brunetti, K. Fredenhagen, M. K\"ohler,
 ``The microlocal spectrum condition and Wick polynomials of free
 fields in curved spacetimes'', Commun. Math. Phys. {\bf 180}, 633 (1996) 
\bibitem{BGLo1} R. Brunetti,  D. Guido, R. Longo, ``Modular
  structure and duality in conformal Quantum Field Theory'',
  Commun. Math. Phys. {\bf 156}, 201 (1993)
\bibitem{BGLo2} R. Brunetti, D. Guido, R. Longo, ``Group 
cohomology, modular theory and space-time symmetries'', 
Rev. Math. Phys., {\bf 7}, 57 (1994)
\bibitem{BGLo3} R. Brunetti,  D. Guido, R. Longo, ``First quantization via BW 
  property'', in progress.
\bibitem{BuSu1} D. Buchholz, S.J. Summers, ``An algebraic
  characterization of vacuum states in Minkowski space'',
  Commun. Math. Phys. {\bf 155}, 442 (1993) 
\bibitem{BDFS1} D. Buchholz, O. Dreyer, M. Florig,
  S.J. Summers, ``Geometric modular action and spacetime symmetry groups'',
  preprint math-ph/9805026
\bibitem{BuFS} D. Buchholz, M. Florig, S.J. Summers, ``Hawking-Unruh
  temperature and Einstein causality in anti-de\,Sitter space-time'',
  hep-th/9905178 
\bibitem{Cla} C.J.S. Clarke, ``A title of cosmic censorship'',
  Class. Quantum Grav. {\bf 11}, 1375 (1994)
\bibitem{Dieck} J. Dieckmann, ``Cauchy surfaces in globally hyperbolic
  spacetimes'', J. Math. Phys. {\bf 29}, 578 (1988)
\bibitem{DimKay} J. Dimock, B.S. Kay, ``Classical and quantum
  scattering theory for linear scalar fields on the Schwarzschild
  metric. I.'' Ann. Phys. (N.Y.) {\bf 175}, 366 (1987)
\bibitem{DHR1} S.~Doplicher, R.~Haag, J.E.~Roberts: ``Fields, observables and 
gauge transformations I'', Commun.\ Math.\ Phys.\ {\bf 13}, 1 (1969)
\bibitem{DR1} S.\ Doplicher, J.E.\ Roberts: ``Endomorphisms of
 $C^*$--algebras,  cross products and duality for compact groups'',
 Ann.\ Math.\ {\bf 130}, 75  (1989)
\bibitem{DR} S.\ Doplicher, J.E.\ Roberts: ``Why there is a field
  algebra with a  compact gauge group describing the superselection
 structure in particle physics'', 
Commun.\ Math.\ Phys.\ {\bf 131}, 51 (1990)
\bibitem{Dri} W. Driessler, ``On the structure of fields and algebras
  on null planes, I'', Acta Phys. Austriaca {\bf 46}, 63 (1977)
\bibitem{FreHa} K. Fredenhagen, R. Haag, ``On the derivation of
  Hawking radiation associated with the formation of a black hole'',
Commun. Math. Phys. {\bf 127}, 273 (1990)
\bibitem{FreReS} K. Fredenhagen, K.-H. Rehren, B. Schroer,
 ``Superselection sectors with braid group statistics and exchange
 algebras''. 1,  Commun. Math. Phys. {\bf 125}, 201 (1989); 2,
 Rev. Math. Phys. {\bf Special Issue}, 111 (Dec. 1992)  
\bibitem{Ger} R. Geroch, ``Domain of dependence'', J. Math. Phys. {\bf
    11}, 437 (1970)
\bibitem{GuLo1} D. Guido, R. Longo, ``Relativistic invariance and
charge conjugation in quantum field theory'',
Commun. Math. Phys. {\bf 148}, 521 (1992)
\bibitem{GuLo2} D. Guido, R. Longo, ``An algebraic spin and statistics
  theorem'', Commun. Math. Phys. {\bf 172}, 517 (1995)
\bibitem{GuLo3}  D. Guido, R. Longo, ``The conformal
spin and statistics theorem'', Commun. Math. Phys.
{\bf 181},  11 (1996)
\bibitem{GLWi1} D. Guido, R. Longo, H.-W. Wiesbrock,
 ``Extensions of conformal nets and superselection structures'',
Commun. Math. Phys. {\bf 192}, 217 (1998)
\bibitem{Haag} R. Haag, Local quantum physics, 2nd ed., Springer, Berlin,
  Heidelberg, New York, 1996
\bibitem{Haw} S.W. Hawking, ``Particle creation by black holes'',
  Commun. Math. Phys. {\bf 43}, 199 (1975)
\bibitem{HawEll} S.W. Hawking, G.F.R. Ellis, The large scale
  structure of space-time, Cambridge University Press, 1973
\bibitem{Kay} B.S. Kay, ``The double-wedge algebra for quantum fields
  on Schwarzschild and Minkowski spacetimes'',
  Commun. Math. Phys. {\bf 100}, 57 (1985)
\bibitem{KayRP} B.S. Kay, ``Quantum fields in curved spacetime: Non
  global hyperbolicity and locality'',
in the Proceedings of the Conference ``Operator algebras and quantum field
theory'' held in Rome, July 1996, S. Doplicher, R. Longo, J. Roberts, 
L. Zsido eds, International Press, 1997
\bibitem{KRW} B.S. Kay, M.J. Radzikowski, R.M. Wald, ``Quantum field
  theory on spacetimes with a compactly generated Cauchy-horizon'',
  Commun. Math. Phys. {\bf 183}, 533 (1997) 
\bibitem{KayWald}  B.S. Kay, R.M. Wald, ``Theorems on the uniqueness
  and thermal properties of stationary, nonsingular, quasifree states
  on spacetimes with a bifurcate Killing horizon'', Phys.Rep. {\bf
    207}, 49  (1991) 
\bibitem{Keyl} M. Keyl, ``Causal spaces, causal complements and their
  relations to quantum field theory'', Rev. Math. Phys. {\bf 8}, 229
  (1996)
\bibitem{Kuc} B. Kuckert, ``A new approach to spin and statistics'',
  Lett. Math. Phys. {\bf 35}, 319 (1995) 
\bibitem{LeuKlSt} H. Leutwyler, J.R. Klauder, L. Streit, ``Quantum
  field theory on lightlike slabs'', Nouvo Cimento {\bf 66A}, 536 (1970)
 \bibitem{Long2} R. Longo, ``Index of subfactors and statistics of 
 quantum fields. I'', Commun. Math. Phys. {\bf 126}, 217 (1989)
\bibitem{Long3} R. Longo, ``On the spin-statistics relation
for topological charges'',
in the Proceedings of the Conference ``Operator algebras and quantum field
theory'' held in Rome, July 1996, S. Doplicher, R. Longo, J. Roberts, 
L. Zsido eds, International Press, 1997
\bibitem{Long4} R. Longo, ``An analogue of the Kac--Wakimoto formula and 
black hole conditional entropy'', Commun. Math. Phys. 
{\bf 186}, 451 (1997)
\bibitem{LoRo} R. Longo, J.E. Roberts, ``A theory of dimension'',
  K-Theory {\bf 11}, 103 (1997) 
\bibitem{Mueg1} M. M\"uger, 
``On soliton automorphisms in massive and conformal theories'',
Rev. Math. Phys. {\bf 11}, 337 (1999)
\bibitem{ONeill} B. O'Neill, Semi-Riemannian geometry, Academic Press,
  New York, 1983
\bibitem{Rad} M.J. Radzikowski, ``Micro-local appraoch to the Hadamard
  condition in quantum field theory in curved space-time'',
  Commun. Math. Phys. {\bf 179}, 529 (1996)
\bibitem{Reh} K.-H. Rehren, ``Algebraic holography'', preprint hep-th/9905179 
\bibitem{R1} J.E.\ Roberts: Net cohomology and its applications to field
theory. In: Quantum fields -- algebras, processes,
 ed.\ L.~Streit, pp.\ 239-268.  Springer, Wien, New York, 1980
\bibitem{R} J.E.\ Roberts: Lectures on algebraic quantum field theory. In: 
The algebraic theory of superselection sectors. Introduction and recent 
results, ed.\ D.\ Kastler, pp.\ 1-112. World Scientific, Singapore,
 New Jersey,  London, Hong Kong 1990
\bibitem{Robe1} J.E. Roberts: ``Some applications of dilation 
invariance to structural questions in the theory of local 
observables'', Commun. Math. Phys. {\bf 37}, 273 (1974)
\bibitem{Rohr} F. Rohrlich, ``Null plane field theory'', Acta
  Phys. Austriaca, Suppl. {\bf 8} (Conference Proc., Schladming 1971),
  227 (1971)  
\bibitem{Sew} G.L. Sewell, ``Quantum fields on manifolds: PCT and
  gravitationally induced thermal states'', Ann. Phys. (N.Y.) {\bf
  141}, 201 (1982)
\bibitem{SumVer} S.J. Summers, R. Verch, ``Modular inclusion, the
  Hawking temperature, and quantum field theory in curved spacetime'',
 Lett. Math. Phys. {\bf 37}, 145 (1996)
\bibitem{Tak} M. Takesaki, Tomita's theory of modular Hilbert-algebras 
and its applications. Lecture Notes in Mathematics Vol.\ 128 
Springer, Berlin-Heidelberg-New York 1970 
\bibitem{Ver1} R. Verch, ``Continuity of symplectically adjoint maps
  and the algebraic structure of Hadamard vacuum representations for
  quantum fields in  curved spacetime'', Rev. Math. Phys. {\bf 9}, 635
  (1997)
\bibitem{Ver2} R. Verch, ``Notes on regular diamonds'', preprint,
  available as ps-file at 
http://www.lqp.uni-goettingen.de/lqp/papers/
\bibitem{WaldI} R.M. Wald, General relativity, University of Chicago
  Press, 1984
\bibitem{Wald2} R.M. Wald, Quantum field theory in curved spacetime
  and black hole thermodynamics, University of Chicago Press, 1994
\bibitem{Wald3} R.M. Wald, ``Gravitational collapse and cosmic
  censorship'', gr-qc/9710068, to appear in ``The Black Hole Trail'', 
  ed. by B. Iyer.
\bibitem{Wies1} H.-W. Wiesbrock, ``Half-sided modular inclusions of
von Neumann algebras''  Commun. Math. Phys. {\bf 157},  83 (1993)

\bibitem{Wies4} H.-W. Wiesbrock, ``Conformal quantum field theory and
half-sided modular inclusions of
von Neumann algebras ''  Commun. Math. Phys. {\bf 158},  537 (1993)
\bibitem{Wies5} H.-W. Wiesbrock, ``Symmetries and modular
  intersections of von Neumann algebras'', Lett. Math. Phys. {\bf 39},
  203 (1997)
\bibitem{Wies6} H.-W. Wiesbrock, ``Modular intersections of von
  Neumann algebras is quantum field theory'', Commun. Math. Phys. {\bf
    193}, 269 (1998)
\bibitem{Zimm} R. Zimmer, ``Ergodic Theory of Semisimple Groups'', 
  Boston-Basel-Stuttgart: Birkh\"auser, 1984 
\end{thebibliography}
\end{document}